\documentclass{article}
\usepackage[utf8]{inputenc}
\usepackage{amssymb}
\usepackage{amsmath}
\usepackage[]{algorithm}
\usepackage{algpseudocode}
\usepackage{natbib}
\usepackage{graphicx}
\usepackage{subcaption}
\usepackage[margin=3cm]{geometry}
\usepackage{comment}
\usepackage{color}
\usepackage{bbm}
\usepackage{url}
\usepackage{ulem} 
\usepackage{tikz}

\title{Guided sequential ABC schemes for intractable Bayesian models}
\author{Umberto Picchini$^1$\footnote{picchini@chalmers.se}, Massimiliano Tamborrino$^2$}
\date{\small $^1$ Department of Mathematical Sciences, Chalmers University of Technology and the University of Gothenburg, 
Sweden.
\\
\small $^2$  Department of Statistics, University of Warwick, UK.
}

\usepackage{amssymb}
\usepackage{amsmath}
\usepackage[]{algorithm}
\usepackage{algpseudocode}
\usepackage{natbib}
\usepackage{graphicx}
\usepackage{subcaption}
\usepackage{comment}
\usepackage{color}
\usepackage{bbm}
\usepackage{url}
\usepackage{ulem} 
\usepackage{tikz}
\usepackage{amsthm}
\usepackage[colorlinks,citecolor=blue,urlcolor=blue,filecolor=blue,backref=page]{hyperref}
\usepackage{pdfpages}
\newtheorem{remark}{Remark}

\newcommand{\red}{\textcolor{black}}   

\newcommand{\brown}{\textcolor{black}}  

\begin{document}
\maketitle

\begin{abstract}
Sequential algorithms such as sequential importance sampling (SIS) and sequential Monte Carlo (SMC) have proven fundamental in Bayesian inference for models not admitting a readily available likelihood function. For approximate Bayesian computation (ABC),  SMC-ABC is the state-of-art sampler. However, since the ABC paradigm is intrinsically wasteful, sequential ABC schemes can benefit from well-targeted proposal samplers that efficiently avoid improbable parameter regions. We contribute to the ABC modeller's toolbox with novel proposal samplers that are conditional to summary statistics of the data. In a sense, the proposed parameters are ``guided'' to rapidly reach regions of the posterior surface that are compatible with the observed data. This speeds up the convergence of these sequential samplers, thus reducing the computational effort, while preserving the accuracy in the inference. We provide a variety of guided Gaussian and copula-based samplers for both SIS-ABC and SMC-ABC easing inference for challenging case-studies, including multimodal posteriors, highly correlated posteriors, hierarchical models with about 20 parameters, and a simulation study of cell movements using more than 400 summary statistics.
\end{abstract}

\noindent
\textbf{Keywords:} Approximate Bayesian computation; copulas; copulas; sequential importance sampling; simulation-based inference; sequential Monte Carlo.

\section{Introduction}

Approximate Bayesian computation (ABC) is arguably the most popular family of Bayesian samplers for statistical models characterized by intractable likelihood functions (\citealp{sisson2018handbook}). By this, we refer to many scenarios where the likelihood $p(y|\theta)$, for a dataset $y\in\mathcal{Y}$ and parameter $\theta$, is not available in closed-form or may be too cumbersome to evaluate computationally or even approximate. However, it is assumed that is feasible to simulate from the data-generating model  to produce  a simulated/synthetic dataset $z\sim p(z|\theta)$. The latter notation means that $z\in \mathcal{Y}$ has been implicitly generated by the likelihood function, that is, $p(z|\theta)$ is unavailable in closed form, but we can obtain simulated data $z$ from it. When the time to generate many simulated dataset is not prohibitive, ABC samplers exploit the information brought by many simulated dataset at different values of $\theta$ to learn an approximation of the posterior distribution. This is attained by comparing the several  synthetic dataset $z$ with the observed $y$, possibly by first reducing the data-dimension via informative summary statistics, and rejecting the values of $\theta$ yielding simulated data that are too different from the observations. 
The most basic ABC sampler is the so-called ABC-rejection sampler. This typically compares \textit{features} of the simulated and observed data by introducing summary statistics $S$ which reduce the dimension of $z$ as well as $y$ via $s:=S(z)$ and $s_y:=S(y)$, respectively (\citealp{pritchard1999population}). Ultimately, $s$ is compared to $s_y$, rather than comparing $z$ and $y$ directly, via a distance $||s-s_y||$ or 
a kernel function $K(||s-s_y||)$, see, e.g. \citealp{sisson2018handbook}. As an example, the ABC-rejection sampler works as follows: (i) a candidate parameter $\theta^*$ is proposed from the prior $\pi(\theta)$; (ii) corresponding simulated data are obtained as $z\sim p(z|\theta^*)$, and summaries $s^*=S(z^*)$ are obtained; (iii) 
accept and store $\theta^*$ if  $||s^*-s_y||<\delta$ for some $\delta>0$, otherwise discard it.
 Steps (i)-(iii) are iterated until $N$ accepted parameters are obtained. For the reader's benefit, a more general version of this algorithm is given in {Supplementary Material K}. 
 Accepted draws from ABC-rejection are then samples from the marginal posterior  
 \[\pi_\delta(\theta|s_y)\propto \int \mathbb{I}_{||s-s_y||<\delta}\cdot p(s|\theta)\pi(\theta)\mathrm{d}s\]
(with $\mathbb{I}_A$ being the indicator function returning 1 if $A$ is true and 0 otherwise).
The ABC-rejection sampler is particularly wasteful, as the prior is used as the parameter proposal. A number of improvements to this basic algorithm has been produced in the past 20 years and are considered in \cite{sisson2018handbook}. The most important alternative samplers are Markov chain Monte Carlo (MCMC) ABC (\citealp{marjoram2003markov,sisson2011likelihood,picchini2014inference}) and sequential Monte Carlo ABC (SMC-ABC, \citealp{sisson2007sequential}, \citealp{beaumont2009adaptive}, \citealp{del2012adaptive}).  SMC-ABC is especially popular due to  its somehow simpler tuning compared to MCMC-ABC. For example, SMC-ABC is the chosen algorithm in recent inference platforms such as \texttt{ABCpy} (\citealp{dutta2017abcpy}) and \texttt{pyABC} (\citealp{pyabc}).
However, ABC algorithms are generally computationally wasteful, making their use computationally challenging when simulating complex systems or inferring high-dimensional parameters. The goal of this work is to construct novel proposal functions for sequential ABC samplers (SMC-ABC and sequential importance sampling ABC), having the unique feature of being rapidly ``guided'' to target the region of the parameter space that is compatible with the observed data, hence considerably reducing the computational effort compared to non-guided proposals, especially in the initial iterations. 
 We achieve this by constructing  several proposal functions generating  parameters conditionally to summaries of the data $s_y$. In particular,  we construct Gaussian and copula-based proposal functions,  which we call ``guided proposals'' due to the explicit conditioning on $s_y$, and show how 
 our {methods notably increase the acceptance rate of proposed parameters while thoroughly exploring the posterior surface}. \\ 
\indent Previous work exploiting information from data summaries $s_y$ to adjust the output of an ABC procedure is, e.g., \cite{beaumont2002approximate}, \cite{blum2010non} and \cite{li2017extending}. However, these approaches  do not use $s_y$ to improve the \textit{proposal} sampler during a run of some ABC  algorithm, but only adjust the final output, thus acting on the already accepted parameters. Instead, our approaches  make use of $s_y$ to guide the parameter proposals while ABC is still running. A first work in this direction is that of \cite{BonassiWestBA2015}, where a joint product kernel for $(\theta,s_y)$ is considered to derive SMC-ABC with adaptive weights, with particles sampled from a proposal (kernel) based on $\theta$ and weights based on the kernel on $s_y$. While computationally convenient, this independence assumption does not reflect the intrinsic dependency between $\theta$ and $s_y$, which we instead consider when constructing samplers for $\theta|s_y$. A more recent work in
this direction is that of \cite{chen2019adaptive}, where ``regression adjustement''  (\citealp{beaumont2002approximate}, \citealp{blum2010non}) is performed by employing neural networks.  In particular, starting  from many pairs $(\theta^{(i)},s^{(i)})$ simulated from the prior-predictive distribution, they regression-adjust the accepted parameters as $\theta^{(i)'}=g(s_y)+\theta^{(i)}-g(s^{(i)})$, where $g(\cdot)$ is a function learned by training a neural-network via early-stopping on the prior-predictive simulations. Then,  they use a ``best subset'' of the adjusted parameters to construct a multivariate Gaussian sampler with mean $g(s_y)$ and an inflated covariance matrix based on the $\theta^{(i)'}$ to further propose parameters, which are then fitted using a Gaussian copula and appropriately transformed, to produce an approximated posterior as final output of the procedure.
\\ \indent While we also propose, among others, copula-based samplers and use the information of $s_y$  while ABC is running, there are several substantial differences with the approach proposed by \cite{chen2019adaptive}. First, we do not need to construct a neural network architecture,  hence we do not have to choose its design and the optimization schemes required for its tuning.   Our approach make use of the accepted draws to construct an effective proposal sampler, without regression adjustments,  as the latter are known to be effective only if the distribution of the residuals from the regression is roughly constant.  Second, for some of our novel guided proposals, we introduce the ability to exploit a-priori knowledge about correlations in blocks of parameters  by proposing from, say, $g(\theta_1,\theta_2|s_y,\theta_3^*,\theta_4^*)$ if $\theta=(\theta_1,...,\theta_4)$ with $(\theta_1,\theta_2)$ known to be highly correlated and $(\theta_3^*,\theta_4^*)$ being the third and fourth component of a previously accepted $\theta^*=(\theta_1^*,...,\theta_4^*)$. We have shown that this considerably ease exploration of highly correlated posteriors.   Third, we do not only consider Gaussian-copulas but also Student's t-copulas, exploring the performance of  a variety of possible suggested (instead of learned via kernel density estimation) marginals such as Gaussian, location-sclae Student's t,  Gumbel, logistic, uniform and triangular, recommending the last two among all. {Interestingly, the copula-based samplers are not derived starting from $(\theta,s_y)$ (as it happens  for the guided Gaussian samplers), but directly constructed for $\theta|s_y$. This makes them more general and flexible than the guided Gaussian proposals, derived assuming that the summary statistics are Gaussian distributed, something which may not always be true (despite this having a limited impact on the inference results, as shown here).}  Fourth, besides constructing guided proposals for SMC-ABC, most of our guided proposals are actually for sequential importance sampling ABC (SIS-ABC), and their performance is competitive, remarkably outperforming the state of art SMC-ABC. 
 Ultimately, we show that our guided proposals can dramatically accelerate convergence to the bulk of the posterior by increasing acceptance rates while preserving accuracy in the posterior inference. Moreover, our methods seem to offer a viable route for inference when the dimension of the summary statistics is large, whereas SMC-ABC without guided proposals may struggle. \\ 
\indent Other contributions for sequential likelihood-free Bayesian inference that are not within the ``standard ABC'' framework (i.e. procedures that are not necessarily based on specifying summary statistics that get compared via a threshold parameter $\delta$) are, e.g.,  \cite{papamakarios2016fast},
\cite{lueckmann2017flexible},
\cite{papamakarios2019sequential}, \cite{greenberg2019automatic}, \cite{wiqvist2021sequential}, with ensemble Kalman inversion (EKI) \citep{chada,duffield2022ensemble} being a further possibility. While these works have notable merits, we do not consider them here, as our goal is to improve the SIS-ABC and SMC-ABC samplers. However, our guided proposals may be incorporated into other samplers, e.g. EKI.  In Section \ref{ref:abc-sss}, we briefly summarize sequential ABC samplers before introducing our original proposal functions in Section \ref{sec:guided-samplers}. Simulation studies are reported in Section  \ref{sec:examples},  where we consider examples with multimodal posteriors (Section \ref{sec:twomoons}), highly non-Gaussian summary statistics (Sections \ref{sec:twomoons} and \ref{sec:recruitment}),  highly correlated posteriors (section \ref{sec:twisted}), hierarchical models with
high-dimensional parameters (Section \ref{sec:g-and-k}), \textcolor{black}{continuous time Markov jump processes with an application in ecology and systems biology (Section \ref{sec:lotka-volterra})}  and 
high-dimensional summaries with up to 400 components (Sections \ref{sec:g-and-k} and \ref{sec:cell}). Finally, additional theoretical considerations, further  results from the examples and setups for the simulation studies are provided as Supplementary Material. Supporting MATLAB code is available at \url{https://github.com/umbertopicchini/guidedABC}.

\section{Sequential ABC schemes}\label{ref:abc-sss}

Throughout this section, we summarize the key features of two of the most important sequential ABC algorithms, i.e.  SIS-ABC and SMC-ABC, since  they are at the core of our novel contributions. 
\subsection{Sequential importance sampling ABC}\label{ref:abc-rej-is}

 An obvious improvement to the basic ABC-rejection considers proposing parameters from a carefully constructed  ``importance sampler'' $g(\theta)$. In this procedure, each accepted parameter receives a weight $\pi(\theta)/g(\theta)$ to correct for the fact that the parameter was not proposed from $\pi(\theta)$, and the weight correction provides (weighted) samples from the corresponding approximate posterior $\pi_\delta(\theta|s_y)$. When $g(\theta)\equiv \pi(\theta)$, importance sampling ABC reduces to the ABC-rejection algorithm.
Importance sampling is particularly appealing when embedded into a series of $T$ iterations, see Algorithm \ref{alg:sis-abc}, which can also suggest ways to employ a decreasing sequence of thresholds $\delta_1<\delta_2<\cdots <\delta_T$ (more on this later). This implies the introduction of $T$ importance distributions $g_1,...,g_T$, with  $g_1(\theta)\equiv \pi(\theta)$. 
The output of the algorithm provides us with weighted samples from the last iteration: $(\theta^{(1)}_T,\tilde{w}^{(1)}_T),\ldots,(\theta^{(N)}_T,\tilde{w}^{(N)}_T)$, where $\theta^{(i)}_T\sim \pi_{\delta_T}(\theta|s_y)$. These can be used to compute the ABC posterior mean as $E(\theta|s)\approx \sum_{i=1}^N w^{(i)}_T\theta^{(i)}_T$ (where the $w_t$ denote normalized weights, i.e. $w^{(i)}_t=\tilde{w}^{(i)}_t/\sum_{j=1}^N \tilde{w}^{(j)}_t$) and posterior quantiles by using eq. (3.6) in \cite{chen1999monte}. Otherwise, perhaps more practically, it is possible to sample $N$ times with replacement from the set $(\theta^{(1)}_T,\ldots,\theta^{(N)}_T)$, using probabilities given by the normalized weights, and then produce histograms, or compute posterior means and quantiles by using the resampled particles (after resampling, all samples have the same normalized weight $1/N$). 
However, Algorithm \ref{alg:sis-abc} does not clarify the most important issue: how to \textit{sequentially} construct the samplers $g_1,...,g_T$. 
Inspired by work on sequential learning  (\citealp{cappe2004population}, \citealp{del2006sequential}), a number of sequential ABC samplers have been produced, such as \cite{sisson2007sequential}, \cite{toni2008approximate}, \cite{beaumont2009adaptive}, \cite{del2012adaptive}. We collectively refer to these methods as SMC-ABC, as described in Section \ref{sec:abc-smc}.

\begin{algorithm}[t!]
\scriptsize
\caption{\bf Sequential importance sampling ABC (SIS-ABC)}
\label{alg:sis-abc}
\begin{algorithmic}[1]
\State\textcolor{black}{\noindent {\it Input:}\\
Observed data $y$, vector of summary statistics $S(\cdot)$, number of kept samples per iteration $N$, prior $\pi(\theta)$, importance sampler $g_t$,  number of iterations $T$ and starting threshold $\delta_1$.}
\State Set $t:=1$.
 \For{$i=1, \ldots, N$}
 \Repeat
\State Sample $\theta^{*}\sim \pi(\theta)$.
\State Generate $z^i\sim p(z|\theta^{*})$ from the model. 
\State Compute summary statistic $s^i=S(z^i)$.
\Until{$||s^i-s_y||<\delta_1$}
\State Set $\theta^{(i)}_1:=\theta^*$ 
\State set $\tilde{w}^{(i)}_1:=1$.
\EndFor
\State Obtain $\delta_2$.
 \For{$t=2, \ldots, T$}
 \For{$i=1, \ldots, N$}
 \Repeat
\State Sample $\theta^{*}\sim g_t(\theta)$.
\State if $\pi(\theta^*)=0$ go to step 15, otherwise continue.
\State Generate $z^i\sim p(z|\theta^{*})$ from the model. 
\State Compute summary statistic $s^i=S(z^i)$.
\Until{$||s^i-s_y||<\delta_t$}
\State Set $\theta^{(i)}_t:=\theta^*$ 
\State set $\tilde{w}^{(i)}_t=\pi(\theta^{(i)})/g_t(\theta^{(i)})$.
\EndFor
\State Decrease the current $\delta_t$.
\EndFor
\State
\noindent {\it Output:}\\
A set of weighted parameters $(\theta^{(1)}_T,\tilde{w}^{(1)}_T),\ldots,(\theta^{(N)}_T,\tilde{w}^{(N)}_T)\sim \pi_{\delta_T}(\theta|s_y)$.
\end{algorithmic}
\end{algorithm} 

\subsection{Sequential Monte Carlo ABC samplers}\label{sec:abc-smc}

At iteration $t$, SMC-ABC constructs automatically tuned proposal samplers by
either considering ``global'' features of the collection of $N$ accepted samples from the previous iteration $(\theta^{(1)}_{t-1},\tilde{w}^{(1)}_{t-1}),\ldots,$ $(\theta^{(N)}_{t-1},\tilde{w}^{(N)}_{t-1})$, or ``local'' features that are specific to each sample $(\theta_{t-1}^{(i)},\tilde{w}_{t-1}^{(i)})$. In this framework, a sample is traditionally named ``particle''.
For example, a global feature could be the (weighted) sample covariance $\Sigma_{t-1}$ of the particles $(\theta^{(1)}_{t-1},...,\theta^{(N)}_{t-1})$, and this could be used in a sampler to propose particles at iteration $t$. More generally, if we consider the importance sampler as $g_t(\theta)\equiv g_t(\theta|\theta_{t-1}^{(1)},...,\theta_{t-1}^{(N)})$, we can set  $g_t(\theta)=q(\theta|\mu_{t-1}, \Sigma_{t-1})$, where $\Sigma_{t-1}$ is the previously defined covariance matrix and $\mu_{t-1}$ is some central location of the particles at iteration $t-1$. A reasonable and intuitive choice is  (with some abuse of notation) $g_t(\theta)=q(\theta|\mu_{t-1}, \Sigma_{t-1})\equiv \mathcal{N}(\mu_{t-1}, \Sigma_{t-1})$, where $\mathcal{N}(a,b)$ is the $N$-dimensional Gaussian distribution with mean $a$ and covariance matrix $b$. However, this would create a global sampler that may only be appropriate if the targeted posterior is approximately Gaussian.\\
\indent 
In SMC-ABC, the game-changer idea is the random sampling of a particle (with replacement) from the $N$ particles with associated normalized weights $(\theta^{(1)}_{t-1},{w}^{(1)}_{t-1}),\ldots$, $(\theta^{(N)}_{t-1},{w}^{(N)}_{t-1})$: call the sampled particle $\theta^*_{t-1}$, and then randomly ``perturb''  it to produce a $\theta^{**}_t\sim q_t(\cdot|\theta^*_{t-1})$ (notice the latter notation does not exclude dependence on other particles as well).  The perturbed proposal $\theta^{**}_t$ may be accepted or not according to the usual ABC criterion. The procedure is iterated until $N$ proposals are accepted at each iteration.
The SMC-ABC algorithm is exemplified in Algorithm \ref{alg:smc-abc}, where the key step of sampling a particle based on its weight is in step 15, and its perturbed version is in step 16. 
\begin{algorithm}[t!]
\scriptsize
\caption{\bf Sequential Monte Carlo ABC (SMC-ABC)}
\label{alg:smc-abc}
\begin{algorithmic}[1]
\State\textcolor{black}{\noindent {\it Input:}\\
Observed dataset $y$, summary statistics $S(\cdot)$, number of kept samples per iteration $N$, prior $\pi(\theta)$, a perturbation sampler $q_t$,  number of iterations $T$ and starting threshold $\delta_1$.}\State Set $t:=1$.
 \For{$i=1, \ldots, N$}
 \Repeat
\State Sample $\theta^{*}\sim \pi(\theta)$.
\State Generate $z^i\sim p(z|\theta^{*})$ from the model. 
\State Compute summary statistic $s^i=S(z^i)$.
\Until{$||s^i-s_y||<\delta_1$}
\State Set $\theta^{(i)}_1:=\theta^*$ 
\State set $\tilde{w}^{(i)}_1:=1$.
\EndFor
\State Obtain $\delta_2$.
 \For{$t=2, \ldots, T$}
 \For{$i=1, \ldots, N$}
 \Repeat
\State Randomly pick (with replacement) $\theta^{*}$ from the weighted set $\{\theta^{(i)}_{t-1},w^{(i)}_{t-1}\}_{i=1}^N$.
\State Perturb $\theta^{**}\sim q_t(\cdot|\theta^*)$.
\State if $\pi(\theta^{**})=0$ go to step 15, otherwise continue.
\State Generate $z^i\sim p(z|\theta^{**})$ from the model. 
\State Compute summary statistic $s^i=S(z^i)$.
\Until{$||s^i-s_y||<\delta_t$}
\State Set $\theta^{(i)}_t:=\theta^{**}$ 
\State set $\tilde{w}^{(i)}_t = \pi(\theta^{(i)}_t)/\sum_{j=1}^N {w}_{t-1}^{(j)}q_t(\theta^{(i)}_t|\theta^{(j)}_{t-1})$.
\EndFor
\State normalize the weights: $w^{(i)}_t:=\tilde{w}^{(i)}_t/\sum_{j=1}^N \tilde{w}^{(j)}_t$.
\State Decrease the current $\delta_t$.
\EndFor
\State
\noindent {\it Output:}\\
A set of weighted parameters $(\theta^{(1)}_T,\tilde{w}^{(1)}_T),\ldots,(\theta^{(N)}_T,\tilde{w}^{(N)}_T)\sim \pi_{\delta_T}(\theta|s_y)$.
\end{algorithmic}
\end{algorithm} 
In summary, an accepted particle $\theta^{(i)}_t$ results out of: (i) a randomly drawn particle from the set at iteration $t-1$ using probabilities (normalized weights) $\{w_{t-1}^{(i)}\}_{i=1}^N$;  (ii) an additional perturbation of the particle sampled in (i). Which means that the proposal distribution for the particle $\theta^{(i)}$ is an $N$-components mixture distribution with mixing probabilities $\{w_{t-1}^{(j)}\}_{j=1}^N$ and components $q_t(\theta^{(i)}|\theta_{t-1}^{(j)})$ ($j=1,...,N$), that is\\ $g_t(\theta^{(i)})=\sum_{j=1}^N {w}_{t-1}^{(j)}q_t(\theta^{(i)}_t|\theta^{(j)}_{t-1})$, which motivates the importance weight 
\begin{equation}
\tilde{w}^{(i)}_t = \pi(\theta^{(i)}_t)/\sum_{j=1}^N {w}_{t-1}^{(j)}q_t(\theta^{(i)}_t|\theta^{(j)}_{t-1}), \qquad i=1,...,N.
\label{eq:importance-weight}
\end{equation}
Notice that, in  \eqref{eq:importance-weight}, the denominator has to be evaluated anew for each accepted $\theta^{(i)}$. One of the measures of the effectiveness of an importance or an SMC sampler is the ``effective sample size'' (ESS), where $1\leq \mathrm{ESS}\leq N$, the larger the ESS the better. At iteration $t$, this is traditionally approximated as $\mathrm{\widehat{ESS}}_t=1/\left(\sum_{j=1}^N (w_t^{(j)})^2\right)$, however alternatives are explored in \cite{martino2017effective}.\\
\indent Some notable SMC-ABC proposal samplers for $\theta_t^{**}\sim q_t(\cdot|\theta_{t-1}^*)$ are detailed in the Supplementary Material L.
Here, we briefly recall the two samplers that provide a useful comparison with our novel methods introduced in Sections \ref{sec:guided-samplers}. Some of the most interesting work for the construction of such samplers is in \cite{filippi2013optimality}, where particles randomly picked from the previous iterations are perturbed with Gaussian proposals with several proposed tuning of the covariance matrix, improving on  \cite{beaumont2009adaptive}. 
As a first approach, \cite{filippi2013optimality} propose the full $\theta^{**}$ by generating at iteration $t$ from the $d_\theta$-dimensional Gaussian $q(\theta|\theta^*)\equiv \mathcal{N}(\theta^*,2\Sigma)$, with $\Sigma$ being the empirical weighted covariance matrix from the particles accepted at iteration $t-1$. This results in the importance weights \eqref{eq:importance-weight} becoming (with the usual notation abuse)
$\tilde{w}^{(i)}_t = \pi(\theta^{(i)}_t)/\sum_{j=1}^N {w}_{t-1}^{(j)}\mathcal{N}(\theta^{(i)}_t;\theta^{(j)}_{t-1},2\Sigma_{t-1})$, where $\mathcal{N}(x;a,b)$ denotes the probability density function of a Gaussian distribution with mean $a$ and covariance matrix $b$ evaluated at $x$. 
In our experiments, this proposal embedded into SMC-ABC is denoted \texttt{standard}, being in some way a baseline approach. Having $\theta^{**}_t\sim \mathcal{N}(\theta^*_{t-1},2\Sigma_{t-1})$ implies that the mean of the proposal sampler exploits ``local'' features, since perturbed draws lie within an ellipsis centred in  $\theta^{*}_t$, while its covariance matrix is still global. To obtain a ``local'' covariance,
\cite{filippi2013optimality}  proposed the  ``optimal local covariance matrix'' (\texttt{olcm}) sampler. The key feature of this sampler is that each proposed particle $\theta^{**}$ arises from perturbing $\theta^*$ using a Gaussian distribution having a covariance matrix that is specific to $\theta^*$ (hence ``local''), rather than being tuned on all particles accepted at $t-1$ as in the \texttt{standard} sampler. To construct the \texttt{olcm},  \cite{filippi2013optimality} define the following weighted set of particles of size $N_0\leq N$ at iteration $t-1$
\begin{equation}
\{\tilde{\theta}_{t-1,l},\gamma_{t-1,l}\}_{1\leq l\leq N_0}=\biggl\{\biggl({\theta}_{t-1}^{(i)}, \frac{w_{t-1}^{(i)}}{\bar{\gamma}_{t-1}} \biggr), \text{ s.t. } ||s^{i}_{t-1}-s_y||<\delta_t, \quad i=1,...,N \biggr\}
,\label{eq:N0-particles}
\end{equation}
where $\bar{\gamma}_{t-1}$ is a normalisation constant such that $\sum_{l=1}^{N_0}\gamma_{t-1,l}=1$.
That is, the $N_0$ weighted particles are the subset of the $N$ particles accepted at iteration $t-1$ (when using $\delta_{t-1}$) having generated summaries that produce distances that are \textit{also} smaller than $\delta_t$.
We used the letter $\gamma$ to denote normalized weights associated to the subset of $N_0$ particles, rather than $w$, to avoid confusion. At iteration $t$, the \texttt{olcm} proposal is 
$q_t(\theta|\theta^*)=\mathcal{N}(\theta^*,\Sigma^{\mathrm{olcm}}_{\theta^*})$, 
where
$
\Sigma^{\mathrm{olcm}}_{\theta^*} = \sum_{l=1}^{N_0}\gamma_{t-1,l}(\tilde{\theta}_{t-1,l}-\theta^*)(\tilde{\theta}_{t-1,l}-\theta^*)'$, where $'$ denotes transposition throughout our work. Accepted particles are then given unnormalized weights
$\tilde{w}^{(i)}_t = \pi(\theta^{(i)}_t)/\sum_{j=1}^N {w}_{t-1}^{(j)}\mathcal{N}(\theta^{(i)}_t;\theta^{(j)}_{t-1},\Sigma^{\mathrm{olcm}}_{\theta^{(j)}})$.
 Note that we are required to store the $N$ distances $||s^{i}_{t-1}-s_y||$ accepted at iteration $t-1$ to be able to determine which indices $i$ have distance smaller than the $t$-th threshold $\delta_t$, i.e. $||s^{i}_{t-1}-s_y||<\delta_t$.
Evaluating $\Sigma^{\mathrm{olcm}}_{\theta^{*}}$ causes some non-negligible overhead in the computations, since it has to be performed for every proposal parameter, and therefore sanity checks are required at every proposal to ensure that it results in a  positive definite covariance matrix, or otherwise computer implementations will halt with an error. We postpone such discussion in Section \ref{sec:guided-smc-abc}, where it is better placed, and introduce now our novel proposal samplers.

\section{Guided sequential samplers}\label{sec:guided-samplers}

Here, we describe our main contributions to improve the efficiency of both SIS-ABC and SMC-ABC samplers by constructing proposals that are conditional on observed summaries $s_y$, to guide the particles by using information provided by the data.  In particular, we  create proposal samplers of type $g(\theta|s_y)$ (when the entire vector parameter is proposed in block) or even conditional proposals of type $g(\theta_k|\theta_{-k},s_y)$, where $\theta_k$ is the $k$-th component of a $d_\theta$-dimensional vector parameter $\theta$ and $\theta_{-k}=(\theta_1,...,\theta_{k-1},\theta_{k+1},...,\theta_{d_\theta})$.
The starting idea behind the construction of these samplers is loosely inspired by \cite{picchini2020adaptive}, where a guided Gaussian proposal is constructed for MCMC inference using synthetic likelihoods (\citealp{wood2010statistical,price2018bayesian}). This initial inspiration is significantly expanded in multiple directions, e.g. by providing strategies to avoid the ``mode-seeking'' behaviour of the basic guided proposals, by constructing the above mentioned proposals of type $g(\theta_k|\theta_{-k},s_y)$ and by introducing non-Gaussian proposal samplers using copulas with a variety of marginal structures, yielding an ABC methodology capable to tackle considerably challenging problems. To help the reader navigate through the several proposals samplers introduced in this work, we classify them in Table \ref{table:samplers}.

\begin{table}[t!]
\footnotesize
\centering
\begin{tabular}{l|c|c|c|c|c}
Proposal sampler & Guided & Family & Global or local & Algorithm & Section  \\
    \hline
    \texttt{standard} & no & Gaussian & local mean, global cov & SMC-ABC & \ref{sec:abc-smc} \\
    \texttt{olcm} & no & Gaussian & local mean, local cov & SMC-ABC & \ref{sec:abc-smc} \\
    \hline
    \texttt{blocked} & yes & Gaussian & global mean, global cov & SIS-ABC & \ref{sec:basic-guided-SIS} \\
    \texttt{blockedopt} & yes & Gaussian & global mean, global cov & SIS-ABC & \ref{sec:guided--local-sis-abc} \\
      \texttt{hybrid} & yes & Gaussian & global mean, global cov & SIS-ABC & \ref{sec:guided--hybrid-sis-abc}\\
\hline 
\texttt{cop-blocked} & yes & Gaussian or t copula & global mean, global cov & SIS-ABC & \ref{sec:copula-sis} \\
\texttt{cop-blockedopt} & yes & Gaussian or t copula & global mean, global cov & SIS-ABC & \ref{sec:copula-sis} \\
      \texttt{cop-hybrid} & yes & Gaussian or t copula & global mean, global cov & SIS-ABC & \ref{sec:copula-sis}\\
      \hline 
  \texttt{fullcond} & yes & Gaussian & local mean, global var & SMC-ABC & \ref{sec:guided-smc-abc}\\
    \texttt{fullcondopt} &  yes & Gaussian & local mean, local var & SMC-ABC & \ref{sec:guided-optimal}\\
    \hline
\end{tabular}
\caption{Proposal samplers considered in this work, categorized according to whether the sampler is guided by data or not, its distributional family, whether the sampler has features that are specific for each particle (local) or common to all particles (global), the type of algorithm employing the proposal sampler and the relevant section introducing it. All listed samplers are novel, except for \texttt{standard} and \texttt{olcm}.}
\label{table:samplers}
\end{table}


In Section \ref{sec:basic-guided-SIS}, we consider  a first approach for a guided Gaussian proposal function for SIS-ABC, whose covariance matrix is in some sense optimized in Section \ref{sec:guided--local-sis-abc}. A hybrid approach combining the two is proposed in Section \ref{sec:guided--hybrid-sis-abc}. Then,  Section \ref{sec:copula-sis} generalizes the guided Gaussian approaches for SIS-ABC to copula-based proposals, {constructing a sampler directly for $\theta|s_y$ instead of deriving it starting from a Gaussian sampler $g(\theta,s_y)$. }{Finally, in Section \ref{sec:guided-smc-abc}, we propose a guided Gaussian SMC-ABC scheme, whose covariance matrix is then optimized in Section \ref{sec:guided-optimal}.} 
\subsection{A first guided Gaussian SIS-ABC sampler}\label{sec:basic-guided-SIS}

In \cite{picchini2020adaptive}, an MCMC proposal sampler was constructed to aid inference via synthetic likelihoods (\citealp{wood2010statistical}, \citealp{price2018bayesian}). There, the idea was to collect the several (say \red{$L$}) model-simulated summary statistics \red{$\{s^i\}_{i=1}^L$} that were generated at a given proposed parameter $\theta$, average them to obtain $\red{\bar{s}=\sum_{i=1}^L s^i/
L}$ so that, by appealing to the Central Limit Theorem, for \red{$L$} ``large'' enough, $\bar{s}$ was approximately Gaussian distributed. \cite{picchini2020adaptive} then observed that if the joint $(\theta,\bar{s}^i)$ is approximately multivariate Gaussian, then a conditionally Gaussian $g(\theta|s=s_y)$ can be constructed and  used as a proposal sampler. \\
\indent Let us now set $d_\theta=\dim(\theta), d_s=\dim(s_y), d=d_\theta+d_s$ and denote by $(\theta^{(i)},s^{(i)})$ a $d$-dimensional particle that has been accepted at iteration $t-1$ of SIS-ABC. Assume for a moment that $(\theta^{(i)},{s}^{(i)})$ is a $d$-dimensional Gaussian distributed vector $(\theta^{(i)},{s}^{(i)})\sim \mathcal{N}_d(m,S)$. We stress that this assumption is made merely to construct a proposal sampler, and does not extend to the actual distribution of $(\theta^{(i)},s^{(i)})$.
We set a $d$-dimensional mean vector $m\equiv(m_\theta,m_s)$ and the $d\times d$ covariance matrix  
\[
{S}\equiv \left[\begin{array}{cc}
{S}_{\theta} & {S}_{\theta s}   \\
{S}_{s\theta} & {S}_{s} \\
\end{array}\right],
\]
where ${S}_{\theta}$ is $d_\theta\times d_\theta$, ${S}_{s}$ is $d_s\times d_s$, ${S}_{\theta s}$ is $d_\theta \times d_s$ and of course ${S}_{s\theta}\equiv {S}_{\theta s}'$ is $d_s \times d_\theta$.
Once all $N$ accepted particles have been collected for iteration $t-1$ of the SIS-ABC sampler, we estimate $m$ and $S$ using the accepted (weighted) particles. That is, denote by  $x^{(i)}_{t-1}:=(\theta^{(i)}_{t-1},{s}^{(i)}_{t-1})$ a $d$-dimensional particle accepted at iteration $t-1$. By using their normalized weights, we have the following estimated weighted mean and weighted covariance matrix
\begin{equation}
\hat{m}_{t-1} = \sum_{i=1}^N{w^{(i)}_{t-1}}x^{(i)}_{t-1},\qquad  
\hat{S}_{t-1} = \frac{1}{(1-\sum_{i=1}^N {w_{t-1}^{(i)}}^2)}\sum_{i=1}^{N}w^{(i)}_{t-1}(x_{t-1}^{(i)}-\hat{m}_{t-1})(x^{(i)}-\hat{m}_{t-1})'\label{eq:m-S}
\end{equation}
 (snippets of vectorized code to efficiently compute \eqref{eq:m-S} are reported in the Supplementary Material {D}).
Once $\hat{m}_{t-1}$ and $\hat{S}_{t-1}$ are obtained, it is possible to extract the corresponding entries  $\hat{m}_\theta$, $\hat{m}_s$ and $\hat{S}_\theta$, $\hat{S}_s$, $\hat{S}_{s\theta}$, $\hat{S}_{\theta s}$, where we have disregarded the ``$t-1$'' iteration subscript to simplify the notation. We can then use well known formulas for the conditional distributions of a multivariate Gaussian, to obtain a proposal distribution for iteration $t$ given by $g_{t}(\theta|s_y)\equiv \mathcal{N}(\hat{m}_{\theta|s_y,t-1},\hat{S}_{\theta|s_y,t-1})$, with 
\begin{align}
    \hat{m}_{\theta|s_y,t-1} &= \hat{m}_{\theta}+\hat{S}_{\theta s}(\hat{S}_s)^{-1}(s_y-\hat{m}_s) \label{eq:mu-conditional}\\
    \hat{S}_{\theta|s_y,t-1} &= \hat{S}_{\theta}-\hat{S}_{\theta s}(\hat{S}_{s})^{-1}\hat{S}_{s \theta},\label{eq:cov-conditional}
\end{align}
and weights \eqref{eq:importance-weight} given by $\tilde{w}^{(i)}_t=\pi(\theta_t^{(i)})/\mathcal{N}(\hat{m}_{\theta|s_y,t-1},\hat{S}_{\theta|s_y,t-1})$, $i=1,...,N$. 
Hence, a parameter proposal for Algorithm \ref{alg:sis-abc} can be generated as $\theta^*\sim \mathcal{N}(\hat{m}_{\theta|s_y,t-1},\hat{S}_{\theta|s_y,t-1})$, which has an explicit ``guiding'' term $(s_y-\hat{m}_s)$. We call $g_t(\theta|s_y)$ a ``guided SIS-ABC sampler''. To distinguish it from other guided SIS-ABC samplers we introduce later, this one is named \texttt{blocked}, since all coordinates of $\theta^*$ are proposed jointly, hence ``in block''.\\
\indent 
Note that the guiding term $(s_y-\hat{m}_s)$ becomes less and less relevant as $\hat{m_s}\approx s_y$, which is supposed to happen when $\delta$ is small enough.
Of course, a concern around the efficacy of such sampler may arise if the joint $(\theta,s)$ is not approximately multivariate Gaussian. The case study in Sections \ref{sec:twomoons} and \ref{sec:recruitment}  (and in Supplementary Material {E and H})  have markedly non-Gaussian summary statistics, but our guided sampler behaves well. 
However, an undesirable feature is that it may occasionally display ``mode-seeking'' behaviour, i.e., it may rapidly approach high posterior density areas, quickly accepting promising particles in the initial iterations, but it may ending up exploring mostly the area around the posterior mode and not necessarily the tails of the targeted distribution. We address this issue in the next section by tuning the covariance matrix of the sampler while preserving its ``guided'' feature.

\subsection{Guided Gaussian SIS-ABC sampler with optimal local covariance}\label{sec:guided--local-sis-abc}

Proposal samplers can be designed according to several intuitions. For example, they could be based on the similarity with the targeted density, as suggested by a small Kullback-Leibler (KL) divergence while maximizing the acceptance probability as in  \citealp{filippi2013optimality}, or based on minimizing the variance of the importance weights while maximizing the acceptance probability as in \cite{alsing2018optimal}, or using guided approaches as those outlined before. Here, we let the mean of a Gaussian sampler to be the same as in  \eqref{eq:mu-conditional}, hence the proposal is guided by the observed summaries, but we construct the proposal covariance matrix following a reasoning inspired by \cite{filippi2013optimality} for SMC-ABC, and extended here to accommodate SIS-ABC.\\
\indent 
For the ``target'' distribution denoted by $q^*_{\delta_{t}}(\theta^*)=\pi_{\delta_t}(\theta^{*}|s_y)$, we wish to determine a proposal $q_{\delta_{t}}(\theta^*)$ by minimizing the KL divergence $KL(q_{\delta_{t}},q^*_{\delta_{t}})$ between its arguments, where
\[
KL(q_{\delta_{t}},q^*_{\delta_{t}}) = \int q^*_{\delta_{t}}(\theta^*)  \log\frac{q^*_{\delta_{t}}(\theta^*)}{q_{\delta_{t}}(\theta^*)}d\theta^*.  
\]
In the context of SMC-ABC, which we discuss later in Section \ref{sec:guided-optimal}, \cite{filippi2013optimality} mention that it is possible to consider a ``multi-objective optimization'' problem where, in addition to minimize the KL divergence, the maximization of an ``average acceptance probability'' is also carried out. By transposing their reasoning to our SIS-ABC context, the multi-objective optimization is equivalent to maximizing the following quantity
\begin{equation}
    Q(q_t,\delta_t,s_y)=\int  \pi_{\delta_t}(\theta^{*}|s_y)\log q_t(\theta^{*})d\theta^*
\end{equation}
 with respect to $q_t$. By considering $q_t(\theta^{*})\equiv \mathcal{N}(\hat{m}_{\theta|s_y,t-1},\Sigma_t)$ for unknown $\Sigma_t$ and fixed $\hat{m}_{\theta|s_y,t-1}$ defined as in \eqref{eq:mu-conditional}, 
the resulting maximization in $\Sigma_t$ leads to (details are in Supplementary Material {A})
\begin{equation}\label{CovSISABC}
\Sigma_t=\int  \pi_{\delta_t}(\theta^{*}|s_y)(\theta^*-\hat{m}_{\theta|s_y,t-1})'(\theta^*-\hat{m}_{\theta|s_y,t-1}) d\theta^*.
\end{equation}
The latter integral can be approximated following the same reasoning for \texttt{olcm}, detailed in the Supplementary Material {A}. Namely,  we can consider the $N$ particles accepted at iteration $t-1$ (which used a threshold $\delta_{t-1}$) and select from those the $N_0\leq N$ particles that produced a distance $||s-s_y||$ that is \textit{also} smaller than $\delta_t$, to obtain
$
\hat{\Sigma}_{t} = \sum_{l=1}^{N_0}\gamma_{t-1,l}(\tilde{\theta}_{t-1,l}-\hat{m}_{\theta|s_y,t-1})(\tilde{\theta}_{t-1,l}-\hat{m}_{\theta|s_y,t-1})'
$,
with $(\tilde{\theta},\gamma)$ defined as in \eqref{eq:N0-particles}.  We call \texttt{blockedopt} a SIS-ABC sampler  with  $g_{t}(\theta|s_y)\equiv \mathcal{N}(\hat{m}_{\theta|s_y,t-1},\hat{\Sigma}_t)$ as guided proposal. Importantly, notice that unlike \texttt{olcm}, the covariance matrix $\hat{\Sigma}_t$ is ``global'', which means that at each iteration only one covariance matrix needs to be computed and used for all proposed particles. This relieves the computational budget from the necessity to ensure (via Cholesky decomposition and in case of numerical issues, a more expensive modified-Cholesky decomposition as in \citealp{higham1988computing}) that the particle-specific covariance matrix is positive definite for each proposed particle. Of course, there is appeal in having a particle-specific ``local'' covariance matrix, and this is explored in Section \ref{sec:guided-smc-abc}.

\subsection{Hybrid guided Gaussian SIS-ABC sampler}\label{sec:guided--hybrid-sis-abc}

 We consider a further type of guided Gaussian SIS-ABC sampler, which we denote \texttt{hybrid}, which is a by-product of the \texttt{blocked} and \texttt{blockedopt} strategies. At iteration $t=1$, the \texttt{hybrid} sampler proposes from the prior, as in all examined strategies. At $t=2$, it proposes using \texttt{blocked} for rapid convergence towards the modal region, while for $t>2$, it uses \texttt{blockedopt}  to correct for possible mode-seeking behaviour.

\subsection{Guided copula-based  SIS-ABC samplers}\label{sec:copula-sis}

The previously proposed guided {Gaussian} SIS-ABC samplers assume {an underlying Gaussian distribution for $(\theta,s_y)$, which is then used to construct} the proposal  $g_t(\theta|s_y)$, which follows a $d_\theta$-dimensional Gaussian distribution with a certain mean vector $m^*$ and covariance matrix $S^*=(S^*_{ij})_{i,j=1}^d$. 
This can be generalised {by dropping the assumption of Gaussianity on $(\theta,s_y)$, making the samplers $g_t(\theta|s_y)$} more flexible, both in terms of joint distribution and marginals. To do this, we use copulas, proposing what we call \texttt{cop-blocked, cop-blockedopt, cop-hybrid}, the copula-based versions of \texttt{blocked, blockedopt} and \texttt{hybrid}, respectively, {with a few key differences though, as discussed in Remarks \ref{Remark1} and \ref{Remark2} below}. The idea of using copula modelling within ABC has been proposed e.g. by  \cite{li2017extending}, who use a Gaussian copula for approximating the ABC posterior for a high-dimensional parameter space, and by \cite{an2018robust} for modelling a high-dimensional summary statistic. Instead, we 
use (Gaussian and t) copulas for the proposal sampler {$g_t(\theta|s_y)$ (and not for the kernel for $(\theta,s_y))$}. \\
\indent
Suppose that the $d_\theta$-dimensional random vector $X=(X_1,\ldots, X_{d_\theta})$ has a joint cumulative distribution function $H$ and continuous marginals $F_1,\ldots, F_{d}$, i.e., $X\sim H(F_1(x_1),\ldots, F_d(x_{d_\theta}))$. Sklar's theorem (\citealp{Sklar1959}) states that the joint distribution $H$ can be rewritten in terms of ${d_\theta}$ uniform marginal distributions and a multivariate copula function $C:[0,1]^{d_\theta}\to [0,1]$ that describes the correlation structure between
them, i.e.
\[
H(x_1,\ldots, x_{d_\theta})=C(u_1,\ldots, u_{d_\theta})=C(F_1(x_1),\ldots, F_{d_\theta}(x_{d_\theta})),
\]
with $u_j=F_j(x_j), j=1,\ldots, {d_\theta}$. Hence, a copula is a multivariate distribution with uniform marginals. Denoting by $h$, $f_j$ and $c$ the densities of the joint distribution $H$, the marginal $F_j$ and the copula $C$, respectively, we have
\begin{equation}\label{copuladensity}
h(x_1,\ldots, x_{d_\theta})=c(u_1,\ldots, u_{d_\theta})\prod_{i=1}^{d_\theta} f_j(x_j), \qquad j=1,\ldots, {d_\theta},
\end{equation}
where 
$
c(u_1,\ldots,{d_\theta})=\frac{\partial^{d_\theta}}{\partial u_1\cdots\partial u_{d_\theta}}C(U_1,\ldots,U_{d_\theta}).
$
As copula families $C$, we consider the Gaussian and the $t$ copulas, corresponding on having the joint distribution $H$ to be a multivariate Gaussian and a Student's $t$, respectively. While the Gaussian copula is fully characterised by a correlation {matrix $R$}, the $t$ copulas depend on both a correlation {matrix} $R$  and the degrees of freedom $\nu$, a hyperparameter which we fix {to five (see details in Supplementary Material {B}).}  Both are members of the class of elliptical copulas, which may also be considered (\citealp{Embrechtsetal}).  
Other copula families (e.g. Archimedean copulas) are available, but  they do not allow for negative correlations in dimensions larger than two.
Vine copulas (\citealp{BedfordCooke2002}) may be used to tackle this, but they are not immediate to construct, limiting their use in this context. 
\textcolor{black}{Notice that the correlation matrix of a multivariate distribution with a Gaussian (or t) copula with correlation parameter $R$ is, in general, not $R$, unless all marginals are normal (or t-distributions with the
same degrees of freedom than the t-copula), in which case the Gaussian (or t) copula model coincides
with a multivariate normal (or t) distribution.} This is because the covariance of $(X_i,X_j)$ is not invariant for strictly monotone functions, and does, thus, depend on the underlying marginals $F_i, F_j$, see \cite{Hoeffding} and \cite{Embrechtsetal}. For this reason, here we consider the Kendall's $\tau$, a rank correlation which has the properties of being invariant under monotone transformation, of depending only on the copulas (\citealp{Embrechtsetal}) and of being easier to compute analytically than the Spearman's rho, another dependent measure having similar properties. In particular, the Kendall's $\tau$ of two random variables $X_i, X_j$ with Gaussian or $t$ copulas with correlation parameter $R_{ij}$ is given by $\tau_{ij}:=\tau(X_i,X_j)=2/{\pi}\arcsin{R_{ij}}$. 
Hence, when choosing the copula and the underlying marginals,  we want to preserve the mean vector $m^*$, {the variances $S^*_{ii}, i=1,\ldots, d_\theta$ and the Kendall's tau dependencies} of the guided Gaussian SIS-ABC samplers. 
To do this, we derive the \brown{Kendall's $\tau$ dependencies from the correlation parameter $R$, derived from $S^*$}, choosing the parameters of the underlying marginal distributions $F_j$ such that $\mathbb{E}[X_j]=m_j^*$ and $\textrm{Var}(X_j)=S_{jj}^{*}$. Hence, a $d_\theta$-dimensional parameter proposal $\theta^*\sim g_t(\theta|s_y)$ with copula $C$, marginals $F_1,\ldots, F_{d_\theta}$, mean vector $m^*$, \brown{variances $S_{ii}^*$ and Kendall's tau rank correlations $\tau_{ij}, i,j=1\ldots, d_\theta, i\neq j$}  for SIS-ABC can be generated as follows:
\begin{enumerate}
    \item Derive the parameters of the marginal distribution $F_j$ such that $\mathbb{E}[X_j]=m_j^*$ and $\textrm{Var}(X_j)=S^{*}_{jj}, j=1,\ldots, d_\theta$.
    \item Compute the correlation matrix $R=(R_{ij})_{i,j=1}^d$ from the covariance matrix $S^*=(S^*_{ij})_{i,j=1}^d$, with entries
    \[
    R_{ij}=\frac{\textrm{Cov}(X_i,X_j)}{\sqrt{\textrm{Var}(X_i)\textrm{Var}(X_j)]}}=\frac{1}{\sqrt{S_{ii}^*S_{jj}^*}}S_{ij}^*.\]
    \item Simulate $(u_1,\ldots, u_{d_\theta})$ from the chosen (Gaussian or t) copula {with correlation $R$.}  
    \item Set $\theta^*_j:= x_j = F_j^{-1}(u_j)$ to obtain the desired parameter proposals {with Kendall's tau $\tau_{ij}=2/\pi\arcsin{R_{ij}},i, j=1,\ldots, d_\theta, i\neq j$}.
\end{enumerate}
For step 1, calculations linking the parameters of the chosen marginal distributions to the marginal mean $m^*_j$ and variance $S^{*}_{jj}$ are reported in Supplementary Material {C}.
Note that step 4 is well defined thanks to the marginal distributions being continuous. Finally, the proposal density needed to derive the weights of the $i$th particle (line \red{21} of Algorithm \ref{alg:sis-abc}) can be computed via  \eqref{copuladensity}, using the copula densities reported in Supplementary Material {B}. \brown{As no information is available on $\theta_j|s_y$, we choose all underlying marginal distributions $F_j$ to be from the same family. Distinct families could be also considered, as discussed in Remark \ref{Remark3} below. Here}, as underlying marginal distributions $F_j$, we choose the location-scale Student's t {(with $\nu$ degrees of freedom, see Supplementary Material C for more details)}, the logistic, the Gumbel, the Gaussian, the 
triangular and the uniform distribution. {The normal distribution is chosen to favour a comparison with the corresponding guided Gaussian SIS-ABC samplers (when choosing a Gaussian copula with the right setting, see Remark \ref{Remark2} below).} The first three marginals (resp. last two) are chosen as they have heavier (resp. lighter) tails than the Gaussian distribution, allowing to sample less (resp. more) around their mean values. {Moreover, the Gumbel distribution is also chosen to evaluate the impact of (positive) skewness on the results, as all other marginals are symmetric.} {We refer to the Supplementary Material C for an extensive discussion and comparison of the marginals.}  
{\begin{remark}\label{Remark1}
The guided Gaussian proposal samplers $g_t(\theta|s_y)$, derived in the previous sections, are constructed under the assumption of $(\theta,s_y)$ being Gaussian, thus requiring 
also the distribution of the summary to be Gaussian, which may not always be true (despite having a limited impact on the estimation procedure, as shown in Section \ref{sec:examples}). Instead, the guided copula-based proposal sampler $g_t(\theta|s_y)$  is directly constructed on $\theta|s_y$, without assumptions on the distribution of the summaries $s_y$ and/or of $(\theta,s_y)$.
\end{remark}
\begin{remark}\label{Remark2}
Sampling from a Gaussian copula with  normal marginals and correlation parameter $R$ corresponds to sampling from a multivariate Gaussian proposal with correlation matrix $R$ (\citealp{Embrechtsetal}). Hence, \texttt{cop-blocked, cop-blockedopt} and \texttt{cop-hybrid} for the conditional sampler $g_t(\theta|s_y)$ with this setting coincide with  \texttt{blocked, blockedopt} and \texttt{hybrid} for $\theta|s_y$, respectively. However, even in this setting, the copula-based distribution of $(\theta,s_y)$ would differ from $N_d(m,S)$, the joint Gaussian distribution used to construct the guided Gaussian proposals, unless $s_y\sim N_{d_s}(m_s,S_s)$. 
\end{remark}
\begin{remark}\label{Remark3}
The guided copula-based SIS-ABC samplers introduce more flexibility than the guided Gaussian ones, e.g. in the choice of the copula, the marginals (which may belong to different distribution families or change across iterations{, see the ``mixed'' marginals introduced in Section \ref{sec:examples}}) and their underlying parameters (e.g. the degrees of freedom of the $t$ copula or of the location-scale Student's t marginals). In the simulation studies considered here, the copula and marginal models are fixed in advance, with the idea of investigating whether a particular combination of copula and marginals outperforms the other consistently across the experiments. {Moreover, it is worth stressing that the marginal proposals are not for $\theta_j$, for which some information may be available (e.g. their support) and used in the choice of the marginals, but for $\theta_j|s_y$, for which less is known (e.g., the support will differ from that of $\theta_j$).}  Hence, unless some prior knowledge is assumed/known about $\theta_j|s_y$, we recommend using marginal distributions assuming also negative values.  
Two alternative possibilities could be to perform a kind of \lq\lq regression adjustment\rq\rq\ within each iteration, e.g. by fitting suitable marginals and copula in the spirit of \citealp{chen2019adaptive}, or run a non-parametric estimation of the copula and the marginals. However, these two approaches would introduce some extra computational costs though, which is why we do not consider them here. 
\end{remark}}

\subsection{Guided Gaussian SMC-ABC samplers}\label{sec:guided-smc-abc}

Here, we wish to make the proposal function $g_t$ more dependent on the local features of important particles from the set obtained at iteration $t-1$, as in standard SMC-ABC proposal samplers. As previously discussed, choosing $\theta^{**}\sim \mathcal{N}(\theta^*,2\Sigma_{t-1})$ in step 16 of Algorithm \ref{alg:smc-abc} as in \cite{filippi2013optimality} yields a particle-specific  mean (local feature) and a covariance matrix which is common to all particles (global feature). 
With reference to Gaussian samplers, we can construct samplers where each proposed particle at iteration $t$ has its own specific mean and a global covariance structure and, additionally, be conditional on observed summaries $s_y$. \\
\indent 
Constructing a guided and local SMC-ABC sampler is immediate given  the reasoning in Sections \ref{sec:basic-guided-SIS}-\ref{sec:guided--hybrid-sis-abc}. Remember that we wish to perturb a sampled particle $\theta^*$ to produce a proposed $\theta^{**}$ in step 16 of Algorithm \ref{alg:smc-abc}.
Once $\theta^*$ has been sampled from the accepted particles at iteration $t-1$, we define as ``augmented data'' the vector $(\theta^*,s_y)$, which we want to use to produce $\theta^{**}$. It is useful to imagine $\theta^*$ as $(\theta^*_k,\theta_{-k}^*)$ if we wish to update one component at a time, using the notation established at the beginning of Section \ref{sec:guided-samplers}, or as $(\theta^*_{k,l},\theta_{-(k,l)}^*)$ if we update a block of two components at a time, etc. In Section \ref{sec:twisted}, we consider an example where some components of $\theta$ are highly correlated in the prior, and a sampler exploiting this fact dramatically helps producing an efficient ABC algorithm.
We illustrate 
 this conditional approach by focusing on perturbing a single component of $\theta$, as the notation is easier to convey the message and extensions to multiple components are immediate.
Hence, here we focus on the augmented data expressed as the column vector $(\theta_{k}^*,\theta_{-k}^*,s_y)'$,
for which we assume a joint multivariate Gaussian distribution allowing us to design a ``perturbation kernel''  $q_t(\theta_k|\theta_{-k}^*,s_y)$. Our goal is proposing for the $k$-th component of $\theta$ conditionally on the remaining $d_\theta-1$ coordinates of $\theta^*$ and the whole $s_y$. This means that the sampler producing $\theta^{**}_k$ does not make use of the value of $\theta_k^*$, but the sampler producing $\theta^{**}_{k'}$ will make use of $\theta_k^*$ ($k'\neq k$). Under the same assumptions of joint Gaussianity considered in Sections \ref{sec:basic-guided-SIS}-\ref{sec:guided--hybrid-sis-abc}, we have that $q_t(\theta_k|\theta_{-k}^*,s_y)$ is a Gaussian sampler, which we now construct. \\
\indent Say that all $N$ accepted $d$-dimensional particles of type $(\theta,s)$ have been obtained for iteration $t-1$. We now compute the weighted sample mean and weighed covariance matrix as in \eqref{eq:m-S}, and from these we extract the following quantities
\begin{itemize}
    \item $\hat{m}_{k}$:  scalar element extracted from $\hat{m}_{t-1}$ and corresponding to component $k$ of $\theta$;
    \item $\hat{S}_{k,-k}$:  $(d-1)$-dimensional row vector extracted from $\hat{S}_{t-1}$ by considering the row corresponding to $\theta_k$ and retaining all columns except that corresponding to $\theta_k$. 
    \item $\hat{S}_{-k,-k}$:  $(d-1)\times (d-1)$-dimensional matrix extracted from $\hat{S}_{t-1}$ by eliminating the row and column corresponding to $\theta_k$. 
    \item $
    \begin{bmatrix}
           \theta^{*}_{-k} \\
            s_y 
         \end{bmatrix}
  $
  :  $(d-1)$-dimensional column vector appending $s_y$ to the particle $\theta^*_{-k}$, where the latter is the $\theta^*$ randomly picked in step 15 of Algorithm \ref{alg:smc-abc}  with its $k$-th component eliminated. \item $
    \begin{bmatrix}
          \hat{m}_{-k} \\
          \hat{m}_s
         \end{bmatrix}
  $:
  $(d-1)$-dimensional column vector concatenating the $d_\theta-1$ (weighted) sample means of the particles for $\theta$, except the $k$-th component, and the $d_s$ (weighted) sample means of the corresponding summary statistics.
  \item $\hat{\sigma}^2_k$:  scalar value  found in $\hat{S}_{t-1}$ correspondingly to the row and column entry for $\theta_k$. 
  \item $\hat{S}_{-k,k}$: $(d-1)$-dimensional column vector extracted from $\hat{S}_{t-1}$ by retaining all rows except that  corresponding to $\theta_k$, and considering only the column corresponding to $\theta_k$. 
\end{itemize}
Then, upon completion of iteration $t-1$, we have
\begin{eqnarray}
    \hat{m}_{k|s_y,t-1}^* &=& \hat{m}_{k}+\hat{S}_{k, -k}(\hat{S}_{-k,-k})^{-1}\biggl(
    \begin{bmatrix}
           \theta^{*}_{-k} \\
            s_y 
         \end{bmatrix}\label{eq:guided-abc-smc-mean}
  -\begin{bmatrix}
          \hat{m}_{-k} \\
          \hat{m}_s
         \end{bmatrix}\biggr), \qquad k=1,...,d_\theta\\
    \hat{\sigma}_{k|s_y, t-1}^2 &=& \hat{\sigma}^2_k-\hat{S}_{k, -k}(\hat{S}_{-k,-k})^{-1}\hat{S}_{-k,k}, \qquad k=1,...,d_\theta.\label{eq:guided-abc-smc-var}
\end{eqnarray}
The asterisk in $\hat{m}_{k|s_y,t-1}^*$ is meant to emphasize that this quantity depends on the specific $\theta^*$, unlike the variance \eqref{eq:guided-abc-smc-var} which is common to all perturbed particles. That is,  the sampler has ``local'' mean features but global covariance, which can also be made local as described in Section \ref{sec:guided-optimal}. 
We define the perturbation kernel at iteration $t$ for component $k$ as $\theta^{**}_k\sim \mathcal{N}(\hat{m}_{k|s_y,t-1}^*,\hat{\sigma}_{k|s_y,t-1}^2)$.  By looping through \eqref{eq:guided-abc-smc-mean}--\eqref{eq:guided-abc-smc-var} and then proposing $\theta^{**}_k\sim \mathcal{N}(\hat{m}_{k|s_y,t-1}^*,\hat{\sigma}_{k|s_y,t-1}^2)$ for all $k=1,...,d_\theta$, we form $\theta^{**}=(\theta^{**}_1,...,\theta^{**}_{d_\theta})$. 
\\ \indent
If we follow the procedure above, then the $q_t(\theta|\theta^*)$  consists in the product of $d_\theta$ 1-dimensional Gaussian samplers each of type $q_t(\theta_k|\theta^*_{-k},s_y)=\mathcal{N}(\hat{m}^*_{k|s_y,t-1},\hat{\sigma}_{k|s_y,t-1}^2)$. While we perturb each component of $\theta^*$ separately from the others, our approach is much different from the component-wise approach of \cite{beaumont2009adaptive}, since correlation between the dimensions of $\theta$ is taken into account and, additionally, we condition on $s_y$. Once $N$ particles $\{\theta^{(i)}\}_{i=1}^N$ have been accepted, the importance weight \eqref{eq:importance-weight} is given by 
\begin{align*}
\tilde{w}^{(i)}_t&= \pi(\theta_t^{(i)})/\biggl\{\sum_{j=1}^N w_{t-1}^{(j)}\prod_{k=1}^{d_\theta}\mathcal{N}(\theta_{t,k}^{(i)};\hat{m}_{k|s_y,t-1}^{*(j)},\hat{\sigma}^2_{k|s_y,t-1})\biggr\}, \qquad i=1,...,N, 
\end{align*}
where $\theta_{t,k}^{(i)}$ is the $k$-th component of the accepted $\theta_{t}^{(i)}$ and $\hat{m}_{k|s_y,t-1}^{*(j)}$ is the guided mean \eqref{eq:guided-abc-smc-mean} obtained from every sampled $\theta^{*(j)}$ ($j=1,...,N$)
\begin{equation}
    \hat{m}_{k|s_y,t-1}^{*(j)} = \hat{m}_{k}+\hat{S}_{k, -k}(\hat{S}_{-k,-k})^{-1}\biggl(
    \begin{bmatrix}
           \theta^{*(j)}_{-k} \\
            s_y 
        \end{bmatrix}
  -\begin{bmatrix}
          \hat{m}_{-k} \\
          \hat{m}_s
         \end{bmatrix}\biggr), \qquad j=1,...,N.
 \label{eq:guided-abc-smc-mean-generic}
\end{equation}
When using the SMC-ABC in Algorithm \ref{alg:smc-abc} with guided sampler $q_t(\theta_k|\theta^*_{-k},s_y)=$\\ $\mathcal{N}(\hat{m}^*_{k|s_y,t-1},\hat{\sigma}_{k|s_y,t-1}^2)$, we call the resulting procedure \texttt{fullcond}, since each coordinate $\theta^{**}_k$ is ``fully conditional'' on all the coordinates of $\theta^*$ (except for $\theta^{*}_k$).
The scheme outlined is just an example, and the procedure opens up the possibility to jointly sample ``blocks'' of elements of $\theta$ conditionally on the remaining components and $s_y$. This turns out to be particularly useful if it is known that some  parameter components are highly correlated in the posterior, as proposing those correlated components in block can ease exploration of the posterior surface. For example, in Section \ref{sec:twomoons} we have two highly correlated parameters where proposing from $q_t(\theta_k|\theta_{-k}^*,s_y)$ is largely suboptimal. However, in Section \ref{sec:twisted} we exploit a-priori knowledge of the high correlation between the first two components (of a five dimensional vector $\theta$), and proposing from $q_t(\theta_{1}^{**},\theta_2^{**}|\theta^*_{-\{1,2\}},s_y)$ results in a very efficient sampler whose structure is in the Supplementary Material {F.1}.
Similarly to the guided SIS-ABC \texttt{blocked} sampler, the constructed SMC-ABC \texttt{fullcond} may suffer from ``mode seeking'' behaviour. In the next section,
we address this issue and construct a version that uses an ``optimal local'' covariance matrix.

\subsection{Guided  SMC-ABC samplers with optimal local covariance}
\label{sec:guided-optimal}

We now follow a similar route to that pursued in Section \ref{sec:guided--local-sis-abc} when optimizing the covariance matrix of a guided SIS-ABC sampler, focusing now on improving the SMC-ABC \texttt{fullcond} proposal. At iteration $t$, denote by $\theta^{**}$ the perturbed version of some  $\theta^*$ obtained from the particles  accepted at $t-1$ according to $\theta^{**}\sim q_t(\cdot|\theta^*)$. We can imagine accepting the couple $(\theta^{**},\theta^*)$ (even though of course we are only interested in  $\theta^{**}$) if and only if $||s^{**}-s_y||<\delta_t$, where $s^{**}\sim p(s|\theta^{**})$. A joint proposal distribution for $(\theta^{**},\theta^*)$ should somehow resemble the target product distribution induced by sampling $\theta^{**}$ and $\theta^{*}$ independently from $\pi_{\delta_t}(\theta|s_y)$ and $\pi_{\delta_{t-1}}(\theta|s_y)$, respectively, and whose density is $\pi_{\delta_t}(\theta|s_y) \pi_{\delta_{t-1}}(\theta|s_y)$. \\
\indent 
The approach detailed in \cite{filippi2013optimality} considers the ``target product'' distribution denoted by $q^*_{\delta_{t-1},\delta_{t}}(\theta^*,\theta^{**})=\pi_{\delta_t}(\theta^{**}|s_y) \pi_{\delta_{t-1}}(\theta^{*}|s_y)$  and determines a proposal $q_{\delta_{t-1},\delta_{t}}(\theta^*,\theta^{**})$ by minimizing the KL divergence $KL(q_{\delta_{t-1},\delta_{t}},q^*_{\delta_{t-1},\delta_{t}})$ between its arguments, where
\[
KL(q_{\delta_{t-1},\delta_{t}},q^*_{\delta_{t-1},\delta_{t}}) = \int q^*_{\delta_{t-1},\delta_{t}}(\theta^*,\theta^{**})  \log\frac{q^*_{\delta_{t-1},\delta_{t}}(\theta^*,\theta^{**})}{q_{\delta_{t-1},\delta_{t}}(\theta^*,\theta^{**})}d\theta^*d\theta^{**}.
\]
This time, the multi-objective criterion to optimize is equivalent to maximizing the following $Q$
\begin{equation}
    Q(q_t,\delta_{t-1},\delta_t,s_y)=\int  \pi_{\delta_t}(\theta^{**}|s_y) \pi_{\delta_{t-1}}(\theta^*|s_y)\log q_t(\theta^{**}|\theta^*)d\theta^*d\theta^{**}\label{eq:global-Q}
\end{equation}
with respect to $q_t$. The above is useful if we want to obtain a proposal sampler that has ``global properties'', that is a sampler that is not specific for a given $\theta^{*}$, and in fact the latter is considered as a variable in the integrands for both the KL and the $Q$ quantities above, so that they result independent of $\theta^*$ (and $\theta^{**}$) since it is integrated out. Instead, here we want to construct a sampler which is optimal \lq\lq locally\rq\rq\ for $\theta^*$, optimizing $q_t$ with respect to an unknown variance that is specific to the sampled $\theta^*$ instead of being global for all particles. When looking at the guided SMS-ABC sampler outlined in Section \ref{sec:guided-smc-abc}, we have that $q_t(\theta^{**}|\theta^*,s_y)\equiv \prod_{k=1}^{d_\theta}\mathcal{N}(\hat{m}_{k|s_y,t-1}^*,\sigma^{*^2}_k)$ with local means \eqref{eq:guided-abc-smc-mean} and unknown $\sigma^{*}_k$ which we wish to maximize $Q$ for. As the particle $\theta^*$ entering in \eqref{eq:guided-abc-smc-mean}  as $\theta^*_{-k}$ is considered \textit{fixed}, \eqref{eq:global-Q} simplifies to $Q(q_t,\delta_t,s_y)=\int  \pi_{\delta_t}(\theta^{**}|s_y)\log q_t(\theta^{**}|\theta^*)d\theta^{**}$ which, optimized with respect to $q_t$, yields (see details in Supplementary Material {A})
\begin{equation}
\sigma^{*^2}_k = \int \pi_{\delta_t}(\theta^{**}_k|s_y)(\theta_k^{**}-\hat{m}_{k|s_y,t-1}^*)^2 d\theta^{**}_{k}.
\label{eq:var_fullcondopt}
\end{equation}
The latter can be approximated via standard Monte Carlo. In fact, even though we only have $N$ samples obtained at iteration $t-1$ (and not at iteration $t$ since these have not been sampled yet), we can use the same argument as for \texttt{olcm}. That is,  from the $N$ particles sampled and accepted at iteration $t-1$ using $\delta_{t-1}$, we subselect the $N_0$ particles, denoted by $\tilde{\theta}_{t-1,l}$, whose distances are also smaller than $\delta_t$, having then  normalized weights $\gamma_{t-1,l}$. These $N_0$ particles are then sampled from $\pi_{\delta_t}(\theta_k|s_y)$ and we can  approximate $\sigma^{*^2}_k$ with
$
\hat{\sigma}^{*^2}_k=\sum_{l=1}^{N_0} \gamma_{t-1,l}(\tilde{\theta}_{t-1,k,l}-\hat{m}_{k|s_y,t-1}^*)^2$, $k=1,...,d_\theta$, 
with $\tilde{\theta}_{t-1,k,l}$ being the $k$-th component of $\tilde{\theta}_{t-1,l}$.
Hence, we have constructed a guided SMC-ABC sampler   having ``local'' features for both the mean and the variance, since they both depend on $\theta^*$, unlike \texttt{fullcond} in Section \ref{sec:guided-smc-abc} where  the variance is global. At iteration $t$, this samples from  $q_t(\theta^{**}|\theta^*,s_y)\equiv \prod_{k=1}^{d_\theta}\mathcal{N}(\hat{m}_{k|s_y,t-1}^*,\hat{\sigma}^{*^2}_k)$, with normalized weights  given by
\begin{align*}
\tilde{w}^{(i)}_t&= \pi(\theta_t^{(i)})/\biggl\{\sum_{j=1}^N w_{t-1}^{(j)}\prod_{k=1}^{d_\theta}\mathcal{N}(\theta_{t,k}^{(i)};\hat{m}_{k|s_y,t-1}^{*(j)},\hat{\sigma}^{*(j)^2}_{k})\biggr\}, \qquad i=1,...,N
\end{align*}
where $\hat{m}_{k|s_y,t-1}^{*(j)}$ is as in \eqref{eq:guided-abc-smc-mean-generic} and
$
\hat{\sigma}^{*(j)^2}_k=\sum\limits_{l=1}^{N_0} \gamma_{t-1,l}(\tilde{\theta}_{t-1,k,l}-\hat{m}_{k|s_y,t-1}^{*(j)})^2$, $j=1,...,N$, $k=1,...,d_\theta.
$
We call \texttt{fullcondopt} the sampler just outlined. An exciting display of a particular version of it is in Section \ref{sec:twisted}, where  the sampler successfully exploits the fact that two components $(\theta_1,\theta_2)$ of a 5-dimensional parameter $\theta$ are highly correlated in the prior. Details on this specific sampler are in Supplementary Material {F.1}.

\section{Examples}\label{sec:examples}  
In this section, we consider several experiments to assess the performances of our guided proposals (cf. Section \ref{sec:guided-samplers}) against those of the non-guided  \texttt{standard} and \texttt{olcm} proposal samplers (cf. Section \ref{sec:abc-smc}). We refer to Table \ref{table:samplers} for an outline/classification of the proposal samplers considered here. The  \texttt{cop-blocked}, \texttt{cop-blockedopt} and \texttt{cop-hybrid} approaches use either Gaussian or t copulas, with triangular, location-scale Student's t \brown{(with five degrees of freedom)}, logistic, Gumbel, uniform or normal marginal distributions{, as discussed in Section \ref{sec:copula-sis} and Supplementary Material C.}  {
Moreover, a ``mixed'' case with uniform marginals at iteration $t=2$ and triangular marginals from $t>2$ is also considered for the copula-based SIS-ABC samplers to tackle the possible mode-seeking behaviour of the samplers with uniform marginals (see Sections \ref{sec:twomoons} and \ref{sec:twisted}). }
Unless differently specified, we launch ten independent runs for each method, fixing the number of particles to $N=1,000$ (except in Section \ref{sec:g-and-k}, where $N=10,000$). We compare the algorithms at either prefixed (cf. Sections \ref{sec:twomoons} and \ref{sec:recruitment}) or online updated $\delta_t$ values (cf. Sections \ref{sec:twisted}, \ref{sec:g-and-k}, \textcolor{black}{\ref{sec:lotka-volterra}} and \ref{sec:cell}), focusing on the quality of the posterior inference, the acceptance rates and the running (wallclock) times. 
When thresholds are automatically updated, we use the following strategy: set an initial threshold $\delta_1<\infty$, then, for $t>1$, $\delta_t$ is automatically chosen as a percentile of all the ABC distances $\rho_{t-1}=||s(z)-s_y||$ produced at iteration $t-1$ (including distances from rejected proposals), say the $\psi$-th percentile. However, occasionally, this may not be enough to produce monotonously decreasing $\delta$'s. We adjust this as follows: call $\tilde{\delta}_t$ the $\psi$-th percentile of $\rho_{t-1}$, then we set ${\delta}_t:=\tilde{\delta}_t$ if $\tilde{\delta}_t<\delta_{t-1}$, otherwise  $\delta_t:=0.95\delta_{t-1}$. All our novel samplers collect, at each iteration, the accepted summaries and reuse them to create our guided proposal samplers. As such, we found that it is best to not allow the initial threshold $\delta_1$ to be too large (e.g. neither $\delta_1=\infty$ nor a $\delta_1$ that accepts almost all proposals), as this would too liberally accept summary statistics that are extremely different from observations. This would allow very poor realizations from the forward model to contribute to the covariance matrices of the summary statistics, thus resulting in a poor initial sampler. 

\subsection{Two-moons model with bimodal posterior and non-Gaussian summaries}\label{sec:twomoons}

\begin{figure}[t!]
\centering
\includegraphics[width=1.0\textwidth]{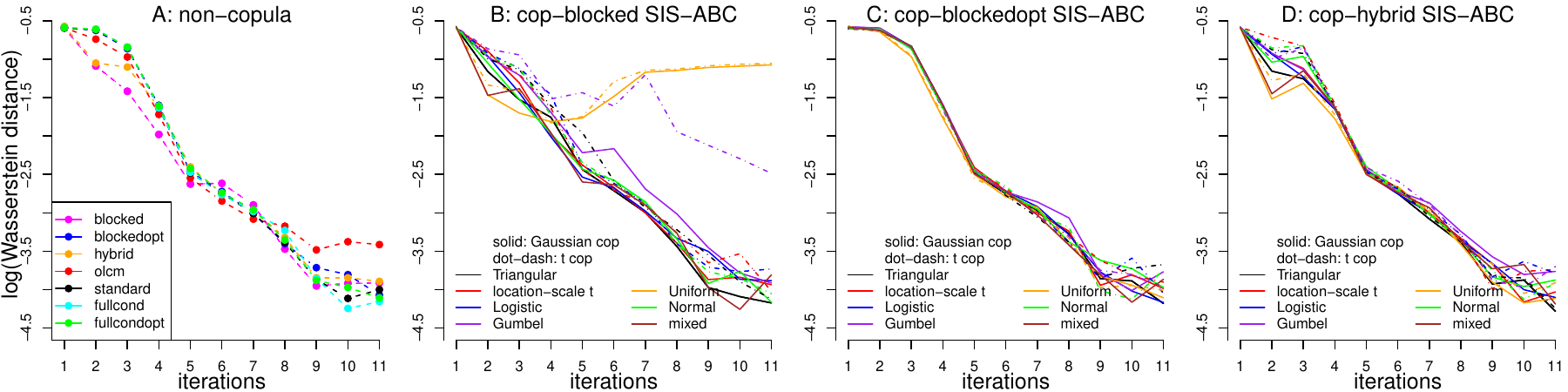} 
\caption{Two-moons model: median log-Wasserstein distances between the \lq\lq exact\rq\rq (MCMC) posterior and the ABC posteriors obtained via the guided or non-guided samplers across ten independent estimations.  {Panel A}:  non-copula-based methods, both guided and non-guided. {Panel B}:  \texttt{cop-blocked}; {Panel C}:   \texttt{cop-blockedopt}; {Panel D}:  \texttt{cop-hybrid}. The
copula-based methods are derived using either Gaussian (solid lines) or t copulas (dot-dashed lines) for different marginals, as
described in the legend.}
\label{fig:twomoons-logwasser}
\end{figure}

\begin{figure}[t!]
\centering
\includegraphics[width=.4\textwidth]{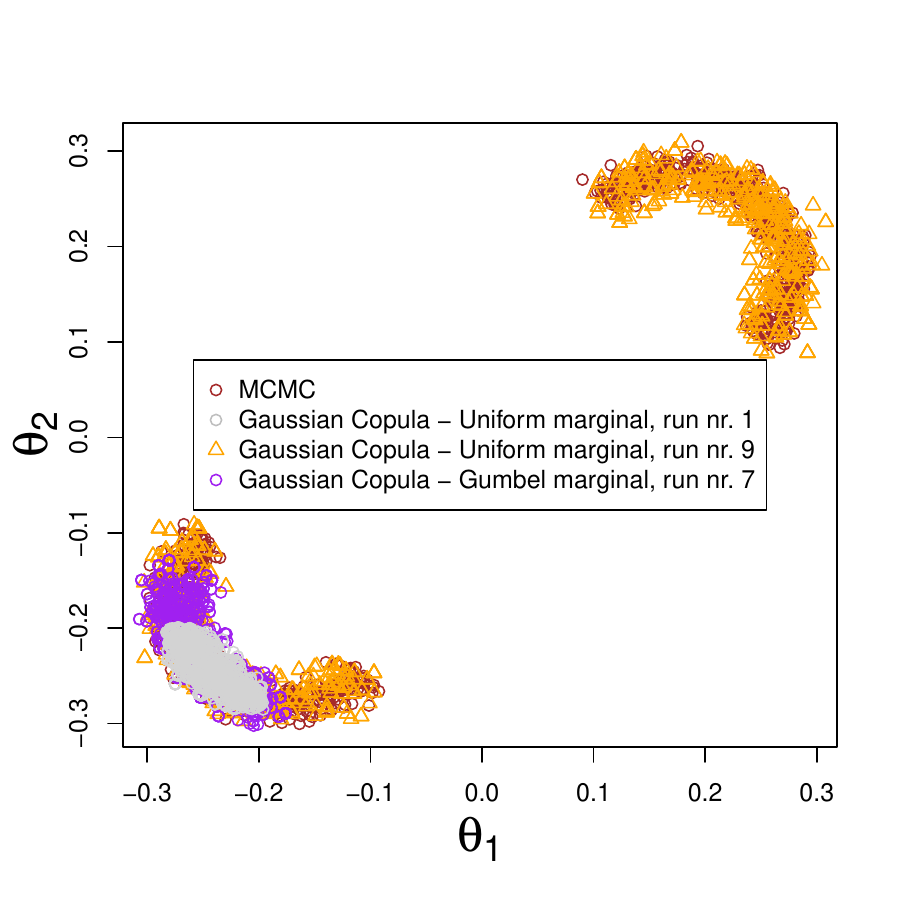}
\caption{Two-moons model: 1,000 MCMC draws  from the \lq\lq exact\rq\rq (MCMC) posterior (brown) and from the ABC posteriors obtained using the \texttt{Gaussian-cop-blocked} sampler with Gumbel marginals (purple, run number 7) and uniform marginals from run number 1 (grey) and 9 (orange). The 1,000 ABC draws obtained from copulas with uniform marginals cover the whole true posterior region only in some runs (here run 9), targeting a small sub-region of one of the two moons in most of the cases (here the first run), while those from the Gumbel marginals {cover a larger sub-region of one of the two moons only in one run, successfully covering both moons in the other runs.}}
\label{fig:twomoons-issues}
\end{figure}

\begin{figure}[t!]
\hspace{-1cm}
\includegraphics[width=1.07\textwidth]{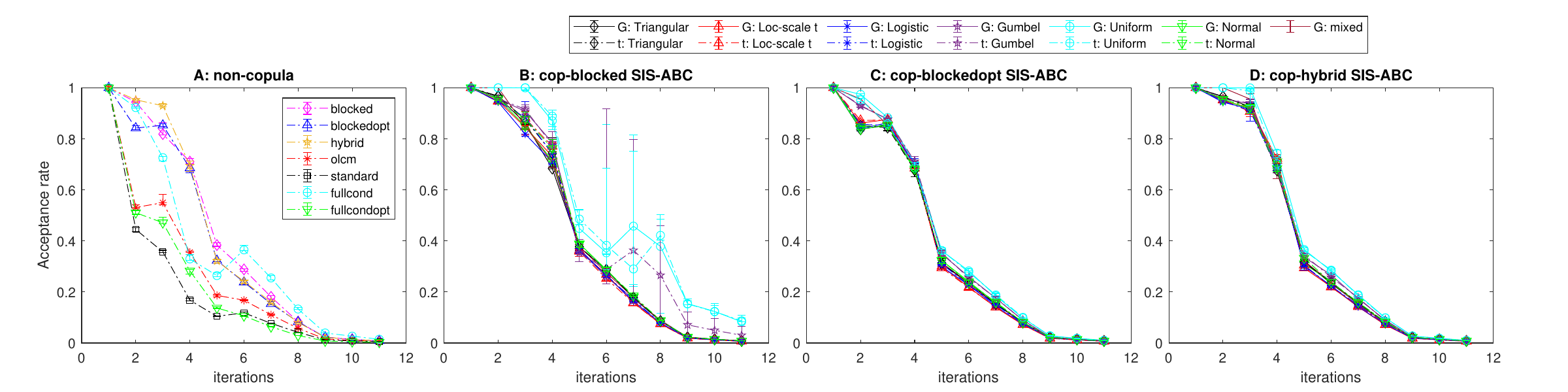}
\caption{Two-moons model: median acceptance rates and error bars with  first and third quartile across ten independent runs at each iteration. Panel A:  Non-copula based methods, both guided and non-guided;  Panel B:  \texttt{cop-blocked}; Panel C:  \texttt{cop-blockedopt}; Panel D: \texttt{cop-hybrid}. The copula-based methods are derived using either Gaussian (G, solid lines) or\mbox{ t copulas (dot-dashed lines) for different marginals, as described in the legend.}}    \label{fig:twomoons-acceptrates}
    
\end{figure}

\begin{figure}[t!]
\centering
\hspace{-1cm}\includegraphics[width=1.08\textwidth]{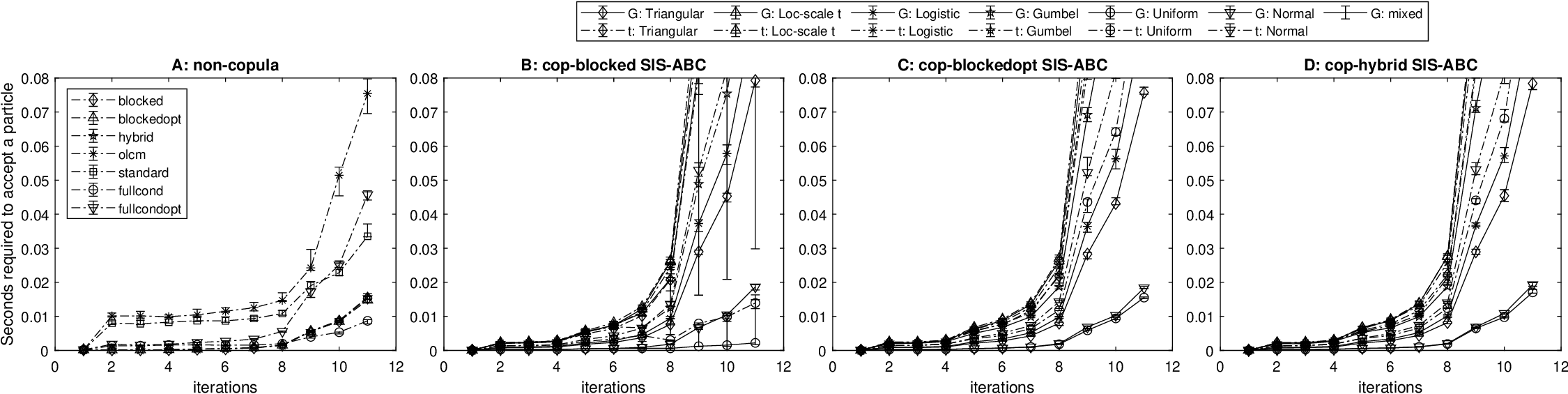}
\caption{Two-moons model: median number of seconds required to accept a particle and error bars with first and third quartile across ten independent runs at each iteration. Panel A:  Non-copula based methods, both guided and non-guided; Panel B: \texttt{cop-blocked}; Panel C:  \texttt{cop-blockedopt};  Panel D:   \texttt{cop-hybrid}. The copula-based methods are derived using either Gaussian (G, solid lines) or t (dot-dashed lines) copulas for different marginals, as described in the legend.}
    \label{fig:twomoons-particles-seconds}
\end{figure}

As some of the novel proposal samplers are multivariate Gaussian distributions derived by exploiting the multivariate Gaussianity of the pair $(\theta,s)$ as in Section \ref{sec:basic-guided-SIS}, it is of interest to assess their performance when such assumption is not met. For this reason, here we consider the \textit{two-moons model}, a bimodal two-dimensional model characterised by highly non-Gaussian summary statistics (cf. Supplementary Material {E}) and a crescent-shaped posterior (under some parameter setting). We consider  the same setup as in \cite{greenberg2019automatic} and \cite{wiqvist2021sequential} for simulation-based inference. We refer to the Supplementary Material {E} for details on the generative model, the priors, and the prefixed values of $\delta_t$. We assume observed data $y=(0,0)$ and consider the identity function  as summary statistics function, i.e. $S(z)=z$.
Here exact posterior inference via MCMC is possible, and therefore we use it as a benchmark. First, we obtain 1,000 MCMC posterior draws via the Python simulator\footnote{The Python code is available at \url{https://github.com/sbi-benchmark/sbibm/tree/main/sbibm/tasks/two_moons}} associated to \cite{lueckmann2021benchmarking}. Then, we compute the order-1 Wasserstein distances (\citealp{SommerfeldMunk2018}) between each of the ABC posteriors and the MCMC posterior using the R package \texttt{transport} (\citealp{transport}). The medians of the log-Wasserstein distances across the ten runs are reported in Figure  \ref{fig:twomoons-logwasser} (quartiles of these distances are not shown to ease the reading of the plot, but the variability across the runs is very small). Interestingly, the guided \texttt{blocked} and \texttt{cop-blocked} approaches have the smallest Wasserstein distances during the first five iterations, that is when approaching the high-posterior probability region. At smaller values of $\delta_t$ (and thus larger iterations), when more precise local information is needed, the methods perform similarly, except at the smallest $\delta_t$, where the non-guided \texttt{standard} and \texttt{olcm} display slightly larger distances. The different types of copulas and marginal distributions yield similar results, except for the \texttt{Gaussian cop-blocked} (and \texttt{t cop-blocked}, figure not shown) with uniform marginals, which performed poorly in most of the runs, with particles sampled from the ABC posterior covering only a sub-region of only one of the two moons (see Figure \ref{fig:twomoons-issues}, first run) and both moons only in few attempts (see Figure \ref{fig:twomoons-issues}, ninth run){, which explains the higher log-Wasserstein distances in Figure \ref{fig:twomoons-logwasser},
panel B. {This is tackled when considering  mixed marginals}, which successfully target both moons and have similar performances as the samplers with triangular marginals (see Figure \ref{fig:twomoons-logwasser}, Panel B, brown line).} 
{The \texttt{Gaussian cop-blocked} with Gumbel marginals performed similarly bad in only one of the runs (see Figure \ref{fig:twomoons-issues}, seventh run), with this happening more often for the \texttt{t cop-blocked} proposal, which explains the higher log-Wasserstein distances (Figure \ref{fig:twomoons-logwasser}, panel B).} {For these marginals, the performance (measured by the log-Wasserstein) can be improved (resp. decreased) by choosing $t$-copulas with higher (resp. smaller) degrees of freedom (e.g. $\nu=10$ vs $3$). Similar results and conclusions hold for the copula-based samplers with location-scale Student's t marginals with degrees of freedom $\nu$ (results not shown), while  other marginals are not affected by this (results not shown). Throughout this work, we choose $\nu=5$, as choosing $\nu=10$
would lead to samplers closer to the Gaussian copulas/Gaussian marginals, with results very similar to those samplers. }\\
\indent Acceptance rates are reported in Figure \ref{fig:twomoons-acceptrates}, and appear very stable across the ten runs, except for \texttt{cop-blocked} with uniform marginals. As also observed in \cite{filippi2013optimality}, \texttt{olcm} is superior to  \texttt{standard}  in terms of acceptance rate, but both are outperformed across all iterations by all our novel guided methods except \texttt{fullcondopt} which is better than \texttt{standard} but slightly worse than \texttt{olcm}.  Importantly, differences between the acceptance rates are large in early iterations, which is where the samples are still very dispersed in the posterior surface. Our methods are of great help in this case, as we wish to spend the least time possible to rule out initial ``bad'' particles, which is especially relevant for models that are computationally intensive to simulate. 
When we consider the wallclock running times, Figure \ref{fig:twomoons-particles-seconds} reveals that \texttt{olcm} requires way more time than any guided method to accept a particle (without improving the final inference though), and the guided \texttt{fullcond}, \texttt{block}, \texttt{blockedopt} and \texttt{hybrid} methods are the fastest in accepting particles. Overall, on our desktop machine (Intel Core i7-7700 CPU 3.60GHz 32 GB RAM) and without using any parallelization, the inference across the ten runs was completed in 5.5 minutes (with \texttt{blocked}), 5.4  minutes (\texttt{blockedopt}), 5.7  minutes (\texttt{hybrid}),  4.8 minutes (\texttt{fullcond}), 17.7  minutes (\texttt{fullcondopt}), 38.7  minutes (\texttt{olcm}) and 23.6 minutes (\texttt{standard}).  These are major time-differences given that this model is particularly simple to simulate.  Copula methods are intrinsically slower in simulating proposals due to the more involved construction of a generic copula sampler and the less optimized numerical libraries compared to those implementing multivariate Gaussian samplers, 
independently on whether the codes are run in, say, R or Matlab (e.g. the \texttt{Gaussian-cop-blocked}, \texttt{Gaussian-cop-blockedopt} and  \texttt{Gaussian-cop-hybrid} with normal marginals take 8.1, 11.4 and 9.4 minutes, instead of 5.5, 5.4 and 5.7 minutes \textcolor{black}{of the corresponding \texttt{blocked}, \texttt{blockedopt} and \texttt{hybrid}}).  
Since the acceptance rates of the guided-copula  samplers are similar to those of the guided Gaussian samplers, the wallclock times of the former will become lower than the alternative non-guided samplers should the implementation of the R/Matlab copula built-in routines be improved (which is outside the scope of this work).

\subsection{Twisted-prior model with an highly correlated posterior}\label{sec:twisted}

We now consider a model with a challenging posterior characterised by a strong correlation between some of the parameters. This case study is particularly interesting, as the likelihood only provides location information about the unknown parameter while the dependence structure in the posterior comes mostly from the prior. For this reason, the posterior dependence
changes direction depending on whether the likelihood locates the posterior in the left or right tail of
the prior (see \citealp{nott} for a graphical illustration). This case study was analysed in an ABC context in \cite{li2017extending}.  The model assumes observations $y=(y_1,...,y_{d_\theta})$ drawn from a $d_\theta$-dimensional Gaussian $y\sim \mathcal{N}(\theta,\Psi)$, with $\theta=(\theta_1,...,\theta_{d_\theta})$ and diagonal covariance matrix $\Psi=\mathrm{diag}(\sigma_0,...,\sigma_0)$. The prior is the ``twisted-normal'' prior of \cite{haario1999adaptive},  with density function proportional to
$
\pi(\theta)\propto \exp\biggl\{-\theta_1^2/200-(\theta_2-b\theta_1^2+100b)^2/2-\mathbbm{1}_{\{d_\theta>2\}}\sum_{j=3}^{d_\theta}\theta_j^2\biggr\},
$
where $\mathbbm{1}_B$ denotes the indicator function of the set $B$.  This prior is essentially a product of independent Gaussian distributions with the exception that the component for $(\theta_1,\theta_2)$ is modified to produce a ``banana shape'', with the strength of the bivariate dependence determined by the parameter $b$.  Simulation from $\pi(\theta)$ is achieved by first drawing $\theta$ from a $d_\theta$-dimensional multivariate Gaussian as $\theta\sim \mathcal{N}(0,A)$, where $A=\mathrm{diag}(100,1,...,1)$, and then placing the value $\theta_2+b\theta_1^2-100b$ in the slot for $\theta_2$. We consider a value for $b$ that induces a strong correlation in the prior between the first two components of $\theta$.  Specifically, we use the same setup as  \cite{li2017extending}, namely $\sigma_0=1$, $b=0.1$ and $d_\theta=5$, i.e. both $y_\mathrm{obs}$ and $\theta$ have length five, with observations given by the vector $y_\mathrm{obs}=(10,0,0,0,0)$. We take the identity function as summary statistic, i.e. $S(y)=y$, set an initial $\delta_1=50$ and let $\delta_t$ automatically decrease across iterations by taking $\psi=1$ (first percentile of the distances), as described at the beginning of Section \ref{sec:examples}, until the updated $\delta$ value gets smaller than 0.25, when the inference is then stopped.
 \begin{figure}[t!]
\includegraphics[width=1.0\textwidth]{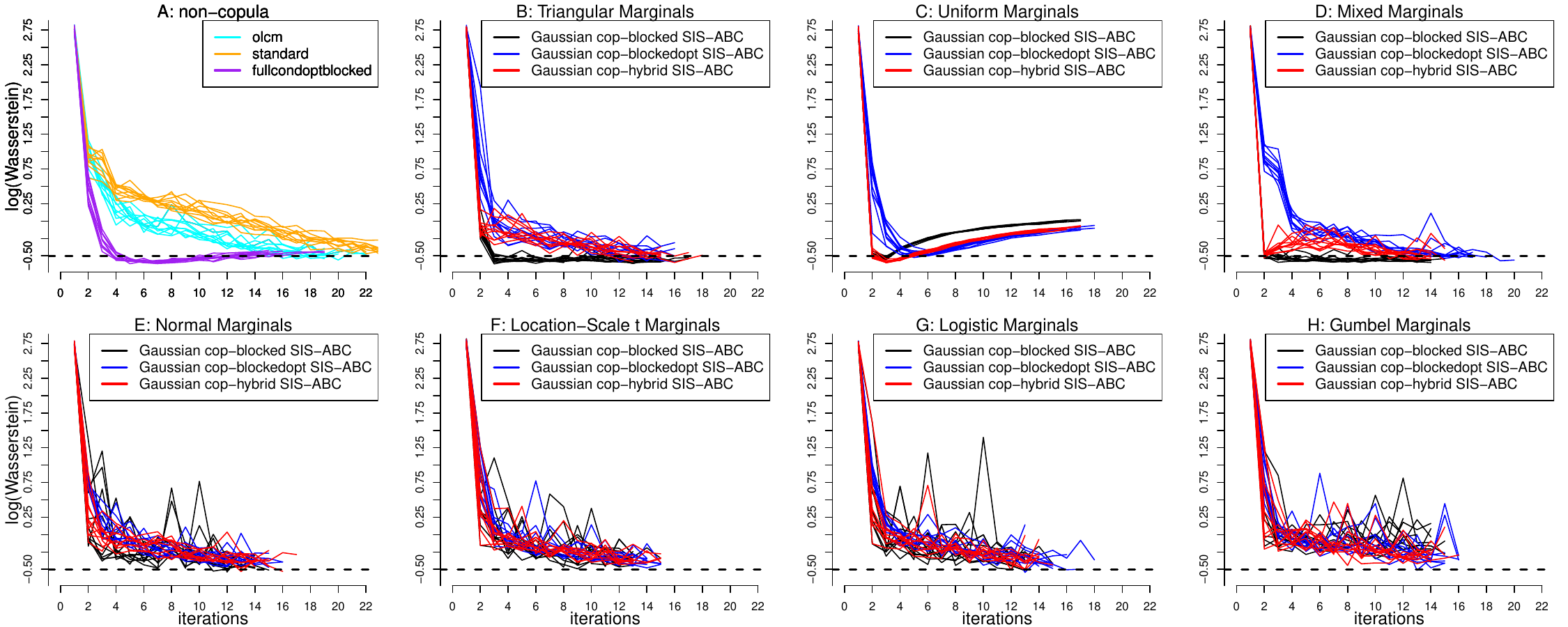}
         \caption{Twisted model: median log-Wasserstein distances between the ``exact'' posterior (MCMC) and the ABC posteriors across ten independent runs. Panel A: non-copula based methods, both guided (\texttt{fullcondopt} with $(\theta_1,\theta_2$) blocked) and non-guided (\texttt{olcm, standard}); Panels B-H:  guided Gaussian copula-based methods with   triangular (B), {uniform (C)}, {``mixed'' (D)}, {normal (E)}, location-scale Student's t {(F)}, logistic {(G)} and {Gumbel (H)} marginals. The dashed lines mark a log-distance value of -0.5 (an arbitrary value only meant to ease eye-comparisons). Some runs are shorter than others depending on how many iterations $t$ it took to reach a value of $\delta_t$ smaller than 0.25. {The results for the t copula-based samplers are almost indistinguishable to whose based on Gaussian copulas, and are thus not reported.}}
     \label{fig:twisted_wass}
\end{figure}
Low values of the ESS for some of the guided-methods (see the Supplementary Material {F.2}) are responsible for a larger variability between different runs, as in this case only few particles are resampled at the last iteration, and these appear to differ between runs. Instead, the ESS of the guided SMC-ABC \texttt{fullcondoptblocked} (Panel A), the guided SIS-ABC  \texttt{cop-blocked} with triangular (Panel B) \brown{or mixed marginals (Panel D)}, or the guided-copulas with uniform marginals (Panel C) are higher than the non-guided ones (see the Supplementary Material {F.2}), being then less sensible to variability across independent runs.  However, the distances obtained from marginal uniforms increase with the iterations as the consequence of the method becoming somehow overconfident, as illustrated in the Supplementary Material {F.2}, where the contour plots of the ABC posteriors at the last iteration of $(\theta_1,\theta_2)$ are reported. \brown{As for the two-moons study, this can be solved by considering mixed marginals (Panel D). Similar increasing distances} happen also to \texttt{fullcondoptblocked}.  Deriving a \lq\lq sanity check\rq\rq to prevent this deterioration in the inference performance, while of interest, is out of the scope of this work. However, a possible suggestion would be to stop the methods once the acceptance rate becomes lower than $1.5\%$ for two consecutive iterations (similarly to \citealp{del2012adaptive}), as it is unlikely that the inference will improve while the computational cost will increase. Additional results using this stopping criterion are in the Supplementary Material {F.2}, showing, indeed, an improved inference. Among all methods, for this case study we recommend using \texttt{cop-blocked} with \brown{either mixed or triangular marginals}, as they have the merits of being robust across several runs and, more importantly, having small log-Wasserstein distances starting from as little as \brown{two or three  iterations, respectively, \mbox{with the mixed marginals having also higher acceptance rates at iteration two.}}

\begin{figure}[t!]
\footnotesize
\center
\begin{tikzpicture}
\node (a) at (-2.4, 1.5) {$\alpha$};
\node (b1) at (-4.2, 0.5) {$A_{1}$};
\node (b2) at (-3.2, 0.5)  {$A_{2}$};
\node (b3) at (-1, 0.5) {$A_{n}$};
\node at (-2, 0.5) {$\cdots$};

\draw [->] (a)--(b1);
\draw [->] (a)--(b2);
\draw [->] (a)--(b3);

\node (x1) at (-4.6, -0.8) {$\begin{Bmatrix}
    x_{11} \\
    x_{12} \\
    \vdots \\
    x_{1J}
  \end{Bmatrix}$};
\node (x2) at (-3.2, -0.8) {$\begin{Bmatrix}
    x_{21} \\
    x_{22} \\
    \vdots \\
    x_{2J}
  \end{Bmatrix}$};
\node (x3) at (-0.7, -0.8) {$\begin{Bmatrix}
    x_{n1} \\
    x_{n2} \\
    \vdots \\
    x_{nJ}
  \end{Bmatrix}$};
\node at (-2, -0.8) {$\cdots$};

\draw [->] (b1)--(x1);
\draw [->] (b2)--(x2);
\draw [->] (b3)--(x3);

\end{tikzpicture}
\caption{Schematic representation of the hierarchical g-and-k model.}
\label{fig:gk}
\end{figure}
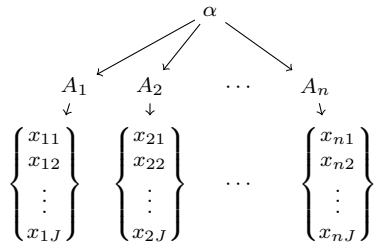
\subsection{Hierarchical g-and-k model with high-dimensional summaries}\label{sec:g-and-k}

\begin{figure}
\begin{subfigure}[b]{0.49\textwidth}\includegraphics[width=1.05\textwidth]{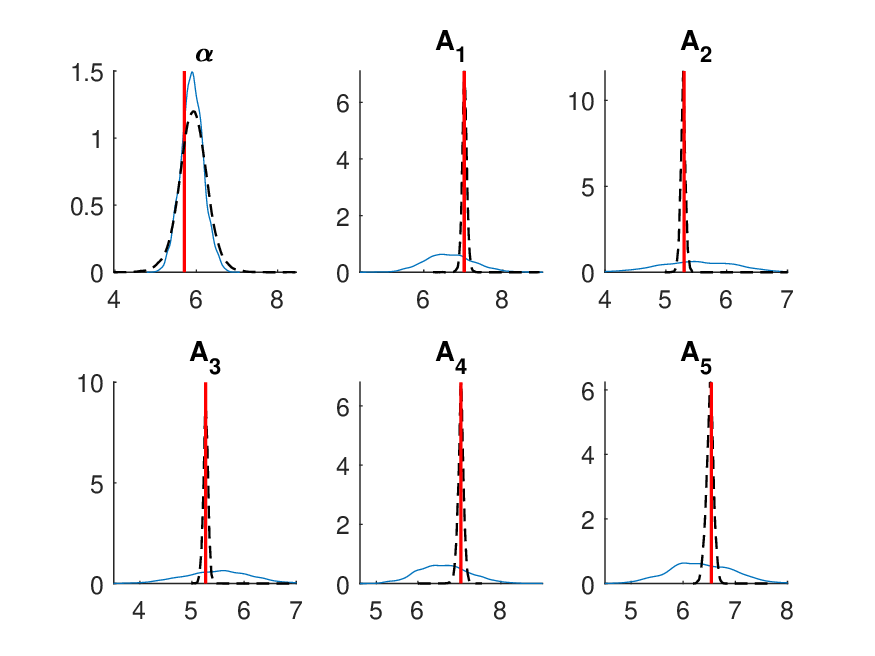}
\caption{\texttt{standard}}
\end{subfigure}
    \begin{subfigure}[b]{0.49\textwidth}
\includegraphics[width=1.05\textwidth]{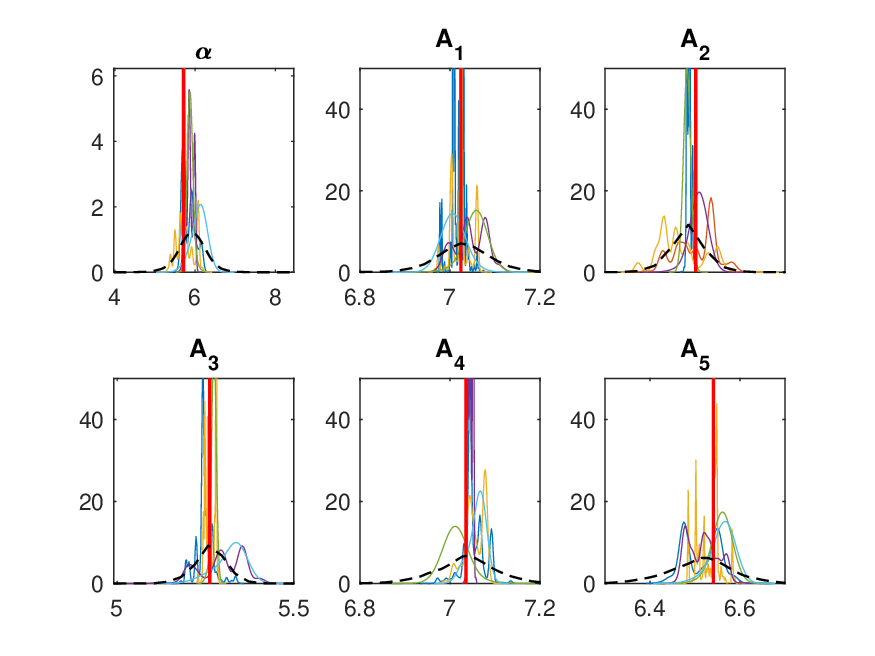}
    \caption{\texttt{hybrid}}
\end{subfigure}
    \caption{Hierarchical g-and-k model: posteriors displayed only for $(\alpha,A_1,...,A_5)$, using: (a) \texttt{standard}  at iteration $t=26$ and $\delta=2.14$ after $14.5\cdot 10^6$ model simulations; 
    (b) five inference runs of \texttt{hybrid},  \mbox{all reaching $\delta=0.62$ in approximately $10^6$ model simulations per run. 
Dashed lines are} \mbox{posteriors from ABC-Gibbs. Red vertical lines mark ground-truth values.}}
    \label{fig:gk-marginals}
\end{figure}

We now consider a high-dimensional model from \cite{clarte2021componentwise}. This is a hierarchical version of the g-and-k model, with the latter being often used as a toy case study in simulation-based inference (e.g. \citealp{fearnhead2012constructing}), since its probability density function is unavailable in closed-form but it is possible to simulate from its quantile function. The  g-and-k distribution is used to model non-standard data
through five parameters $\theta=(A,B,g,k,c)$, though in practice $c$ is often fixed to 0.8 (\citealp{gk}), as we do here. Details about simulating $x\sim gk(A,B,g,k)$ from a g-and-k distribution are in the Supplementary Material {G}.   
Same as \cite{clarte2021componentwise}, we assume to have observations $x_{ij}$ ($i=1,...,n, j=1,...,J$) sampled from a hierarchical g-and-k model, where each $J$-dimensional vector $x_i=(x_{i1},...,x_{iJ})$ is characterised by its own parameter $A_i$, while $(B,g,k)$ are common to all $n$ units  and are assumed known. We  assume 
$A_i\sim \mathcal{N}(\alpha,1)$ for $\alpha$ unknown and $x_{ij}\sim gk(A_i,B,g,k)$, see Figure \ref{fig:gk}. We also assume $\alpha\sim \mathrm{Unif}(-10,10)$ and, ultimately, we  infer  $\theta=(\alpha,A_1,...,A_n)$. Here, we generate data with $n=20$ and $J=1,000$ by using  $(B,g,k,\alpha)=(0.192,0.622,0.438,5.707)$. Same as \cite{clarte2021componentwise}, summary statistics for $x_i$ are the vector of nine quantiles $\mathrm{quant}(x_i,l/8)$ ($l=0,1,...,8$), where $\mathrm{quant}(x,p)$ is the $p$-th quantile of sample $x$.
This setting leads to a challenging inference problem for ABC, as the vector $\theta$ to infer is 21-dimensional and summary statistics are a vector of length $9n=180$.  \\
\indent 
Since an exact posterior is not available here, {we obtain 1,000 posterior samples from the ABC-Gibbs sampler\footnote{We appropriately modified the code at \url{https://github.com/GClarte/ABCG} to work with our data and model settings.} of \cite{clarte2021componentwise} to produce a ``reference  posterior'', as this method is designed (and thus especially suited) for hierarchical models. Here, our focus is on comparing the performance (with respect to the reference ABC-Gibbs) and running times of $N=10^4$ draws sampled from our proposed guided \texttt{hybrid} and \texttt{cop-hybrid} SIS-ABC versus the SMC-ABC \texttt{olcm} and \texttt{standard}.} 
For sequential ABC methods, we set an initial $\delta_1=50$ and let $\delta_t$ automatically decrease across iterations as described at the beginning of Section \ref{sec:examples} by taking $\psi=25$ (25-th percentile of all simulated distances), stopping the methods as soon as the updated $\delta$ value gets smaller than 0.62. However, most methods did not manage to reach this value in reasonable time and had to be halted, as the number of seconds required to accept a particle became rapidly larger than that of \texttt{hybrid}, which succeeds in drastically decreasing the threshold $\delta_t$ at iteration 3 compared to the non-guided approaches, see the Supplementary Material {G}. In terms of running times, the non-guided  \texttt{standard} was very slow, 
with 14.5 million model simulations (for \textit{a single} run) to attain a threshold around $2.14$  in 42 hours (at iteration $t=26$). Moreover,  \texttt{standard} resulted in quite uninformative marginal posteriors for the $A_i$ parameters, see Figure \ref{fig:gk-marginals} (for ease of display, we only report the ABC posteriors of $\alpha$ and $A_1,...,A_5$). Hence, not much is being learned except for $\alpha$, despite the large number of simulations. On the contrary, hybrid locates the high posterior mass region with higher precision and much more rapidly. In each of the ten runs, \texttt{hybrid} reached $\delta=0.62$ with $9\cdot 10^5-1\cdot 10^6$ model simulations in 2--3 hours, a speedup of \textit{at least} fourteen times compared to \texttt{standard}. In fact, the total speedup is likely to be much larger, as we do not know how longer \texttt{standard}  should have run to reach satisfying inference. Notice that, for this example, the guided \texttt{hybrid} is plagued by a very small ESS (see the Supplementary Material {G}),  so the posterior variability is different between runs. A possibility to tackle this may be to incorporate guided proposal samplers into the SMC-ABC of \cite{del2012adaptive} (or vice versa), which is designed to progressively, albeit slowly, reduce $\delta$ while maintaining a reasonably high ESS value. \red{Doing this would, on the one hand, open  some theoretical  questions (e.g. whether the resulting proposal kernel satisfies the detailed balance equation) and, on the other hand, introduce a MCMC step which would modify all weights, something which goes beyond the scope of this work.}
Note however that, despite the low ESS in this example, the bulk of the posteriors resulting from \texttt{hybrid} resembles that from the ABC-Gibbs method. 
 The non-guided \texttt{olcm} yields more satisfactory results than \texttt{standard} (see the Supplementary Material {G}), but the corresponding threshold value after 1 million model simulations was still $\delta=1.87$, so it was still fairly spread compared to \texttt{hybrid}. Moreover, \texttt{olcm} took about 55 hours to reach $\delta=0.70$ compared to about 2--3 hours for \texttt{hybrid} to reach $\delta=0.62$, with a median speedup of 20 times. {ABC-Gibbs was the fastest one (with a running time of 30 minutes), which is not surprising, as that scheme is especially suited for hierarchical models.}

\subsection{Recruitment boom-and-bust model with highly skewed summaries}\label{sec:recruitment}

We now challenge our guided proposals samplers, whose non-copula (more precisely, non-$t$-copula) approaches have been constructed assuming joint normality of the particle pairs $(\theta,s)$, on another example characterised by highly non-Gaussian summary statistics, namely, the \textit{recruitment boom-and-bust model}. This is a  stochastic discrete time model which may be used to describe the fluctuation of population sizes over time. The model is characterised by four parameters $\theta=(r,\kappa,\alpha,\beta)$, with small values of $\beta$ (as used here to generate our data) giving rise to highly non-Gaussian summary statistics.
This case study, which we fully describe and analyze in the Supplementary Material {H},  was also considered in \cite{fasiolo2018extended}, \cite{an2018robust} and \cite{picchini2020adaptive} in the context of MCMC via  synthetic likelihood,  to test how that methodology, constructed under approximately Gaussian distributed summary statistics, performed.  
As a reference gold-standard posterior is unavailable, here we compare guided and non-guided ABC approaches with  the robustified semiparametric (Bayesian) synthetic likelihood approach of \cite{an2018robust}, denoted semiBSL.  Note that this case study uses twelve summary statistics, three times as many as the number of parameters to infer, a setting where ABC is expected to struggle with its curse-of-dimensionality. In fact, the ABC posteriors are more spread than the posterior returned by semiBSL, which concentrates around the true parameter values thanks to its (semi)parametric nature. Among our proposed approaches, guided methods without ``optimised'' covariances occasionally appear mode-seeking, as discussed and observed elsewhere. However,  some of the copula-methods (notably Gaussian copulas with triangular marginals) behave quite similarly to semiBSL. More generally, guided approaches show higher acceptance rates, similar (if not better) performances and quicker runtimes than  the considered traditional non-guided ABC approaches.

\subsection{Lotka-Volterra}\label{sec:lotka-volterra}

\textcolor{black}{We now test our guided proposals on a case-study which is often considered in the likelihood-free inference literature (see for example \citealp{owen2015likelihood,owen2015scalable}), namely the Lotka-Volterra model, a process (here expressed as a continuous time Markov jump process) used in population dynamics and system biology to describe the interactions between species and chemical components, respectively. In the chosen formulation, the model is characterised by three reactions and three parameters, as detailed 
in Supplementary Section I. There we also report all inferential results obtained when comparing our guided method \texttt{blokedopt} with the best non-guided method \texttt{olcm} on 10 independent runs. 
Not only \texttt{blockedopt} had higher acceptance rates than \texttt{olcm} in most of the iterations and run, with the latter requiring approximately twice as many model simulations than \texttt{blockedopt} and a total three hours longer runtime (12.8h with \texttt{olcm} vs 9.4h with \texttt{blockedopt}), but led to smaller Wasserstein distances (with respect to a reference posterior) and thus more accurate inference than \texttt{olcm}.
}

\subsection{Cell motility and proliferation with high-dimensional summaries}\label{sec:cell}

Finally, we consider a simulated, yet realistic, study of cell movements characterised by high dimensional summary statistics, a well-known challenge for ABC. We initially considered 145 summary statistics, then 289 and finally 433. Model details and inferential results are reported in the Supplementary Material {J}. \textcolor{black}{While our proposed methods have not been designed to specifically deal with high-dimensional summary statistics (as they would require the inversion of high-dimensional covariance matrices of summary statistics), it is still interesting to test their performance with such feature.} The main take away is that guided and non-guided SMC-ABC  are able to deal with such large dimensionalities, whereas for summaries of size 289 (and larger), the Bayesian synthetic likelihood MCMC sampler of \cite{price2018bayesian} (the original version, we did not consider more recent developments) was unable to mix even for starting parameters set at the ground truth values.  With guided and non-guided SMC-ABC, we managed to perform inference using summaries with up to dimension 433, with the former requiring many fewer model simulations \textcolor{black}{(when using 145 summaries, we required 117,496 model calls with our guided \texttt{hybrid} with Gaussian proposal, versus the 132,963 of non-guided \texttt{olcm}; when using 433 summaries, we required 213,984 model calls for \texttt{hybrid} vs 241,57 for \texttt{olcm}) and a lower running time than the latter (in particular, for a single run, 6.1 vs 8.1 minutes with 145 summaries and 25.2 vs 32.4 minutes with 433 summaries).}

\section{Discussion}\label{sec:discussion}

We introduced a range of multivariate Gaussian and 
copula-based (both Gaussian and t {copulas}, with six considered distributions of
the marginals of the copula) proposal samplers to accelerate inference when using sequential ABC methods, notably sequential importance sampling (SIS-ABC) and sequential Monte Carlo (SMC-ABC). The acceleration is implied by the construction of proposal samplers that are made conditional to the summary statistics of the data 
(which is why we called them ``guided''),  such that the proposed draws rapidly converge to the bulk of the posterior distribution. We challenged our samplers by considering posteriors with multimodal surfaces (two-moons, Section \ref{sec:twomoons}), highly non-Gaussian summary statistics (two-moons, Section \ref{sec:twomoons} and recruitment boom-and-bust, Section \ref{sec:recruitment}), high dimensional parameter space (hierarchical g-and-k, Section \ref{sec:g-and-k}), high dimensional summaries with hundreds of components (hierarchical g-and-k, Section \ref{sec:g-and-k} and cell motility model, Section \ref{sec:cell}), and highly correlated posteriors (twisted model, Section \ref{sec:twisted}). In all these case studies, on the one hand, our methods obtained satisfactory ABC inference, similar, if not better, to non-guided sequential ABC schemes, 
in particular compared to the most commonly implemented
proposal sampler (found in all the most used ABC software packages and papers considering
SMC-ABC), which in fact we named \texttt{standard} (\citealp{beaumont2009adaptive}, but we also extensively compared with the \texttt{olcm} sampler of \citealp{filippi2013optimality}).  On the other hand, thanks to being guided, they returned inference much more rapidly than the non-guided proposal samplers. In particular, 
 for a fast to simulate generative model with a low-dimensional parameter space (two-moons, Section \ref{sec:twomoons}), our methods were already 4--6 times faster than customary non-guided SMC-ABC schemes, suggesting the possibility of even higher accelerations for more expensive simulators and/or higher dimensional parameter/summary statistics spaces. This was indeed observed in a challenging case study (hierarchical g-and-k model, Section \ref{sec:g-and-k}), where the non-guided SMC-ABC sampler \texttt{standard}, which is the typical default option in most software implementing SMC-ABC, was \textit{at least} fourteen times slower than our  guided ABC samplers in approaching a gold-standard ABC posterior. For guided methods showing an ESS lower
than \texttt{standard} and \texttt{olcm}, the corresponding “optimised” version managed to considerably increase
the ESS values.\\
\indent
Among the introduced guided samplers, the copula-based samplers are general and flexible. Among them, those based on Gaussian copulas may be preferred, yielding higher ESS values (for some marginals higher than the non-copula guided samplers) and being slightly faster than the $t$ copulas, while yielding similar performances. {However guided non-copula Gaussian samplers are also competitive, and perform notably better than the non-guided samplers.} Overall, the best copula-based sampler is the \texttt{cop-blocked} with triangular {or mixed} marginal distributions, followed by either \texttt{cop-blockedopt} or \texttt{cop-hybrid} with uniform marginals. However, copula-based samplers involve more operations to produce a proposal and may use less optimized numerical libraries compared to multivariate Gaussian samplers. This difference is negligible if the model simulator is expensive, as in this case the computational bottleneck will be the forward model simulation, but may be less so for particularly simple simulators (e.g. the two-moons), for which the copula-based samplers may then be slower than the guided multivariate Gaussians. This is not a disadvantage if the geometry of the posterior is such that the copula-model better adapts to its exploration, as it happens for the twisted model (for \texttt{cop-blocked} with triangular marginals). \\
\indent 
The proposed approach opens to a number of possible avenues of investigation, e.g.  guided copula-based SMC-ABC samplers, {non-parametric guided copula-based samplers, where the copula and the marginals are fitted non-parametrically from the available data,} or \lq\lq fully\rq\rq\ copula-based sequential samplers, where {a copula is placed on $(\theta,s)$ instead of $\theta|s$}. Embedding our guided proposals into ensemble Kalman inversion (EKI, see \citealp{chada} for a recent review) would also be of interest. Overall, our guided proposals are easy to construct and are rapidly computed from accepted parameters and summary statistics, thus not introducing any substantial overhead. For example, there is no need to construct and train a deep neural network (unlike in the guided method of \citealp{chen2019adaptive}){, or perform a high-dimensional non-parametric optimization}. We believe that the simplicity, and effectiveness, of our proposal samplers makes them appealing and easy to incorporate into the user's toolbox.

\section*{Supplementary Material}
The supplementary material contains both additional methodological details (e.g. the construction of the optimized covariance matrices $\Sigma_t$ \eqref{CovSISABC} and optimized variances $\sigma^{*2}_k$ \eqref{eq:var_fullcondopt}, the derivation of the underlying parameters of the marginal distributions for the guided copula-based SIS-ABC sampler), and further results/figures for the considered simulation studies.

\subsubsection*{Acknowledgments}
UP acknowledges support from the Swedish Research Council (Vetenskapsrådet 2019-03924) and the Chalmers AI Research Centre (CHAIR).

 \bibliographystyle{abbrvnat}
\bibliography{biblio}

\newpage
\appendix

\begin{center}
\LARGE
\title{Supplementary Material to
``Guided sequential ABC schemes for intractable Bayesian models''}
\end{center}

\normalsize

\section{Construction of optimised variances/covariance matrices}\label{appendix0}
Here, we first detail the derivation of the ``optimized covariance matrix'' $\Sigma_t$ for the guided SIS-ABC (eq. (7) in the main text), and afterwards we consider the derivation of the variance ${\sigma^*_k}^2$ for \texttt{fullcondopt} (eq. (13) in the main text).
By considering $q_t(\theta^{*})\equiv \mathcal{N}(\hat{m}_{\theta|s_y,t-1},\Sigma_t)$ for unknown $\Sigma_t$ and fixed $\hat{m}_{\theta|s_y,t-1}$ defined as in eq. (4) in the main text, we have that
\begin{equation}
   Q(q_t,\delta_t,s_y)=\int  \pi_{\delta_t}(\theta^{*}|s_y)\biggl(-\frac{d_\theta}{2}\log(2\pi)-\frac{1}{2}\log|\Sigma_t|-\frac{1}{2}(\theta^*-\hat{m}_{\theta|s_y,t-1})'\Sigma^{-1}_t(\theta^*-\hat{m}_{\theta|s_y,t-1})  \biggr) d\theta^*.\label{eq:Q-blockedopt} 
\end{equation}
 To maximize $Q$ with respect to $q_t$ (i.e. with respect to $\Sigma_t$), we can disregard several terms in \eqref{eq:Q-blockedopt} that do not depend on $\Sigma_t$. By also noting that $\int \pi_{\delta_t}(\theta^{*}|s_y)d\theta^*=1$, we obtain
 \begin{eqnarray*}
   Q(q_t,\delta_t,s_y) & \propto -\frac{1}{2}\log|\Sigma_t|-\frac{1}{2} \int  \pi_{\delta_t}(\theta^{*}|s_y)(\theta^*-\hat{m}_{\theta|s_y,t-1})'\Sigma^{-1}_t(\theta^*-\hat{m}_{\theta|s_y,t-1}) d\theta^*\\
&=\frac{1}{2}\log|\Sigma_t^{-1}|-\frac{1}{2} \int  \pi_{\delta_t}(\theta^{*}|s_y)(\theta^*-\hat{m}_{\theta|s_y,t-1})'\Sigma^{-1}_t(\theta^*-\hat{m}_{\theta|s_y,t-1}) d\theta^*,
\end{eqnarray*}
where in the latter equality we have used that for a squared matrix $A$, $|A^{-1}|=|A|^{-1}$.
 Now, consider a symmetric matrix $M$ and define $\Sigma_t=M^{-1}$. Note that maximizing $Q$ to find the optimal $\Sigma_t$ is equivalent to maximizing the real-valued function $g(M)$ with respect to $M$, where
\[
g(M)=\log|M|-\int  \pi_{\delta_t}(\theta^{*}|s_y)(\theta^*-\hat{m}_{\theta|s_y,t-1})'M(\theta^*-\hat{m}_{\theta|s_y,t-1}) d\theta^*.
\]
The resulting maximization leads to eq. (7) in the main text.

As for the derivation of ${\sigma^*_k}^2$, by getting rid of terms not depending on any $\sigma_k^{*}$, we have that $Q(q_t,\delta_t,s_y)=\int  \pi_{\delta_t}(\theta^{**}|s_y)\log q_t(\theta^{**}|\theta^*)d\theta^{**}$ is proportional to
\begin{align*}
Q(q_t,\delta_t,s_y) &\propto -\int \pi_{\delta_t}(\theta^{**}|s_y)\biggl(\sum_{k=1}^{d_\theta}\log\sigma_k^*+\sum_{k=1}^{d_\theta}\frac{(\theta_k^{**}-\hat{m}_{k|s_y,t-1}^*)^2}{2\sigma^{*^2}_k}\biggr)d\theta^{**}_1\cdots d\theta^{**}_{d_\theta}\\
&= -\sum_{k=1}^{d_\theta}\log\sigma_k^* -\frac{1}{2\sigma_k^{*^2}}\int \pi_{\delta_t}(\theta^{**}|s_y)\sum_{k=1}^{d_\theta}(\theta_k^{**}-\hat{m}_{k|s_y,t-1}^*)^2d\theta^{**}_1\cdots d\theta^{**}_{d_\theta}.\
\end{align*}
By setting equal to zero the first derivative of $Q$ with respect to a generic $\sigma_k^*$, we get the result.

\section{Gaussian and t copula distributions  and densities}\label{appendixB}
Here, we report the copula distribution functions $C(u_1,\ldots, u_{d_{\theta}})$ and the copula densities $c(u_1,\ldots, u_{d_{\theta}})$ of the Gaussian and t copulas, starting from the former. Let $R \in [-1,1]^{d_{\theta}\times d_{\theta}}$ be a correlation matrix, let $\Phi_R$ denote the joint cumulative distribution function (cdf) of a multivariate Gaussian distribution with mean vector 0 and correlation matrix $R$ and $\Phi^{-1}$ be the inverse cdf of an univariate standard normal distribution. The Gaussian copula $C^{\textrm{Gaussian}}_R$ is then given by
\[
C^{\textrm{Gaussian}}_R(u_1,\ldots,u_{d_{\theta}})=\Phi_R\left(\Phi^{-1}(u_1),\ldots,\Phi^{-1}(u_{d_{\theta}})\right).
\]
Denote $\eta=(\eta_1,\ldots,\eta_{d_\theta})^T$, with $\eta_i=\Phi^{-1}(u_i), i=1,\ldots, u_{d_\theta}$. Then the Gaussian copula density is given by
\[
c^{\textrm{Gaussian}}_R(u_1,\ldots,u_{d_\theta})=\frac{1}{|R|^{1/2}}\exp\left(-\frac{1}{2}\eta^T(R^{-1}-I)\eta\right),
\]
 where $|R|$ denotes the determinant of $R$ and $I$ is the identity matrix.
 The $t$-copula is defined similarly. Let $t(\nu,0,R)$ denote a $d_\theta$ dimensional t distribution with zero mean vector and correlation matrix $R \in [-1,1]^{d_{\theta}\times d_{\theta}}$. Let $F^t_{R,\nu}$ and $f^t_{R,\nu}$ denote the cdf and probability density function of such distribution. Let $F^{-1}_{t,\nu}$ be the inverse cdf of a univariate $t$-distribution with $\nu$ degrees of freedom. The t-copula $C^t_{R,\nu}$ is then given by
\[
C^{\textrm{t}}_{R,\nu,\nu}(u_1,\ldots,u_{d_{\theta}})=F^t_{R,\nu}\left(F^{-1}_{t,\nu}(u_1),\ldots,F^{-1}_{t,\nu}(u_{d_{\theta}})\right),
\]
with copula density given by
\[
c^{\textrm{t}}_{R,\nu,\nu}(u_1,\ldots,u_{d_{\theta}})=\frac{f^t_{R,\nu}\left(F^{-1}_{t,\nu}(u_1),\ldots,F^{-1}_{t,\nu}(u_{d_{\theta}})\right)}{\prod_{i=1}^nf^t_{\nu}(F^{-1}_{t,\nu})}.
\]
 \textcolor{black}{Note that the degrees of freedom could possibly vary between the Student's t marginals, i.e. $\nu\neq \nu_i, i=1,\ldots, d_\theta$, obtaining what is known as a $t$-meta distribution. As no a-priori information is known on $\theta|s_y$ (see Remark 3 in Section 3.4 of the main text), here we consider all degrees of freedom to be the same, obtaining a multivariate $t$ distribution when all marginals are also Student's t distributions. Note that, for a fixed correlation $R_{ij},i,j=1,\ldots, d_\theta,i\neq j$, the underlying pairwise (upper/lower) tail dependence decreases in $\nu$ (\citealt{Embrechtsetal}), reaching $0$ as $\nu\to\infty$, when the $t$-copula becomes a Gaussian copula. Throughout the experiments, we fix $\nu=5$, representing a compromise between more (resp. less) heavy-tailed coupling obtained for $\nu=3$ (resp. $\nu=10$). The impact of $\nu$ on the results is illustrated for the two-moon simulation study in the Supplementary Material D.}
\begin{figure}[t!]
    \centering
    \includegraphics[scale=.4]{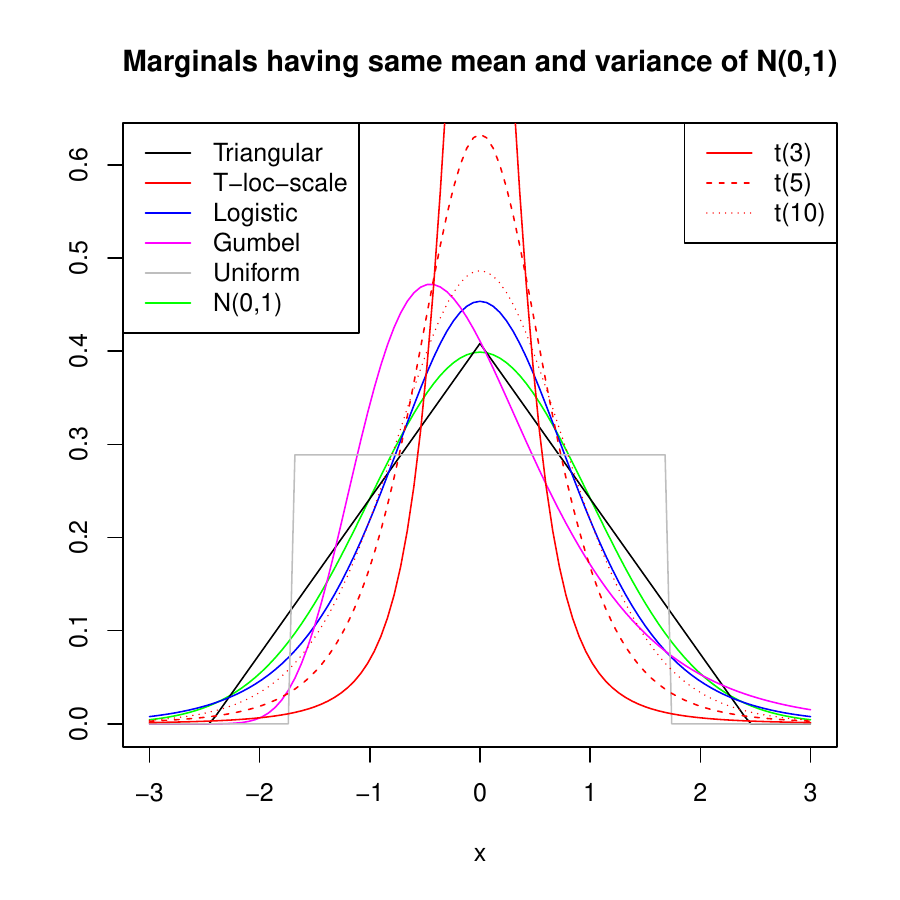}
    \caption{Probability density plots of the triangular (black), location-scale $t$ (red), logistic (blue), Gumbel (magenta) and uniform (gray) distributions with same mean and variance of the standard normal (green).}
    \label{fig:marginals}
\end{figure}

\section{Marginal distributions for the guided copula-based SIS-ABC samplers}\label{AppendixA}
\textcolor{black}{Throughout this work, we propose six marginal distribution families, choosing their parameters such that the random variables have the same mean and variance of the guided Gaussian SIS-ABC proposals, see Section 3.4 of the main text. A description of how this is achieved is reported below. }\\
{\bf Location-scale Student's t distribution.} If $X_j$ follows a location-scale Student's t distribution with degrees of freedom $\nu>0$, location parameter $\mu\in\mathbb{R}$ and scale parameter $\sigma>0$, its mean and variance are given by $\mu, \textrm{ for } \nu>1$ and $\sigma^2\frac{\nu}{\nu-2},\textrm{ for }\nu >2$, respectively.  Setting $\mathbb{E}[X_j]=m^*_j$ and $\textrm{Var}(X_j)=S^*_{jj}$, we get $
\mu=m^*_j$,  $\sigma^2=(\nu-2)S^*_{jj}/\nu$, 
where $\nu$ can be chosen arbitrarily. \textcolor{black}{In our work, we fix $\nu=5$ in all simulation studies. This guarantees that
the kurtosis is finite (as $\nu>4$) and the distribution is different enough from the normal distribution, as the former approaches the latter when $\nu\to\infty$ \textcolor{black}{(see also Figure \ref{fig:marginals})}.
}\smallskip

\noindent {\bf Logistic distribution.} If $X_j$ follows a logistic distribution with location parameter $\mu\in\mathbb{R}$ and scale $s>0$, its mean and variance are given by $\mu$ and $s^2\pi^2/3$, respectively. Setting $\mathbb{E}[X_j]=m^*_j$ and $\textrm{Var}(X_j)=S^*_{jj}$, we get $\mu=m^*_j,  s=\sqrt{3S^*_{jj}}/\pi$. \smallskip

\noindent
{\bf Normal distribution.} If $X_j$ follows a Gaussian distribution with parameters $\mu_j, \sigma^2_j$, we simply set $\mu_j=m^*_j$, $\sigma^2_j=S^*_{jj}$.\smallskip

\noindent
{\bf Triangular distribution.} If $X_j$ follows a triangular distribution in $[a,b]$, $a\leq b$ with mode $c\in[a,b]$, its mean and variance are given by $
\mathbb{E}[X_j]=(a+b+c)/3, \textrm{Var}(X_j)=(a^2+b^2+c^2-ab-ac-bc)/18$. 
Setting $c=m^*_j$, $\mathbb{E}[X_j]=m^*_j$ and $\textrm{Var}(X_j)=S^*_{jj}$ we get $a=m^*_j-\sqrt{6S^{*}_{jj}}, b=m^*_j+\sqrt{6S^{*}_{jj}}, c=m^*_j$. \smallskip

\noindent
{\bf Uniform distribution.} If $X_j$ follows a uniform distribution in $[a,b]$, its mean and variance are given by $\mathbb{E}[X_j]=(a+b)/2$, $\textrm{Var}(X_j)=(b-a)^2/12$. Setting $\mathbb{E}[X_j]=m^*_j$ and $\textrm{Var}(X_j)=S^*_{jj}$, we get $a=m^*_j-\sqrt{3S_{jj}^*},  b=m^*_j+\sqrt{3S_{jj}^*}$.

\noindent
{\bf Gumbel distribution.} \textcolor{black}{If $X_j$ follows a Gumbel distribution with location parameter $\mu\in\mathbb{R}$ and scale parameter $\beta>0$, its mean and variance are given by $\mathbb{E}[X_j]=\mu+\beta\gamma$ and $\textrm{Var}(X_j)=\pi^2\beta^2/6$, where $\gamma$ denotes the Euler-Mascheroni constant. Setting $\mathbb{E}[X_j]=m_j^*$ and $\textrm{Var}(X_j)=S_{jj}^*$, we get $\mu=m_j^*-\gamma\sqrt{6S_{jj}^*}/\pi,  \beta=\sqrt{6S_{jj}^*}/\pi.$ Note that Matlab uses a different parametrization for which $\mathbb{E}[X_j]=\mu-\beta\gamma$, which is why $\mu=m_j^*+\gamma\sqrt{6S_{jj}^*}/\pi$ on the GitHub code.}
\textcolor{black}{
\subsection{Qualitative comparison between the different marginals}
In Figure \ref{fig:marginals}, we report the probability density functions of the different marginals with the parameters obtained as described before. 
Among all choices, the triangular distribution, followed by the uniform, yields the best performance throughout the considered case studies, as seen in Section 4. The intuition is the following.
 By choosing the parameters of the uniform/triangular distributions such that the random variables have the same mean $\mu^*_i$ and variance $S^*_{ii}$ of the guided Gaussian samplers for $\theta_i|s_y$, we obtain that their support is $\mu_i\pm \sqrt{3}\sigma_i$ and $\mu_i\pm \sqrt{6}\sigma_i$, respectively, covering $91.67\%$ and $98.57\%$ of the corresponding Gaussian distribution $X_i\sim N(\mu^*_i,S^*_{ii})$, i.e. $\mathbb{P}(X_i\in \mu^*_i\pm \sqrt{3}S^*_{ii})=0.9167$ and $\mathbb{P}(X_i\in \mu^*_i\pm \sqrt{3}S^*_{ii})=0.9857$, respectively. Note also that, being symmetric, these distributions have a mode in $\mu_i$, with the density of the triangular distribution in $\mu_i$ being $1/\sqrt{6\sigma_i^2}$, very close to that of the normal distribution ($1/\sqrt{2\pi\sigma^2_i}$). Hence, with this parameter choice, the triangular distribution closely approximates a normal distribution (see also Figure \ref{fig:marginals}), cutting off only a minimal part of the tails. The uniform distribution, instead, has the advantage of being less peaked than the Gaussian/triangular, with the drawback of a smaller support and thus an even further cut of the tails. This may be beneficial at initial iterations, when the sampler is moving away from the prior (t=1), but may be leading to over-confident results at later stages, as observed in Section 4.1 and 4.2. The other marginals have tails heavier than the Gaussian distribution, which may be why they do not introduce the same speed-up as the other two marginals.}

\section{Vectorized computation of the weighted covariance matrix for $(\theta,s)$}

The weighted covariance matrix $S$ given in equation (3) of the main text can be efficiently computed in a vectorized way as follows. Following the same notation established in that section, say that $x^{(i)}=(\theta^{(i)},s^{(i)})$ is $d$-dimensional and $i=1,...,N$. Now, let \texttt{x} be the $d\times N$ matrix whose $i$-th column is $x^{(i)}$, use ' to denote transposition and \texttt{w} to denote the $(1\times N)$-dimensional vector of normalized weights. Use \texttt{'} to denote transposition, \texttt{*} to denote matrix multiplication and \texttt{.*} for element-wise multiplication.
Then, we have the following Matlab code, which can easily be adapted to other languages:  
\begin{verbatim}
weigh_mean = w * x';              
S = x' - repmat(weigh_mean, N, 1);           
S = S' * (S .* repmat(w', 1, d));      
S = S ./ (1-sum(w.^2));             
S = 0.5 * (S + S');                 \end{verbatim}
The resulting \texttt{S} is of course a $d\times d$ matrix. The last line in the code is optional but typically useful when small floating-point approximations can make \texttt{S} not exactly symmetric.

\section{Two-moons model}
In this section, we report the definition, the simulated summaries distribution, ESS and the running-times for the copula-based samplers for the two-moons. The output of the two-moons model is a two-dimensional vector $z\in \mathbb{R}^2$ generated via
\begin{align*}
    a &\sim U(-\pi/2,\pi/2),\quad  r\sim \mathcal{N}(0.1,0.01^2)\\
    p &= (r \cos(a)+0.25, r \sin(a)), \quad z = p+\biggl(-\frac{|\theta_1+\theta_2|}{\sqrt{2}}, \frac{-\theta_1+\theta_2}{\sqrt{2}}\biggr),
\end{align*}
see the Supplementary Material in \cite{greenberg2019automatic} for further discussion. Here, we take priors $\theta_j\sim U(-1,1)$ ($j=1,2$), assume observed data $y=(0,0)$ and consider the identity function  as summary statistics function, i.e. $S(z)=z$, so the summary statistics are the data themselves. We fix the number of particles to $N=1000$, and produce ten independent estimation runs for each method, comparing them at eleven prefixed $\delta_t$ values, namely  $\delta_t\in \{4, 3, 2, 1, 0.5, 0.4, 0.3, 0.2, 0.1, 0.08, 0.06\}$.

\begin{figure}[t!]
    \centering
    \includegraphics[scale=0.4]{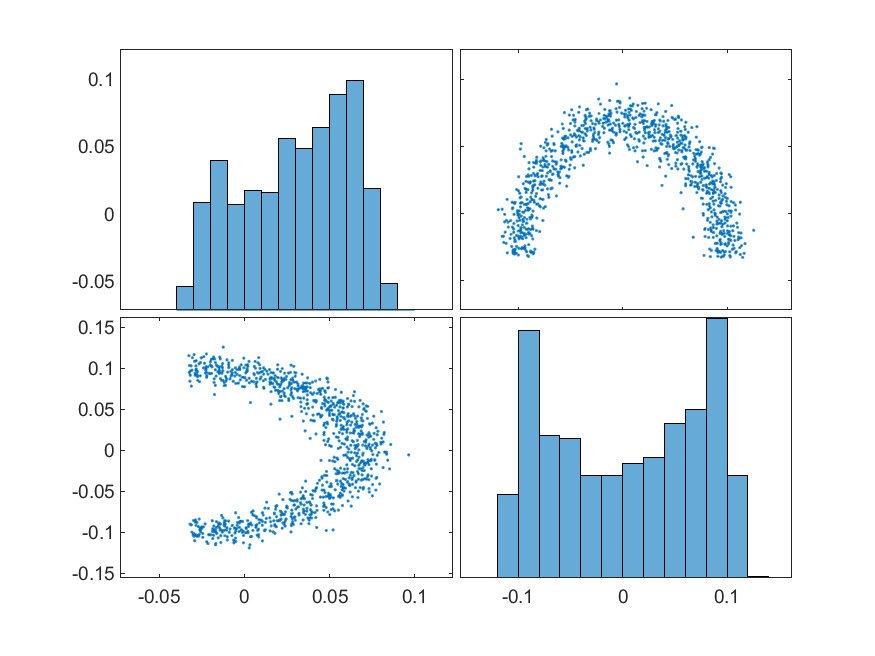}
    \caption{Two-moons: scatter plot matrix of 1,000 summaries simulated with $\theta_1=0.2$ and $\theta_2=0.2$. The diagonal reports the histogram for each dimension of the summary statistics and the other entries give the pairwise associations.}
    \label{fig:twomoons-simsum-scatter}
\end{figure}

\begin{figure}[t!]
\hspace{-.5cm}
\includegraphics[width=1.05\textwidth]{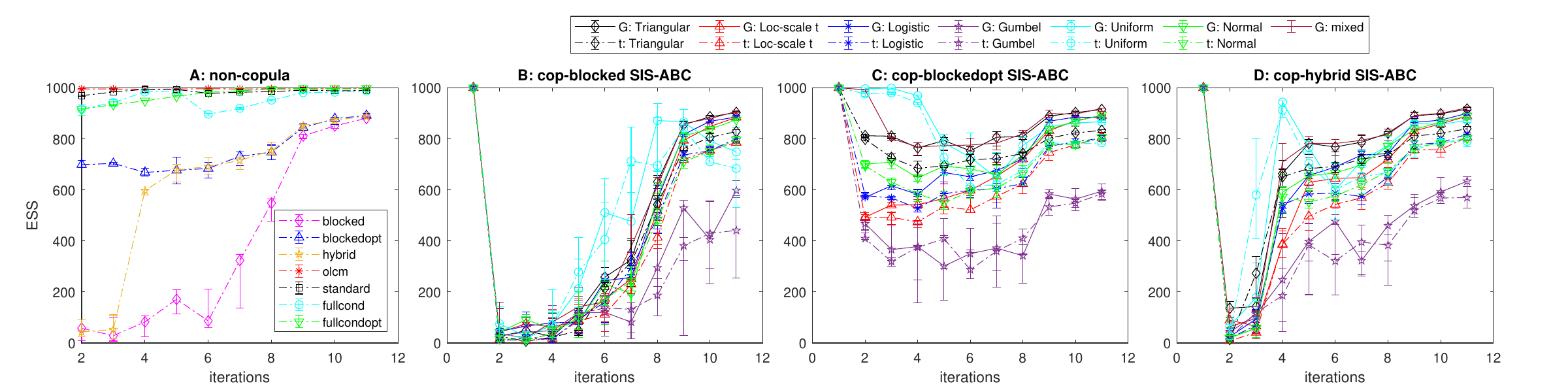}
    \caption{Two-moons: median ESS and error bars with  first and third quartile across ten independent runs at each iteration. Panel A: Non-copula based methods, both guided and non-guided;
 Panel B: \texttt{cop-blocked}; Panel C: \texttt{cop-blockedopt};  Panel D: \texttt{cop-hybrid}. The copula-based methods are derived using either Gaussian (G) or t copulas for different marginals, as described in the legend.}
        \label{fig:ESS_twomoon}
\end{figure}

The two-dimensional summary statistics (i.e. the simulated data) are reported in Figure \ref{fig:twomoons-simsum-scatter}, suggesting a non-Gaussian joint distribution. In Figure \ref{fig:ESS_twomoon}, we report the median ESS at each iteration obtained on ten independent runs for all methods. The guided SMC-ABC \texttt{fullcond} and \texttt{fullcondopt} have the highest ESS among the guided methods, closely resembling those of the non-guided approaches (particularly \texttt{fullcondopt}). Among the guided SIS-ABC, \texttt{cop-blockedopt} with Gaussian copula and triangular \textcolor{black}{or mixed} marginals (black and brown solid lines, respectively) have the highest ESS, with  \texttt{cop-hybrid} behaving similarly well starting from iteration five. Moreover, in Table \ref{TableTimestwo-moon}, we report the wallclock times of the copula-based methods to produce ten runs with the Gaussian and t copulas and five considered marginals, with the Gaussian copula-based samplers being faster than the t copula ones. \textcolor{black}{The wallclock times of the Gumbel and mixed marginals are not reported as we ran them on a different machine.} 

\begin{table}
\begin{tabular}{|c|ccccc|ccccc|}
\hline 
& \multicolumn{5}{|c|}{Gaussian copula} & \multicolumn{5}{c|}{t copula}\\ \hline
& Tr & LoScT & Log  &  Unif& Norm& Tr & LoScT t& Log  &  Unif& Norm\\ \hline
\texttt{cop-blocked}&35.0& 52.5 &43.1& 2  3.4 & 8.1& 35.5& 56.2& 48.3  & 3.2 & 11.3\\
\texttt{cop-blockedopt}& 34.1& 49.8& 43.8 & 9.2 & 11.4 &43.7 & 71.2&  59.6 &   12.0  &13.8\\
\texttt{cop-hybrid} & 37.4 &54.6& 45.9 &  8.5  &9.4& 39.1& 54.9& 48.5&   11.2  &12.8\\ \hline
\end{tabular}
\caption{Two-moons: Total wallclock times (minutes) for ten runs of the \texttt{cop-blocked}, \texttt{cop-blockedopt} and \texttt{cop-hybrid} approaches with triangular (tr), location-scale t (LoScT), logistic (Log),  uniform (Unif) and normal (Norm) marginals.}\label{TableTimestwo-moon}
\end{table}

\section{Twisted-prior model}
\subsection{Derivation of \texttt{fullcondopt} with joint sampling of two components}\label{sec:fullcondoptblocked}

Here, we illustrate how to construct a version of \texttt{fullcondopt} for the 5-dimensional parameter $\theta$ considered in the twisted-prior model, incorporating a-priori knowledge of a high correlation between $\theta_1$ and $\theta_2$. In particular, we aim to sample $\theta_1$ and $\theta_2$ jointly, while the other components  $\theta_3$, $\theta_4$ and $\theta_5$ are proposed from corresponding 1-dimensional Gaussians, as illustrated in Section 3.5 of the main paper. 
In particular, for given $\theta^*=(\theta^*_1,...,\theta^*_5)$ we wish to propose $\theta_{(1,2)}^{**}=(\theta_1^{**},\theta_2^{**})$ jointly from $q(\theta^{**}_{1,2}|\theta^{*}_{3,4,5},s_y)$, that is $\theta_{(1,2)}^{**}$ is proposed in block, and this is why in the main paper we called this version \texttt{fullcondoptblocked}. We have the $2\times 1$ column vector (all column vectors are stacked)
\begin{equation}
    \hat{m}_{(1,2)|s_y,t-1} = \begin{bmatrix}
          \hat{m}_{1} \\
          \hat{m}_2
         \end{bmatrix}+\hat{S}_{(1,2), -(1,2)}(\hat{S}_{-(1,2),-(1,2)})^{-1}\biggl(
    \begin{bmatrix}
           \theta^{*}_{3} \\
           \theta^{*}_{4} \\
           \theta^{*}_{5} \\
            s_y 
        \end{bmatrix}
  -\begin{bmatrix}
          \hat{m}_{3} \\
          \hat{m}_{4} \\
          \hat{m}_{5} \\
          \hat{m}_s
         \end{bmatrix}\biggr).
\end{equation}
Then, we have the $2\times 2$ covariance matrix
\[
\hat{\Sigma}_{(1,2),t} = \sum_{l=1}^{N_0}\gamma_{t-1,l}(\tilde{\theta}_{t-1,(1,2),l}-\hat{m}_{(1,2)|s_y,t-1})(\tilde{\theta}_{t-1,(1,2),l}-\hat{m}_{(1,2)|s_y,t-1})',
\]
leading to  $q(\theta^{**}_{1,2}|\theta^{*}_{3,4,5},s_y)\equiv \mathcal{N}(\hat{m}_{(1,2)|s_y,t-1},\hat{\Sigma}_{(1,2),t})$. Overall, the full perturbed $\theta^{**}=\theta_{(1,...,5)}^{**}$ is proposed from (with notation abuse) \[q(\theta^{**}|\theta^*,s_y)=\mathcal{N}(\hat{m}_{(1,2)|s_y,t-1},\hat{\Sigma}_{(1,2),t})\times \prod_{k=3}^5 \mathcal{N}(\hat{m}^*_{k|s_y,t-1},\hat{\sigma}_{k|s_y,t-1}^2),\] where the first Gaussian in the product is bivariate, while the others $\mathcal{N}(\hat{m}^*_{k|s_y,t-1},\hat{\sigma}_{k|s_y,t-1}^2)$ are univariate as in Section 3.5 of the main text.

\subsection{Acceptance rates, seconds per particle, contour plots and ESS}

\begin{figure}[h!]
    \hspace{-1cm} \includegraphics[width=1.1\textwidth]{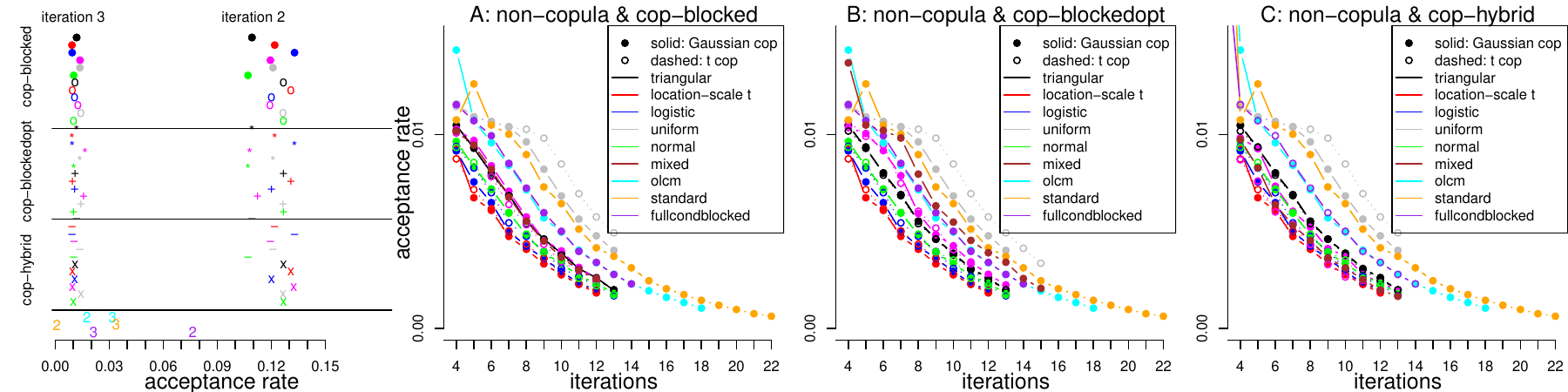}
    \caption{Twisted model: Median acceptance rates across ten independent runs. \textbf{Notice the x-axes are different between figures:} the left panel focuses on iterations 2 and 3, while the other panels on iteration 4 and onward. \textcolor{black}{Non-copula methods, namely \texttt{olcm} (cyan), \texttt{standard} (orange) and \texttt{fullcondoptblocked} (purple), are reported in all panels, together with \texttt{cop-blocked} (Panel A),} \texttt{cop-blockedopt} (Panel B) and \texttt{cop-hybrid} (Panel C), with Triangular (black), Location-scale t (red), Logistic (blue), Uniform (gray), Normal (green) and mixed (brown) marginals  distributions  and Gaussian copulas (top five points in each strip in the left panel, solid lines in the other panels) or t copulas (bottom five points in each strip in the left panel, dashed lines in the other panels). 
    }
    \label{fig:twisted_acceptrate}
\end{figure}

\begin{figure} 
\includegraphics[width=1.\textwidth]{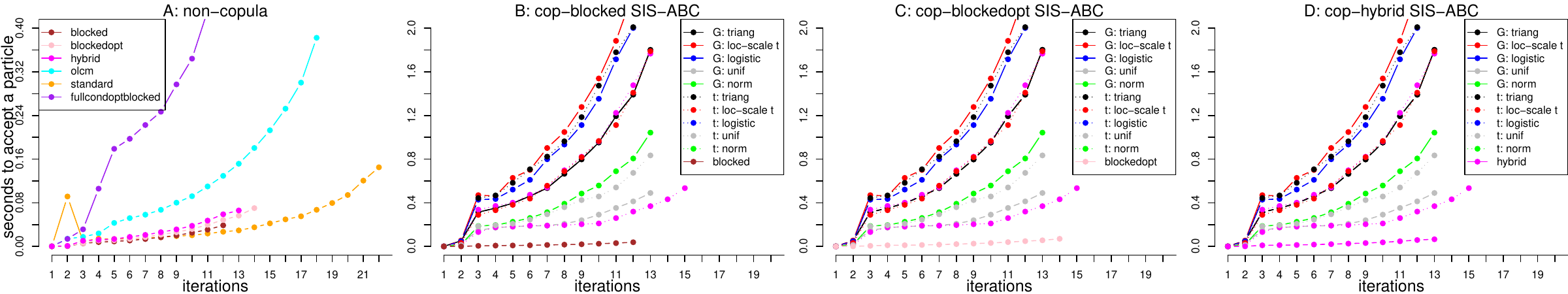}
    \caption{Twisted model: Median number of seconds required to accept a particle across ten independent runs. Panel A: non-copula based methods, both guided and non-guided; Panel B: \texttt{cop-blocked}; Panel C: \texttt{cop-blockedopt}; Panel D:  \texttt{cop-hybrid}. The copula-based methods are derived
using either Gaussian (G) or t copulas for different marginals, as described in the legend. \textcolor{black}{The results for the Gumbel and mixed marginals are not reported, as they ran on a different machine.} Note: we
had to choose different scales of the $y$-axes for comparing the number of seconds required to accept a
particle for the non-copula vs copula-based methods, in order to make the latter more readable.}
    \label{fig:twisted_seconds}
\end{figure}

\begin{figure}[h!]
\includegraphics[width=\textwidth]{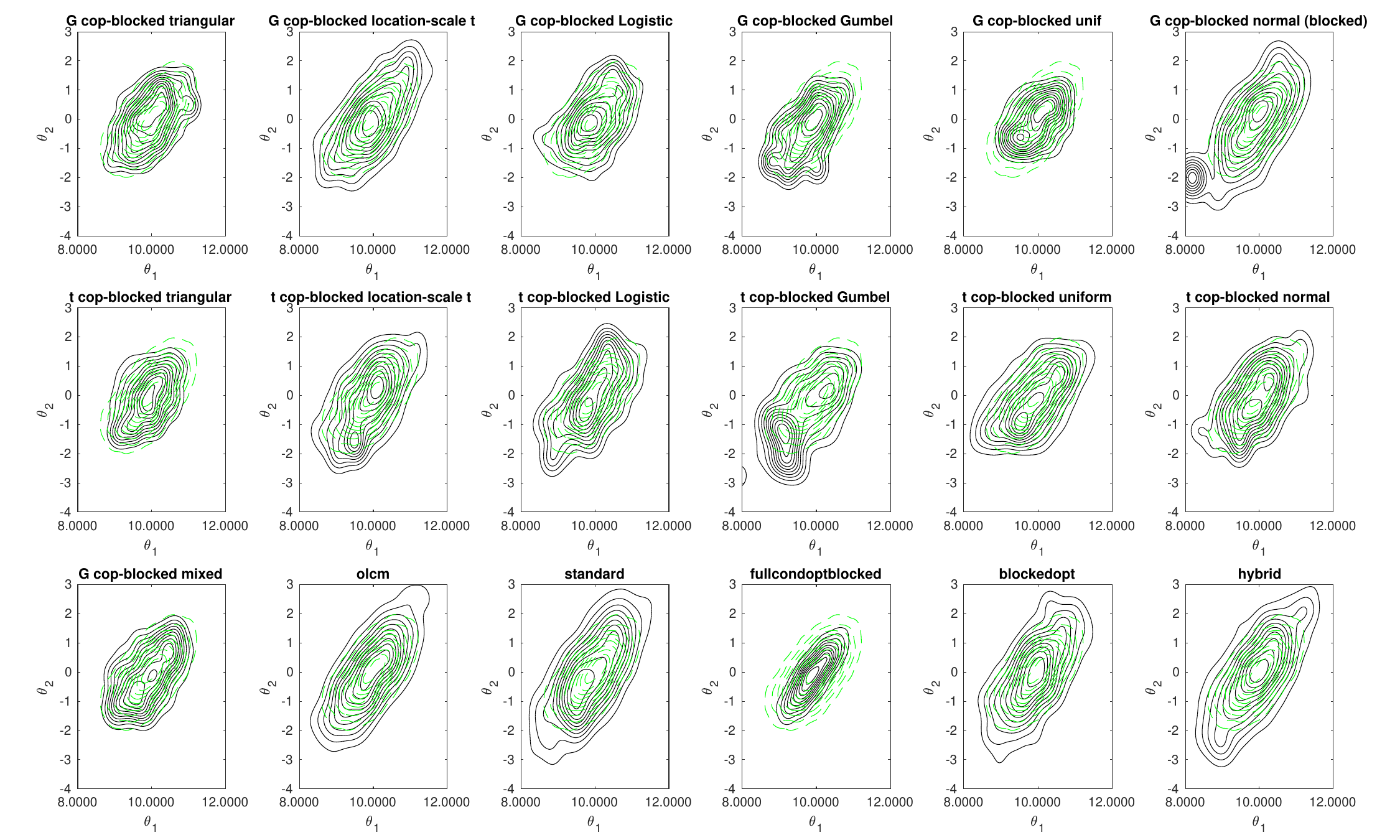}
\caption{Twisted model: Contour plots of the marginal ABC posteriors (black lines) and the exact MCMC posteriors (dashed green lines) of $(\theta_1,\theta_2)$ at the iteration yielding the smallest log-Wassertein distance, see Figure 5 in the main text.}\label{twisted_contouropt}
\end{figure}

\begin{figure}[t!]
\includegraphics[width=\textwidth]{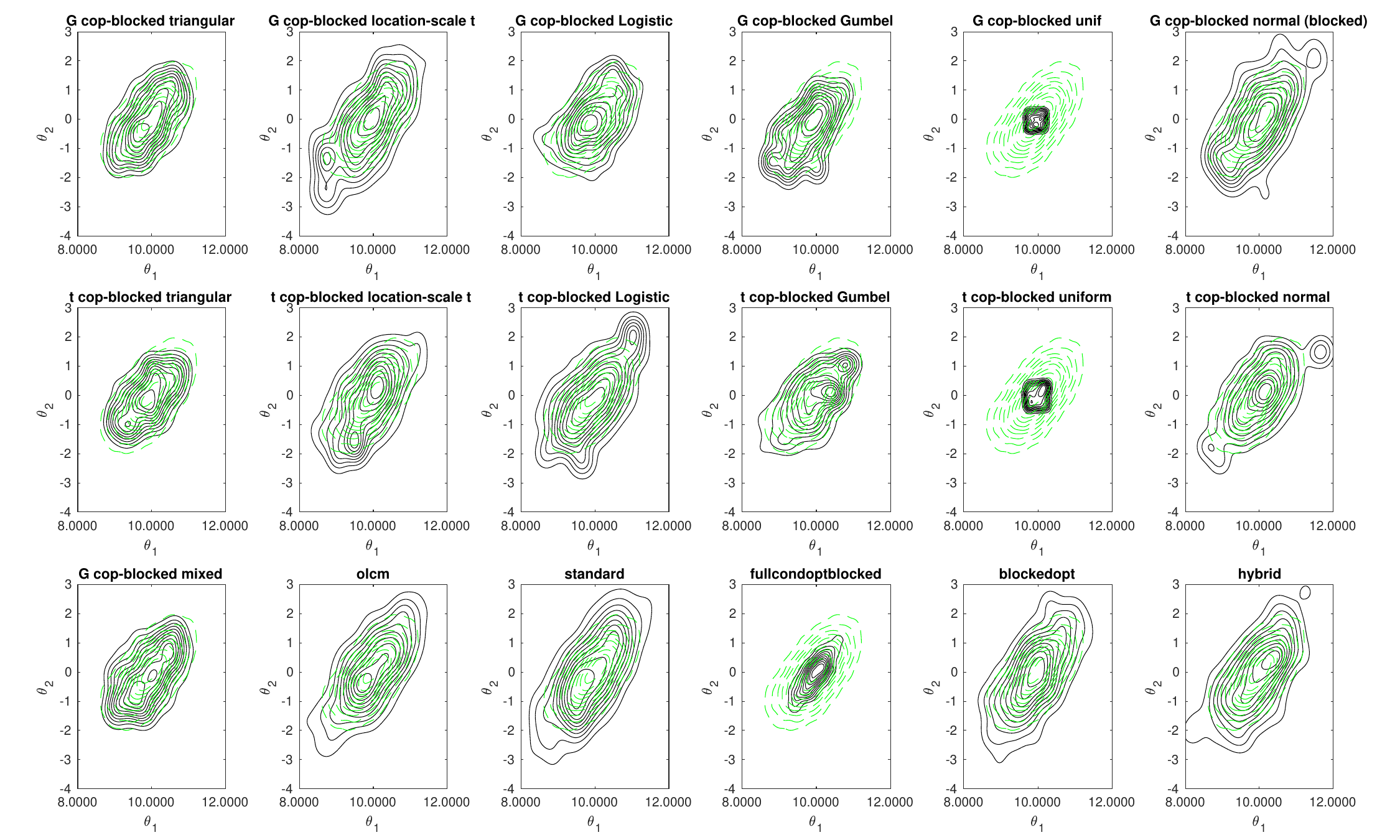}
\caption{Twisted model: Contour plots of the marginal ABC posteriors (black lines) and the exact MCMC posteriors (dashed green lines) of $(\theta_1,\theta_2)$ at the last iteration in one of the ten runs. Note that, for some methods, the results obtained at the last iteration do not necessarily correspond to the best inference (according to the Wasserstein distance). This happens for example for \texttt{fullcondoptblocked} (i.e. \texttt{fullcondopt} with blocked $(\theta_1,\theta_2)$, third row and third column) and \texttt{cop-blocked} with uniforma marginals, as seen comparing it with Figure 5 in the main text  and Figure \ref{twisted_contour015} here.}\label{twisted_contour}
\end{figure}

\begin{figure}[t!]
\includegraphics[width=\textwidth]{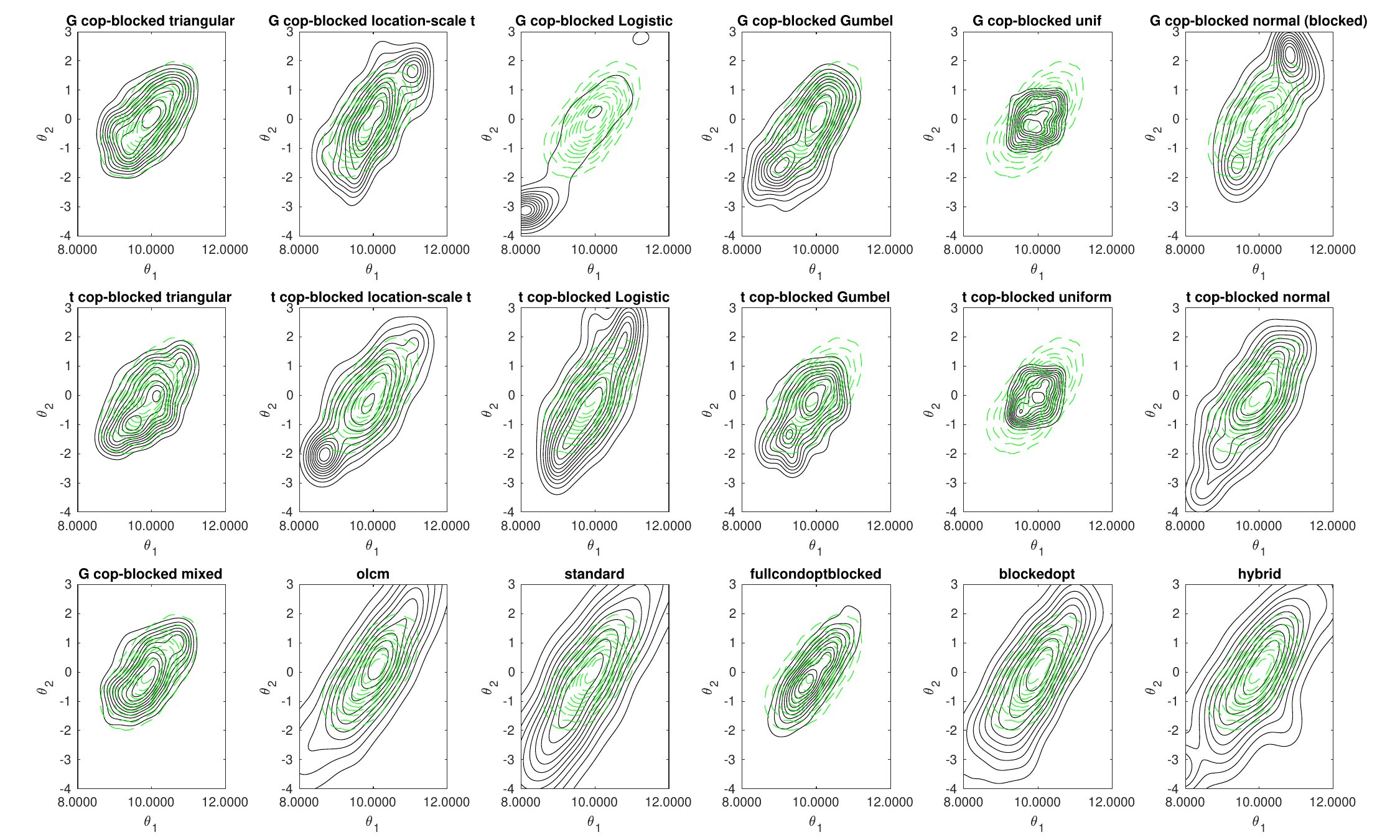}
\caption{Twisted model: Contour plots of the marginal ABC posteriors (black lines) and the exact MCMC posteriors (dashed green lines) of $(\theta_1,\theta_2)$ at the iteration when the acceptance rate goes below 1.5\% for the second consecutive time.}\label{twisted_contour015}
\end{figure}

In Figure \ref{fig:twisted_acceptrate}, we report the median acceptance rates at each iteration across ten independent runs, for all methods. At the first iteration, all approaches are equivalent, as accepted draws are all proposed form the prior. At the second iteration, \textit{all} guided approaches have higher acceptance rates than non-guided ones, around 8-13\%, see left panel in Figure \ref{fig:twisted_acceptrate}, with the non-guided \texttt{olcm} and \texttt{standard} having around 2\% and 0.1\% acceptance rates, respectively. Obviously, this impacts on the number of seconds required to accept a particle, see Panel A in Figure \ref{fig:twisted_seconds}. The second iteration is particularly crucial in terms of performance, as particles are still likely very widely scattered. 
 At iteration 3, guided methods have  1.5-2\% acceptance rates, while the non-guided recover towards a 3\% acceptance rate, after which all methods stabilize towards similar values, with the copula-based approaches with uniform, triangular and mixed marginals having slightly higher acceptance rates than the corresponding normal ones. Importantly, in Figure \ref{fig:twisted_seconds}, we see that the copula-based methods and  \texttt{fullcondoptblocked} appear to require much higher time to accept a particle than the other approaches.  However, some of the copula-based approaches (namely those with uniform, triangular or mixed marginals) and \texttt{fullcondoptblocked} (i.e. \texttt{fullcondopt} with $(\theta_1,\theta_2)$ proposed in block using the method described in the Supplementary Section \ref{sec:fullcondoptblocked}) do not 
 actually require to run for all iterations, as they reach the minimal Wasserstein distance way before the other methods, between 3 to 6 iterations, see Figure 5 in the main text and Figure \ref{twisted_contouropt} here, where we report the corresponding contour plots of the marginal posteriors of $(\theta_1,\theta_2)$.
Obviously, in general we cannot stop an inference run at the iteration minimising the Wasserstein distance, as the true/reference marginal posteriors are assumed unknown. In the main text, we suggested to stop the inference once the acceptance rates turned smaller than 1.5\% for two consecutive iterations. Using this stopping criterion, \texttt{fullcondoptblocked}  returned a much improved inference (see Figure \ref{twisted_contour015}) than that in Figure \ref{twisted_contour}, the latter portraying  the marginal posteriors of $(\theta_1,\theta_2)$ at the last iteration (and thus at the smallest target threshold). Similar results are obtained when stopping the methods once the  acceptance rate goes below $1\%$ for the second consecutive time (figure not shown).

\begin{figure}[h!]
\includegraphics[width=\textwidth]{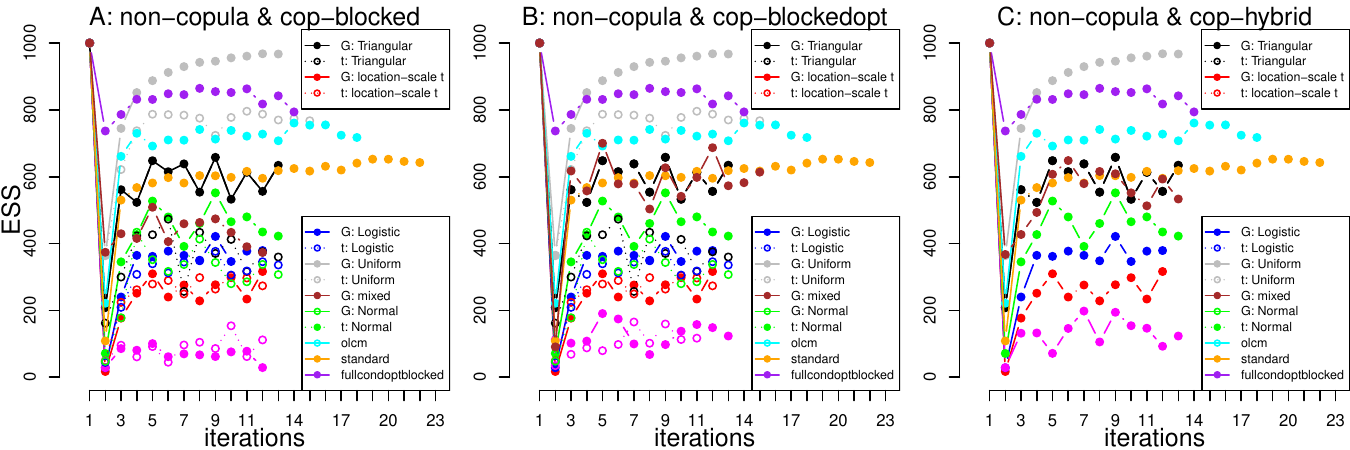}
\caption{Twisted model:  median ESS  across ten independent runs at each iteration. Left panel: \texttt{cop-blocked}; centre panel: \texttt{cop-blockedopt}, right panel: \texttt{cop-hybrid}. The copula-based methods are derived using either Gaussian (G, continuous lines) or t copulas (t, dashed lines) for different marginals, as described in the legend. In all panels, we also report the non-guided \texttt{olcm} (cyan) and \texttt{standard} (orange), and the guided \texttt{fullcondoptblocked} (purple).}
\label{fig:ESStwisted}
\end{figure}

When looking at the ESS in Figure \ref{fig:ESStwisted}, we see that 
\texttt{fullcondoptblocked} (purple lines) has higher ESS than the non-guided \texttt{olcm} and \texttt{standard}, with the latter having similar ESS to the \texttt{cop-blockedopt} and \texttt{cop-hybrid} approaches with Gaussian copula and triangular or mixed marginals. The copula-based guided approaches with uniform marginals have the highest ESS among all methods, but suffer from somehow overconfident inference, as observed from the contour plots in Figure \ref{twisted_contouropt}-\ref{twisted_contour015} and discussed in the main text. \textcolor{black}{This is successfully tackled by the mixed marginals, as visible in the contour plots, as well as in Figure 5, Panel D in the main text.}


\section{Hierarchical g-and-k model}

The g-and-k distribution is defined by its quantile
function $F^{-1}(z;\theta)$, where $F^{-1}(z;\theta):[0,1]\rightarrow \mathbb{R}$ is given by 
\begin{equation}
F^{-1}(z;\theta)= A+B\biggl[1+c\frac{1-\exp(-g\cdot r(z))}{1+\exp(-g\cdot r(z))}\biggr](1+r^2(z))^kr(z),
\label{eq:g-k-inverse}
\end{equation}
where $r(z)$ is the $z$th standard normal quantile, $A\in\mathbb{R}$ and $B>0$ are location and scale parameters, respectively, and $g\in\mathbb{R}$ and $k>-0.5$ are related to the distribution's skewness and kurtosis, respectively. 
An evaluation of \eqref{eq:g-k-inverse} returns a draw ($z$th quantile) from the g-and-k distribution. Alternatively, a  realization $x\sim gk(A,B,g,k)$ from this distribution can be   simulated by sampling a standard Gaussian draw $r^*\sim \mathcal{N}(0,1)$ and then plugging it in place of $r(z)$ in \eqref{eq:g-k-inverse}.   As we are interested into a \textit{hierarchical} g-and-k model, a single realization can be obtained as $x_{ij}\sim gk(A_i,B,g,k)$, ($i=1,...,n, j=1,...,J$).

\begin{figure}[h!]
\hspace{-1cm}    \begin{subfigure}[b]{0.6\textwidth}
  \includegraphics[scale=0.56]{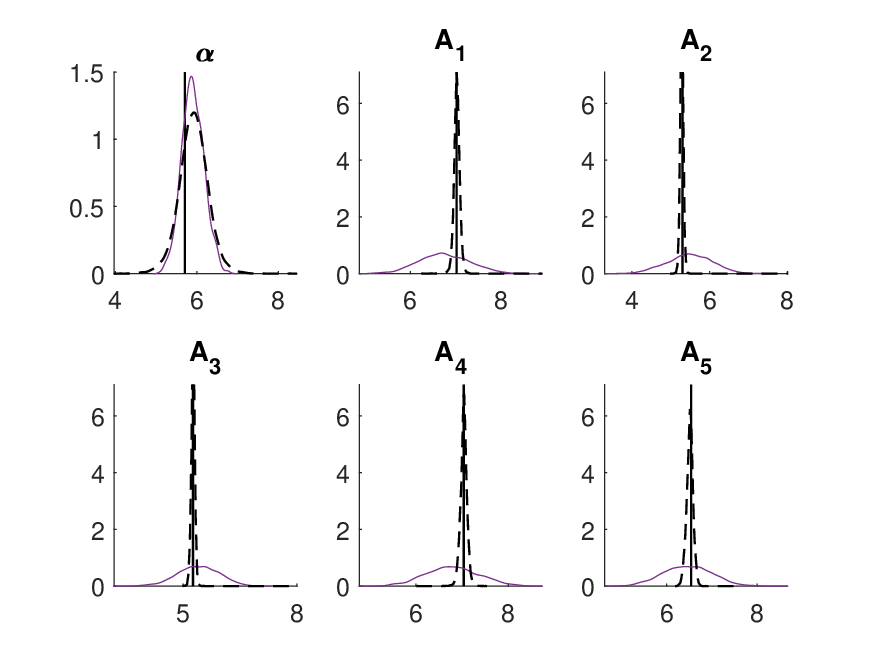}
\caption{Inference after 1 million model simulations.}    \end{subfigure}
    \begin{subfigure}[b]{0.5\textwidth}
\hspace{-2cm}    
\includegraphics[scale=0.43]{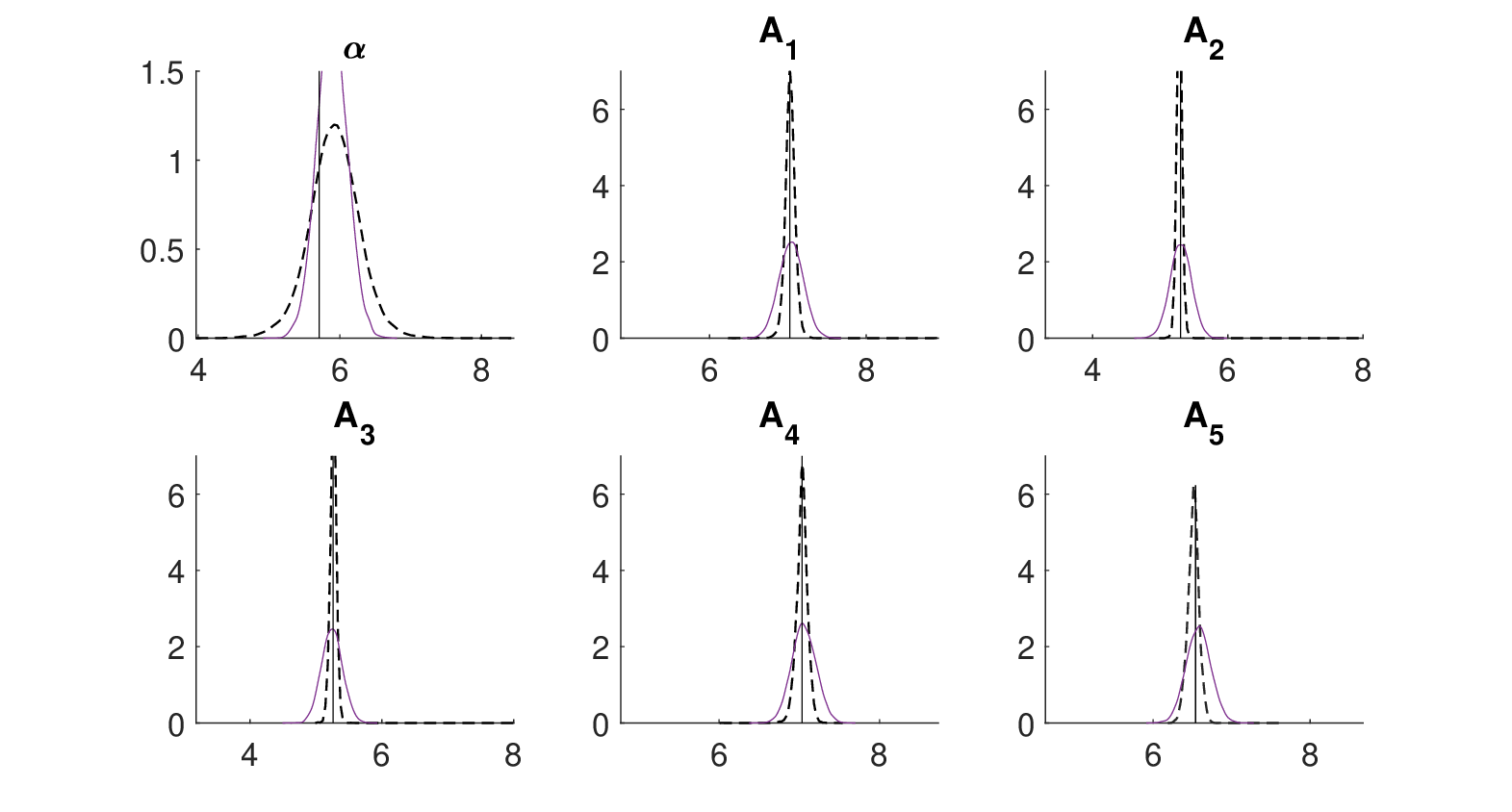}
\caption{Inference after 16 million model simulations.}    \end{subfigure}
    \caption{Hierarchical g-and-k model: marginal posteriors for $(\alpha,A_1,...,A_5)$ from \texttt{olcm} after: (a) 1 million model simulations (at iteration $t=22$, with $\delta=1.87$), which took about 9 hours to run; (b) 16 million model simulations (at iteration $t=41$ with $\delta=0.70$),  which took about 55 hours to run. Dashed lines are marginals from ABC-Gibbs. Black vertical lines mark ground-truth values.}
    \label{fig:gk_olcm}
\end{figure}

\begin{figure}[h!]
    \includegraphics[width=1.\textwidth]{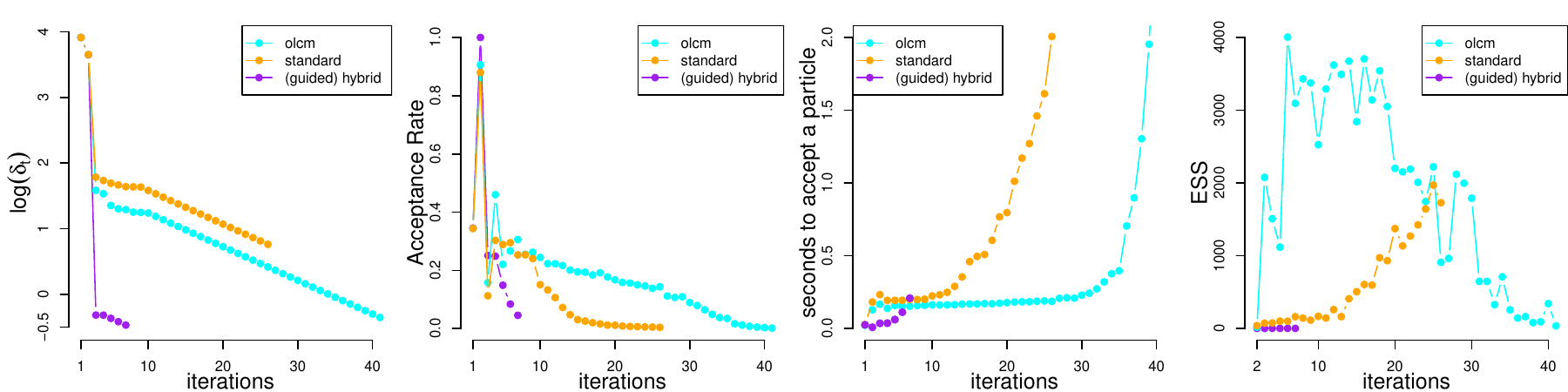}
    \caption{Hierarchical g-and-k model: $\log(\delta_t)$ (left panel), acceptance rate (center-left panel), number of seconds required to accept a particle (center-right panel) and  ESS (right panel) for \texttt{olcm, standard} and non-copula  \texttt{hybrid}. As \texttt{hybrid} is notably faster than the non-guided approaches, producing ten independent runs with this method was feasible, and hence for \texttt{hybrid} we report median quantities across the ten runs.}
    \label{fig:all_gk}
\end{figure}

In Figure \ref{fig:gk_olcm}, we report the marginal posteriors of $(\alpha,A_1,\ldots, A_5)$ from \texttt{olcm} obtained using about (a) 1 million simulations (same as \texttt{hybrid})  and 9 hours of computation; (b): 16 million iterations and 55 hours of computation.  While there is an improvement over the results from \texttt{standard}, the \texttt{olcm} marginal posteriors are wider than those from  ABC-Gibbs (and \texttt{hybrid},  as shown in the main text). Finally, the logarithm of the thresholds $\delta$, acceptance rates, seconds required to accept a particle and ESS are reported in Figure \ref{fig:all_gk}, with a discussion in the main text.

\section{Recruitment boom-and-bust with highly skewed summaries}\label{sec:boom-bust}

\begin{figure}[t!]
    \centering
    \includegraphics[scale=0.8]{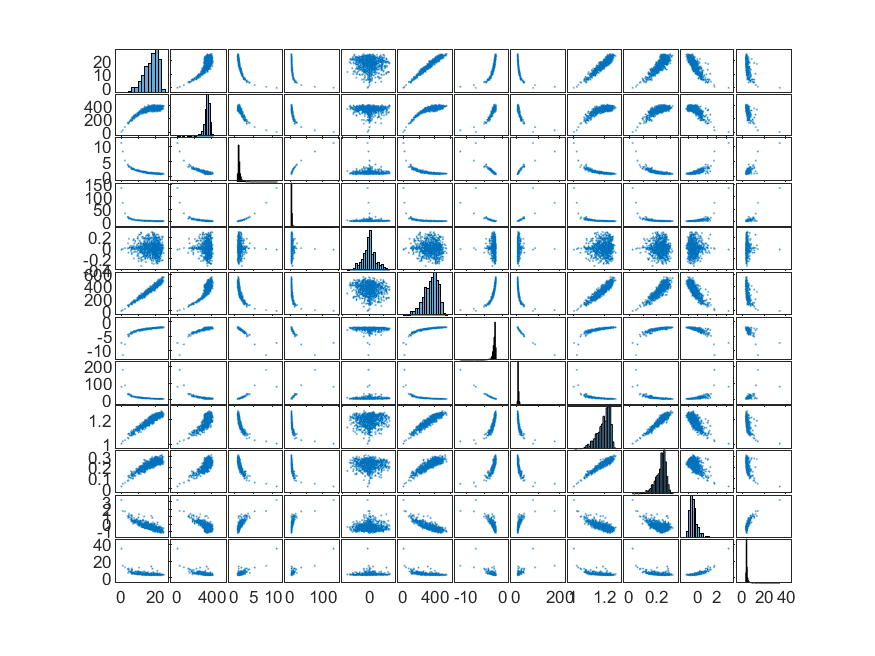}
    \caption{Boom-and-bust model: scatter plot matrix of 1,000 summaries simulated with $r=0.4$, $\kappa=50$, $\alpha=0.09$ and $\beta=0.05$. The diagonal reports the histogram for each dimension of the summary statistics and the other entries give the pairwise associations.}
    \label{fig:recruitment-simsum-scatter}
\end{figure}

 \begin{figure}[t!]
\hspace{-2cm}   \includegraphics[width=1.2\textwidth]{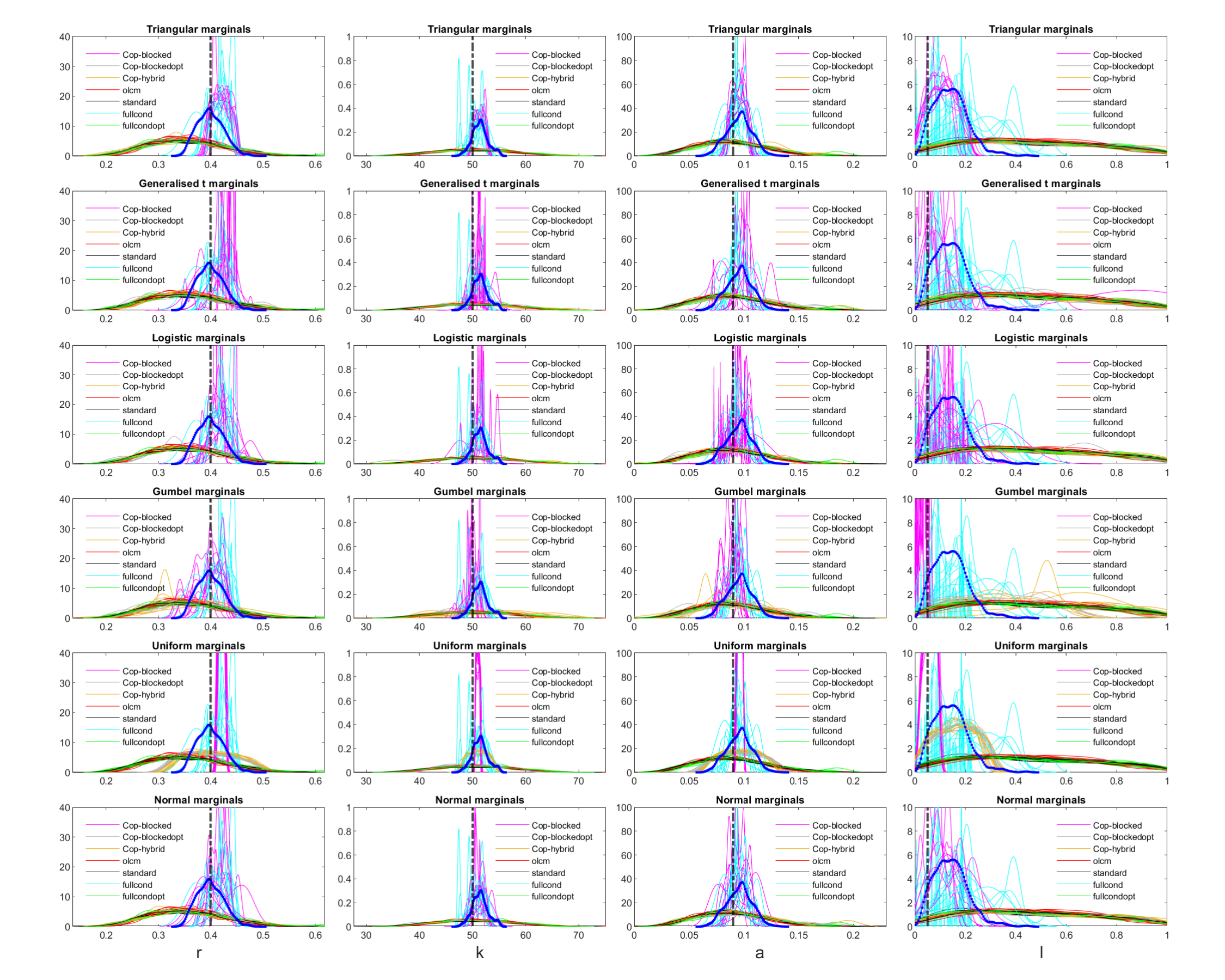}
\caption{Boom-and-bust model: marginal posterior densities at the last iteration in all ten estimation runs obtained with all methods and Gaussian copulas with all considered marginals. True parameters value are denoted by the vertical lines. Robust semiBSL (blue); \texttt{copula-blocked} (magenta); \texttt{copula-blockedopt} (gray); \texttt{copula-hybrid} (orange); \texttt{olcm} (red); \texttt{standard} (black); \texttt{fullcond} (cyan); \texttt{fullcondopt} (green). Remember that \texttt{copula-blocked} (magenta),  \texttt{copula-blockedopt} (gray), \texttt{copula-hybrid} (orange) with Gaussian copula and Gaussian marginals correspond to  \texttt{blocked}, \texttt{blockedopt} and  \texttt{hybrid}, respectively, \textcolor{black}{see Remark 2 in Section 3.4 of the paper}.}
    \label{fig:recruitment-marginals}
\end{figure}

\begin{figure}[t!]
    \includegraphics[width=1.\textwidth]{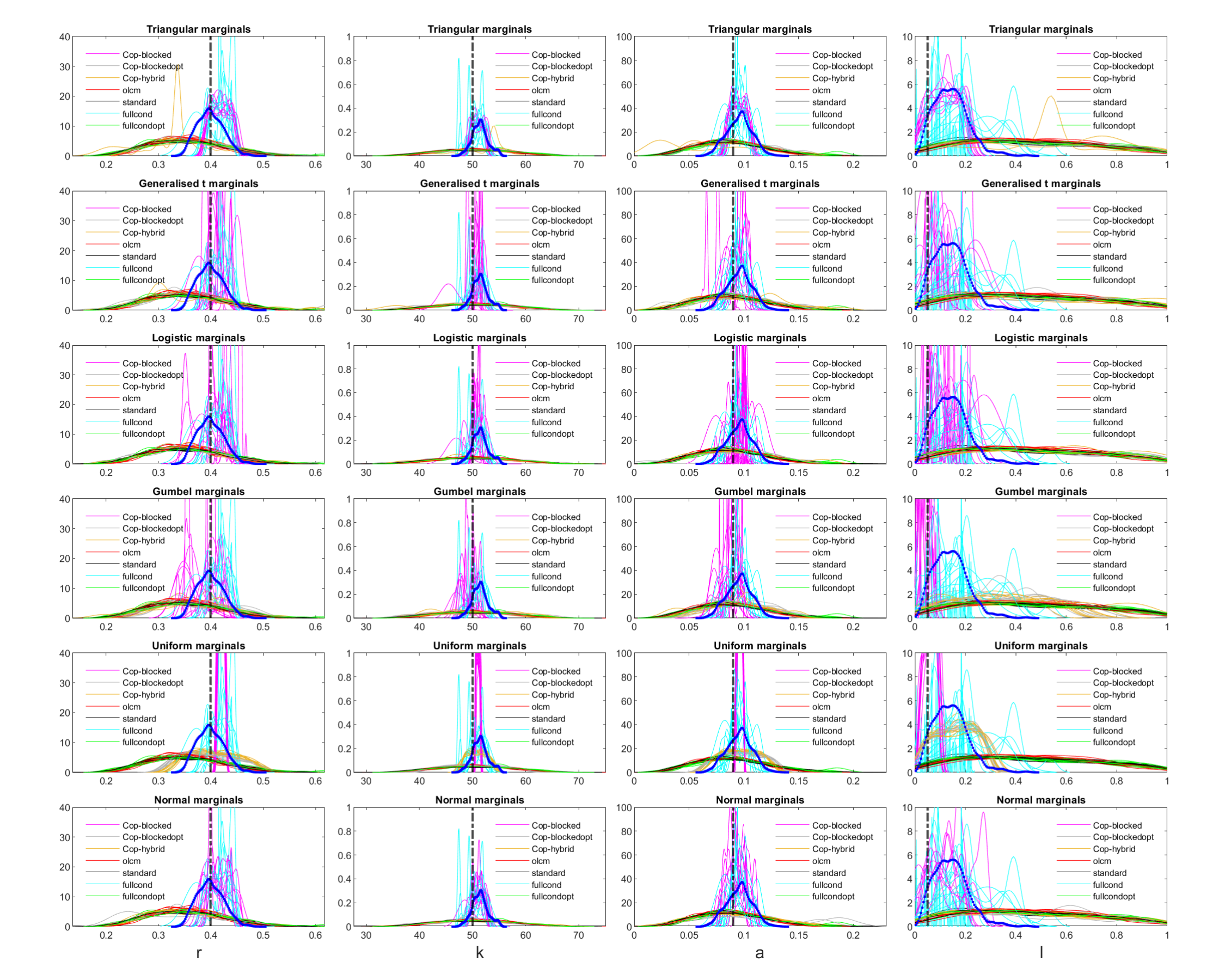}
\caption{Boom-and-bust model: marginal posterior densities at the last iteration in all ten estimation runs obtained with all methods and t copulas with triangular, uniform and normal marginals. True parameters value are denoted by the vertical lines. Robust semiBSL (blue); \texttt{copula-blocked} (magenta); \texttt{copula-blockedopt} (gray); \texttt{copula-hybrid} (orange); \texttt{olcm} (red); \texttt{standard} (black); \texttt{fullcond} (cyan); \texttt{fullcondopt} (green).} 
    \label{fig:recruitment-marginalst}
\end{figure}

We now challenge our guided proposals samplers, constructed assuming joint normality of the particle pairs $(\theta,s)$, on another example characterised by highly non-Gaussian summary statistics. This example was also considered in \cite{fasiolo2018extended}, \cite{an2018robust} and \cite{picchini2020adaptive} in the context of synthetic likelihood, to test how that methodology, constructed under approximately Gaussian distributed summary statistics, performed. The \textit{recruitment boom-and-bust model} is a  stochastic discrete time model  $(N_t)_{t\in\mathbb{N}}$ that 
can be used to describe the fluctuation of population sizes over time. Given the population size $N_t$ and parameter $\theta=(r,\kappa,\alpha,\beta)$, with $r>0$, $\kappa>0$, $\beta>0$, $\alpha\in(0,1)$, the population size $N_{t+1}$ at time $t+1$ follows the following distribution
\begin{align*}
		N_{t+1} \sim 
		\begin{cases}
			\mathrm{Poisson}(N_t(1+r)) + \epsilon_t, & \text{ if }\quad N_t \leq \kappa \\
			\mathrm{Binom}(N_t,\alpha) + \epsilon_t, & \text{ if}\quad N_t > \kappa
		\end{cases},
	\end{align*}
	where $\epsilon_t \sim \mathrm{Pois}(\beta)$ is a stochastic term. The population oscillates between high and low sizes for several cycles. Same as in \cite{an2018robust} and \cite{picchini2020adaptive}, true parameters are set to $r=0.4$, $\kappa=50$, $\alpha=0.09$ and $\beta=0.05$ and we assume a fixed and known initial size $N_1=10$. The chosen value of $\beta$ gives rise to highly non-Gaussian summaries, 
as illustrated in Figure \ref{fig:recruitment-simsum-scatter}.

We have the same modelling and data generation setup as in the previously cited references: the priors are set to $r\sim\mathrm{U}(0,1)$, $\kappa\sim\mathrm{U}(10,80)$, $\alpha\sim\mathrm{U}(0,1)$, $\beta\sim\mathrm{U}(0,1)$ and a dataset $y$ with 250 observations is obtained by simulating  values for the $(N_t)$ process for 300 steps and then discarding  the first $50$ (as in \citealp{fasiolo2018extended,picchini2020adaptive}). For a dataset $y$, we define differences and ratios as $\mathrm{diff}_{y} = \{y_i - y_{i-1} ; i=2,\ldots,250\}$ and  $\mathrm{ratio}_{y} = \{(y_i+1) / (y_{i-1}+1) ; i=2,\ldots,250\}$, respectively. We then consider the sample mean, variance, skewness and kurtosis of $y$,  $\mathrm{diff}_{y}$ and $\mathrm{ratio}_{y}$ as our summary statistic, for a total of twelve  summaries. 
	
Same as for the two-moons example, here we produce comparisons between methods based on prefixed values for $\delta_t$. In particular, we use tolerances $(\delta_1,..,\delta_{15})=(3.5,3,2.8,2.6,2.4,2.2,2.1,2,1.9,1.8,\\1.7,1.6,1.5,1.4,1.3)$ and always consider $N=1,000$ particles. In all experiments,  before starting any ABC algorithm, we first  collect 5,000 simulated summaries produced from the prior predictive distribution and then compute the mean-absolute-deviation (MAD) of each of the twelve-dimensional simulated summaries, which we use to standardise the difference between observed and simulated summaries, i.e. the ABC distances are   $(\sum_{j=1}^{12}((s_j^*-s_{y,i})/\textrm{MAD}_j)^2)^{1/2}$.
The ABC posteriors for ten independent runs, obtained with different methods, are reported in Figure  \ref{fig:recruitment-marginals}. As the true posterior is unavailable, and since ABC is notoriously underperforming when the size of the summaries is larger than a handful, here we consider the posterior obtained via the robustified semiparametric Bayesian synthetic likelihood (semiBSL)  of \cite{an2018robust}  as reference. This is because (i) BSL (\citealp{price2018bayesian}) can better handle summaries having large dimension, and (ii) compared to BSL, semiBSL has been shown to be more robust to departures from Gaussianity, as it is the case here (the highly non-Gaussian distribution of the summaries is in Figure \ref{fig:recruitment-simsum-scatter}). 

\begin{figure}[h!]
\hspace{-2cm}    \includegraphics[width=1.2\textwidth]{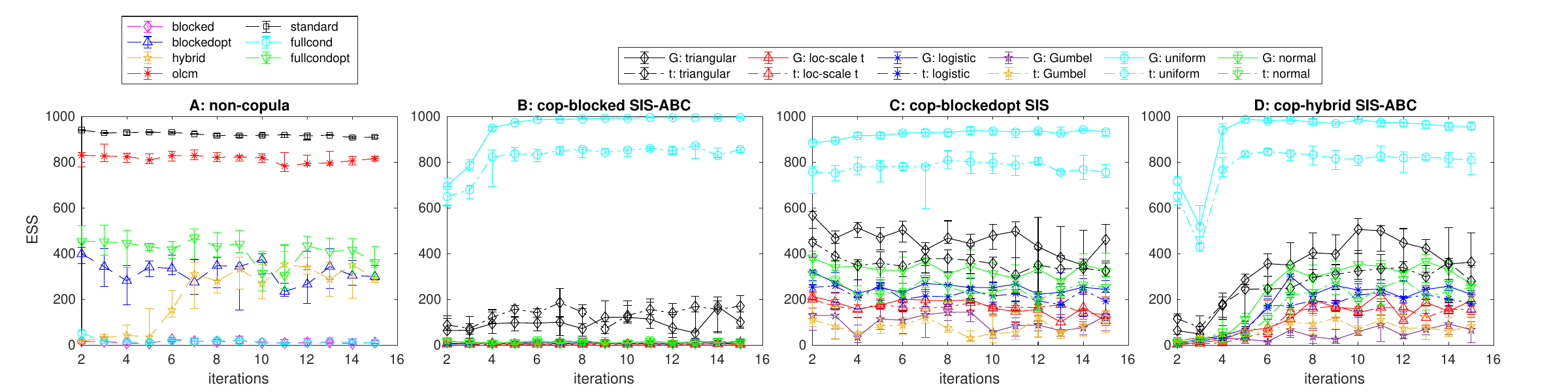}
    \caption{Boom-and-bust model: median ESS and error bars with  first and third quartile across ten independent runs at each iteration. Panel A:  Non-copula based methods, both guided and non-guided;
 Panel B:  \texttt{cop-blocked}; Panel C:  \texttt{cop-blockedopt}; Panel D: \texttt{cop-hybrid}. The copula-based methods are derived using either Gaussian (G) or t copulas for different marginals, as described in the legend.}
    \label{fig:ESS_recruitment}
\end{figure}

We run semiBSL for 3,000 MCMC iterations using 200 model simulations at each proposed parameter and we then disregard the first 1,000 iterations as burn-in. Compared to semiBSL, inference via ABC is expected to return wider marginals due to the nonparametric nature of ABC and the ``curse of dimensionality'' when the summaries' dimension increases. This does not happen for \texttt{blocked}, \texttt{copula-blocked} and \texttt{fullcond} (see Figure \ref{fig:recruitment-marginals} for Gaussian copula and the Figure \ref{fig:recruitment-marginalst} for t copula). However, this is because both \texttt{blocked}, its copula variant and \texttt{fullcond} can be ``mode-seeking'', that is they may overly explore the region around the mode, neglecting the tails, which motivated the creation of the ``optimized versions'' \texttt{blockedopt} and \texttt{fullcondopt}, as described in Sections 3.2 and 3.6 of the main text, respectively. However, the issue is less evident when choosing \texttt{cop-blocked} with triangular marginals as sampler (with either Gaussian or t copulas) \textcolor{black}{ or mixed marginals (results not shown)}, which benefit  from having a wider support than the uniform distribution (which is too ``overconfident'') and a higher probability of sampling within three standard deviations from the mean compared to other marginals, \textcolor{black}{as discussed in Supplementary Material \ref{AppendixA}.} Something interesting happens for \texttt{cop-blocked} and \texttt{cop-hybrid} with uniform marginals. While the former yields the most peaked ABC posterior marginals, the latter leads to ABC posteriors which are closer to those of the semiBSL than the other methods, benefiting from the mode-seeking behavior at iteration two  (coming from \texttt{cop-blocked}) followed by \texttt{cop-blockedopt}. All other methods have posteriors that are wider than those from semiBSL, as expected, and are also more consistent across the multiple independent runs, as observed in Figure \ref{fig:recruitment-marginals} and Figure \ref{fig:recruitment-marginalst}. Methods producing more consistent (i.e. less variable) posteriors across runs also return higher ESS values, see Figure \ref{fig:ESS_recruitment}.  

\begin{figure}[h!]
\hspace{-.5cm}\includegraphics[width=1.1\textwidth]{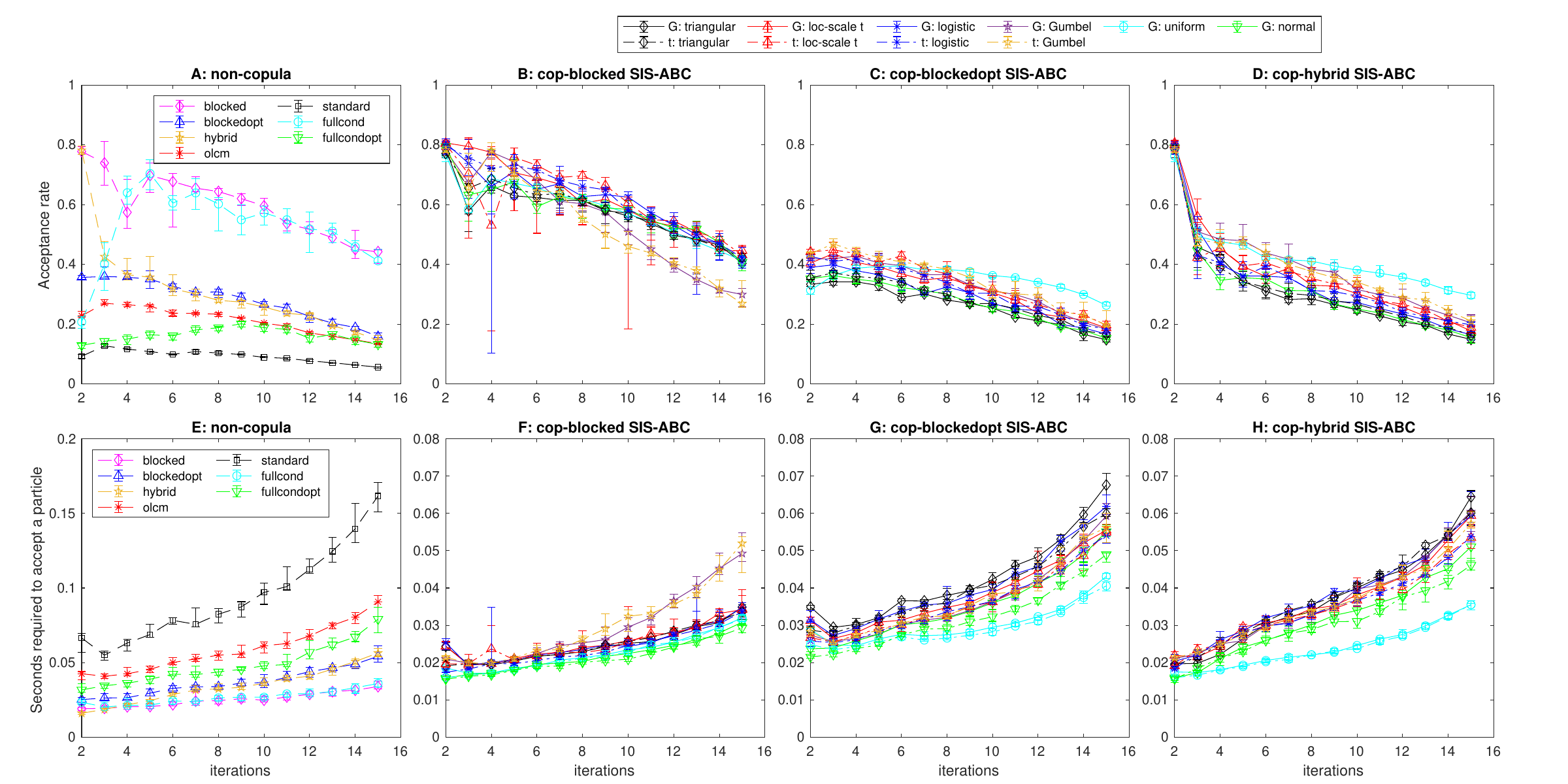}
   \caption{Boom-and-bust model:
median acceptance rates and error bars with  first and third quartile (top figures) and number of seconds required to accept a particle (bottom figures)  across ten independent runs. Panels A, E: non-copula based methods, both guided
and non-guided; Panels B, F: \texttt{cop-blocked}; Panels C, G: \texttt{cop-blockedopt}; Panels D, H: \texttt{cop-hybrid}. The copula-based methods are derived using either Gaussian (G) or t copulas for different
marginals, as described in the legend. Note: we had to choose different scales of the $y$-axes for comparing the number of seconds required to accept a particle for the non-copula vs copula-based methods, to make the latter more readable.}
    \label{fig:boombust-acceptrates}
\end{figure}
Note that narrower ABC posteriors could be obtained  by further reducing $\delta$, incurring in a higher computational cost though. Guided approaches show higher acceptance rates and are also faster in terms of runtime compared to the considered traditional approaches. This can be observed in Figure \ref{fig:boombust-acceptrates}, where the medians, the first and third quartile of the acceptance rates (top panels) and the number of seconds required to accept a particles (bottom panels) are reported. 
While we find that the guided \texttt{blocked}, \texttt{cop-blocked} and \texttt{fullcond} are uniformly the best ones in terms of both higher acceptance rate and lower runtimes, we should not forget that this is a consequence of the ``mode seeking'' behaviour of these methods, and the optimized  \texttt{blockedopt} (and its by-product \texttt{hybrid}) and \texttt{fullcondopt} may generally be preferred. The locally optimal guided \texttt{fullcondopt} and non-guided \texttt{olcm} have similar acceptance rates for small values of $\delta$, with the latter being slower than the former. Finally,  \texttt{standard} has the lowest acceptance rate, yielding then the largest runtime. Unlike for the two-moon case study, not only the wallclock times of the guided-samplers based on copulas are comparable to their corresponding Gaussian variants, but \texttt{cop-blockedopt} and \texttt{cop-hybrid} with uniform marginals yield higher acceptance rates and smaller runtimes than \texttt{blockedopt} and \texttt{hybrid}, respectively, see Figure \ref{fig:boombust-acceptrates}, third and fourth column.

\section{Lotka-Volterra}

\textcolor{black}{
The Lotka-Volterra model is a standard benchmark example for likelihood-free inference. It is typically used in population dynamics and systems biology to model the interaction between two animal species and two chemical compounds, respectively.
Its definition varies, depending on whether it is defined via
an ordinary differential equation (as in \citealp{lueckmann2021benchmarking}), a continuous time Markov Jump Process (MJP, \citealp{owen2015likelihood,owen2015scalable}) or a continuous time  approximation to the MJP given by a stochastic differential equation (as in \citep{golightly2011bayesian}). Moreover, it could be seen as a model involving three (as in \citealp{owen2015likelihood}) or four (as in \citealp{papamakarios2016fast}) reactions.
}

\textcolor{black}{Here, we consider a MJP Lotka-Volterra model, with the interaction between two \lq\lq species\rq\rq given by three-reactions involving three parameters $\theta=(\theta_1,\theta_2,\theta_3)$,  with a setup similar to \cite{owen2015likelihood}, though without perturbing the solution paths of the MJP with an additional measurement error. Denote by $(X^1_t)$ and $(X^2_t)$ the stochastic processes describing the population size of the two species evolving continuously at time $t\in [0,31]$. Available observations are the interpolations of the solution processes  at $n=32$ equidistant discrete time points $\{0,1,2,...,31\}$. 
Same as in \cite{owen2015likelihood}, data generating parameters are $\theta=(1,0.005,0.6)$, and uniform priors are set on the log-scale as $\log\theta_j\sim U(-6,2)$ ($j=1,2,3$), with deterministic starting values for the population sizes given by $X^1_0=50$ and $X^2_0=100$.
We refer to \cite{owen2015likelihood} for the specification of other quantities such as the stoichiometry matrix and the hazard function. Data are simulated exactly via the algorithm in \cite{gillespie1977exact}. 
As summary statistics for the available data of $(X^j_t)$, $j=1,2$, we used the nine summaries suggested in \cite{blog-wilkinson}, 
that is, the sample mean and auto-correlations at lags 1 and 2 of $(X_t^j)$, the sample log-variance of $(X_t^j+1)$ (1 is added to avoid taking the logarithm of zero) and  the cross-correlation between $(X_t^1)$ and $(X_t^2)$.
}

\begin{figure}[t!]
  \begin{subfigure}[b]{0.5\textwidth}
    \centering
    \includegraphics[width=\textwidth]{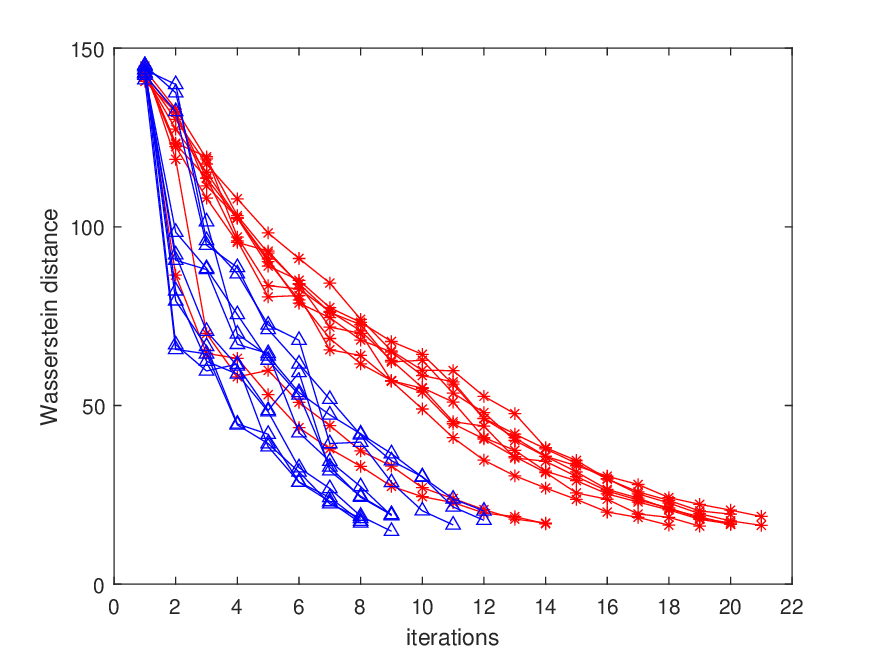}
    \caption{Wasserstein distances}
    \end{subfigure}
    \begin{subfigure} [b]{0.5\textwidth} 
    \centering
\includegraphics[width=\textwidth]{LV_acceptance.pdf}
    \caption{Acceptance rates}
    \end{subfigure}
    \caption{\textcolor{black}{Lotka-Volterra model. Panel (a): Wasserstein distances across iterations. Distances are between the reference posterior (obtained with a long run of the \texttt{standard} SMC-ABC sampler) and the ABC posteriors obtained with \texttt{blockedopt} (blue lines) and \texttt{olcm} (red lines) 
    for each of the 10 independent runs. Panel (b): 
    Acceptance rates across iterations for each of the 10 independent runs for \texttt{blockedopt} (blue lines) and \texttt{olcm} (red lines) over the ten runs.}}
    \label{fig:lv_wass-rates}
\end{figure}

\textcolor{black}{Here, we compare the guided method \texttt{blockedopt} (using a Gaussian proposal) with the non-guided \texttt{olcm} sampler. We used $N=2,000$ particles in all ten independent runs. Since exact inference is not possible,  when computing the Wasserstein distances we use as reference posterior the one
returned by the \texttt{standard} ABC-SMC sampler after a very long run. \texttt{standard} was let run until it reached a threshold smaller than 1 (namely  $\delta=0.97$), which required 45 iterations,  579,338 model simulations.  Instead, in each of the ten runs, \texttt{blockedopt} and \texttt{olcm} were halted once we obtained $\delta<3$ . 
The Wasserstein distances between the \texttt{standard} ABC posterior and the ABC posterior from \texttt{blockedopt} (blue lines) and \texttt{olcm} (red lines) are reported in Figure \ref{fig:lv_wass-rates}(a), with the corresponding acceptance rates in Figure \ref{fig:lv_wass-rates}(b).
These results show how \texttt{blockedopt} yields higher acceptance rates after a few iterations while having smaller Wasserstein distances from the very beginning. Moreover, \texttt{blockedopt} needs fewer iterations than \texttt{olcm} to reach the stopping criterion, leading to a much smaller number of model simulations, as observed in   Table \ref{tab:lv_metrics}, with
the proposed guided scheme needing less than half the number of the guided SMC-ABC in eight of the ten runs (namely runs 1, 3, 4, 5, 6, 7, 8, 9). This is also reflected in the runtime, as \texttt{blockedopt} is always faster than \texttt{olcm} other than in the second run (the higher time in the second run, despite the smaller number of simulations, may have several causes such as a longer time required for some model simulations due to sampled parameters yielding more events in the MJP, or simply some glitches in our desktop computer), with the inference across the ten runs completed in 564 minutes (with \texttt{blockedopt}) or 769 minutes (with \texttt{olcm}).
}

\begin{table}
    \centering
    \textcolor{black}{
    \begin{tabular}{clrrrr}
        run & & \texttt{blockedopt} & \texttt{olcm} & \texttt{blockedopt} & \texttt{olcm}\\
        \hline
        1 & num simulations: & \textbf{62,152} & 114,365
          &  runtime: \textbf{46} & 61\\
        2 & num simulations: & \textbf{66,883} & 94,782
          &  runtime: 201 & \textbf{187}\\
        3 & num simulations: &\textbf{46,888} & 114,035
          &  runtime: \textbf{37} & 61\\
        4 & num simulations: &\textbf{72,605} & 122,860
          &  runtime: \textbf{52} & 65\\
        5 & num simulations: &\textbf{48,110} & 99,242
          &  runtime: \textbf{35} & 52\\
        6 & num simulations:& \textbf{51,724} & 120,097
          &  runtime: \textbf{36} & 64\\
        7 & num simulations: &\textbf{68,752} & 113,712
          &  runtime: \textbf{51} & 60\\
        8 & num simulations: &\textbf{53,031} & 105,919
          &  runtime: \textbf{37} & 59\\
        9 & num simulations: &\textbf{49,793} & 114,643
          &  runtime: \textbf{37} & 93\\
        10 & num simulations:& \textbf{51,768} & 86,309
           & runtime: \textbf{41} & 67\\
        \hline
   \end{tabular}
    \caption{Lotka-Volterra: number of model simulations until a threshold $\delta$ smaller than 3 is obtained for the guided \texttt{blockedopt} and the non-guided \texttt{olcm} methods, and corresponding runtime in minutes.
    Best performances are in bold font.}
    \label{tab:lv_metrics}
    }
\end{table}

\section{Cell motility and proliferation}

\cite{price2018bayesian} show that their Bayesian synthetic likelihood (BSL) methodology, which is based on MCMC sampling, produces more accurate inference than an MCMC-ABC sampler when both are initialized very close to ground-truth parameters, a greatly simplified scenario. Here, we show that, even without initializing the samplers at the true parameters, our SMC-ABC guided samplers, on the one hand yeld inference very similar to BSL, and on the other hand are not effected by the considered large size of the summaries. Here, we use  the same model simulator, summary statistics and data as \cite{price2018bayesian}, providing here a short description of the model and data, referring to their paper for further details. 

Cell motility and proliferation are important parts of many biological processes. Cell motility causes random movement which, together with proliferation or reproduction, may cause tumors to spread or wounds to heal. In real-life experiments, to create the observed data, the cells can be placed on a two-dimensional discrete lattice using image analysis software and some manual intervention. However,  here we consider simulated data. Cells are motile, with the ability to randomly move to a neighboring lattice site, so they are equally likely to attempt a movement in any of the four directions of a two-dimensional surface (north, east, south, west). If the attempted movement is to a vacant site, then the motility event is successful. The movement is modelled using random walks in the four directions, and can be parameterized via the probability of motility and proliferation, $P_m$ and $P_p$, respectively, which are our parameter of interest. Let $X^t_{x,y}\in\{0,1\}$ be an indicator that defines whether a cell is present at position $(x,y)$ for $x\in\{1,...,R\}$, $y\in\{1,...,C\}$ at time index $t\in\{0,1,...,144\}$, which means every 5 minutes for 12 hours. Here,  $R$ and $C$ are the number of rows and columns in the lattice, respectively. Denote the matrix of indicators at time index $t$ as $X^t$. \cite{price2018bayesian} consider  the Hamming distance between $X^t$ and $X^{t-1}$, given by
\[
s_t = \sum_{x=1}^R\sum_{y=1}^C |X^t_{x,y}-X^{t-1}_{x,y}|,
\]
 as an informative summary statistic regarding motility. 
The summary statistic used to provide information regarding the proliferation is the total number of cells at the end of the experiment, denoted as $K$. Thus, the vector of summary statistics is given by $s=(s_1,...,s_{144},K)$ and has dimension 145.  The data are simulated with $P_m=0.36$ and $P_p=0.001$, with $R=27$ and $C=36$. Initially, 110 cells are placed randomly in the rectangle with positions $x\in\{1,2,...,13\}$ and $y\in\{1,2,...,36\}$. 

\begin{figure}[t]
    \centering
    \begin{subfigure}[b]{0.4\textwidth}
         \centering
         \includegraphics[width=\textwidth]{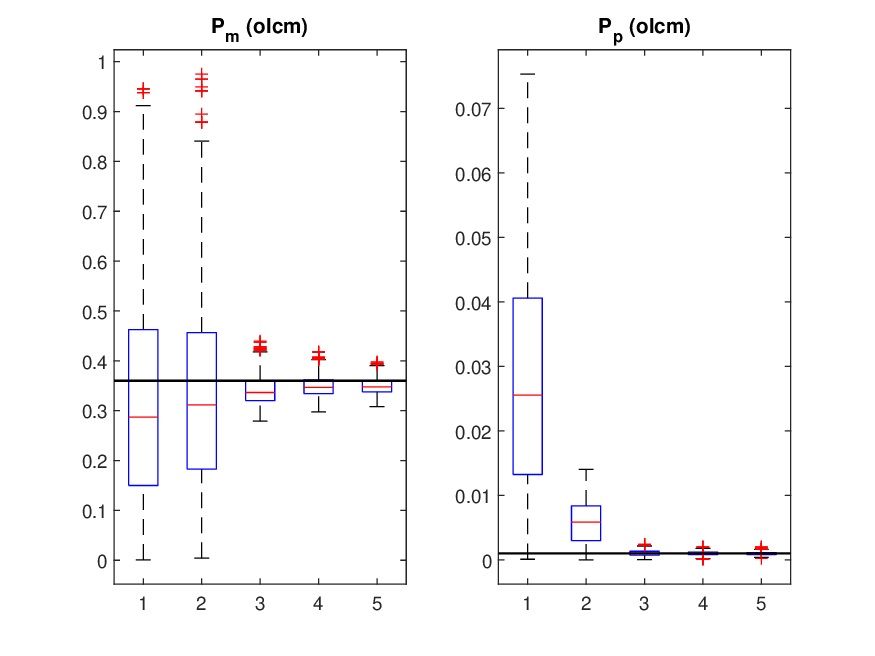}
         \caption{olcm}
     \end{subfigure}
        \begin{subfigure}[b]{0.4\textwidth}
         \centering
         \includegraphics[width=\textwidth]{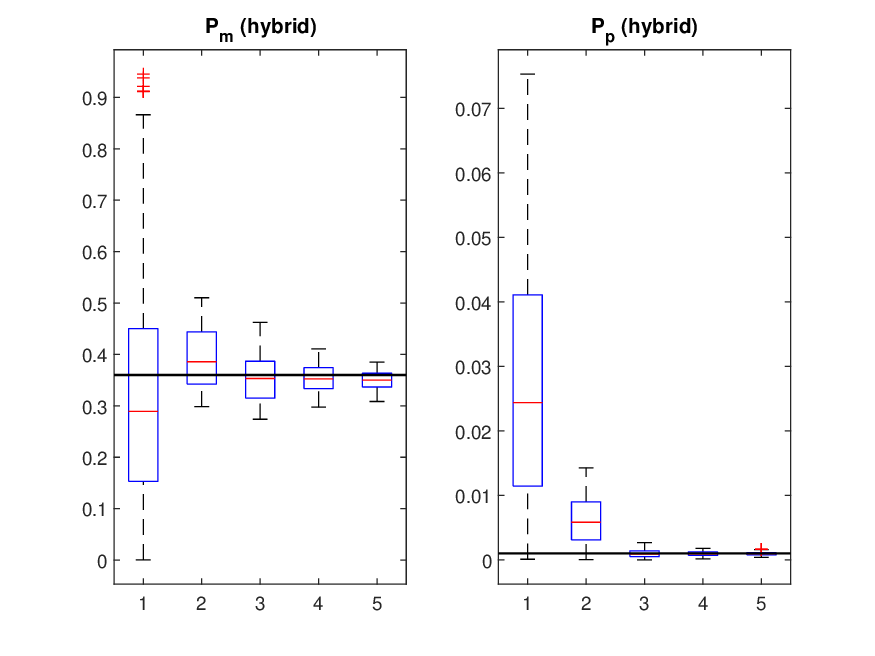}
         \caption{hybrid}
     \end{subfigure}
    \caption{Cell model using 145 summaries:  evolution of the marginal posteriors across five iterations of SMC-ABC when using (a) olcm; (b) hybrid. The horizontal lines are ground-truth values.}
    \label{fig:cell_boxplots}
\end{figure}

\begin{figure}[htbp]
    \centering
    \includegraphics[scale=.5]
    {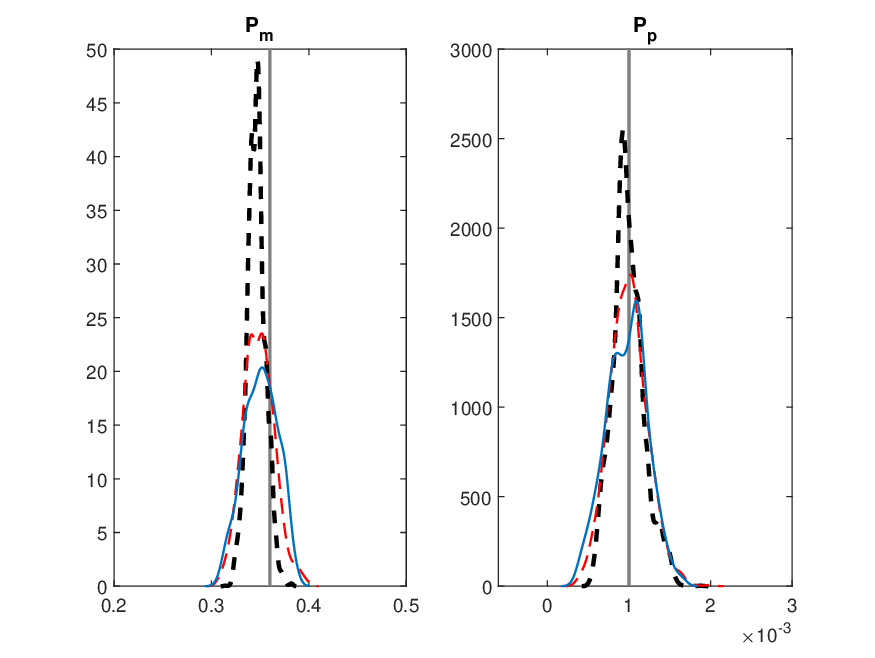}
    \caption{Cell model using 145 summaries:  marginal posteriors for BSL (black dashed, after a total of 25 million model simulations), olcm (red, after a total 132,963 simulations), guided hybrid (blue, after a total 117,496 model simulations). Vertical lines mark ground truth parameters.}
    \label{fig:cell_marginals}
\end{figure}

We compare our guided schemes with \texttt{olcm} and BSL (specifically the version denoted uBSL in \citealp{price2018bayesian}, but we keep writing BSL for simplicity). BSL is run using their provided code, where at each proposed parameter 5,000 model simulations are independently simulated. We run BSL for 5,000 iterations (hence a total 25 million simulations from the model), starting it at $P_m=0.30$ and $P_p=0.001$, with a code running in parallel on a computer node with four cores. In all experiments, we set priors $P_m\sim U(0,1)$ and $P_p\sim U(0,1)$. Here, we run the original (vanilla) version of BSL to replicate the results in \cite{price2018bayesian}, but it should be possible to reduce the computational demands, see below. 
Same as in the other case studies, before starting the ABC procedures,  we simulate summaries from the prior predictive (10,000 times in this example), and compute their MAD. No further tuning is applied to our algorithms, and in all cases the initial threshold was set to $\delta_1=788$ and was automatically decreased as in previous sections, using $\psi=1$ as percentile and $N=1,000$ particles. \textcolor{black}{We run SMC-ABC
with \texttt{olcm} and the guided Gaussian \texttt{hybrid}, and the report results corresponding to an acceptance rate
of approximately 1\%. For both \texttt{olcm} and \texttt{hybrid}, this happens at iteration 5, with automatically attained thresholds equal to $\delta_5=98.3$ and $\delta_5=97.1$ for \texttt{olcm} and \texttt{hybrid}, respectively. Despite reaching a smaller threshold, \texttt{hybrid} required overall
fewer model simulations, namely 117,496, versus the 132,963 of \texttt{olcm}, with a runtime of 6.1 minutes (\texttt{hybrid}) vs 8.1 minutes (\texttt{olcm}). While these differences are not striking, it is important to stress that they are for a single run only, and that they become more remarkable when we increase the size of the summary statistics in next section. Overall, we believe that having only two parameters to infer makes the problem ``simple enough'' for \texttt{olcm} not to struggle. Despite the ``curse of dimensionality'' afflicting ABC algorithms when the size of $s_y$ is large, the inference at the last iteration appears satisfying for both methods, when compared to the marginal returned by BSL, see Figure \ref{fig:cell_marginals}. Figure \ref{fig:cell_boxplots} shows the performance of SMC-ABC across five iterations of \texttt{olcm} and the \texttt{hybrid} method. Note the effect of the guiding mechanism kicking-in at iteration two for hybrid, which is especially evident for $P_m$.}

The considered setup for BSL required 16.6 hours to run (starting from the ground-truth parameters), however there exists  literature towards significantly reducing its computational demands, see for example \cite{everitt2017bootstrapped}, \cite{priddle2019efficient}, \cite{an2019accelerating} and \cite{picchini2020adaptive}. For example, by applying an appropriate adjustment to the original BSL procedure, \cite{an2019accelerating} managed to use 500-1,000 model simulations per iteration instead of 50,000. Here, the focus of our comparison with BSL is in terms of the produced posterior inference rather than accurate time comparisons with the best possible version of BSL. 

\subsection{Cell model: results with more than 400 summaries}\label{sec: cell-large-summaries}

\begin{figure}[h!]
\includegraphics[width=.5\textwidth]{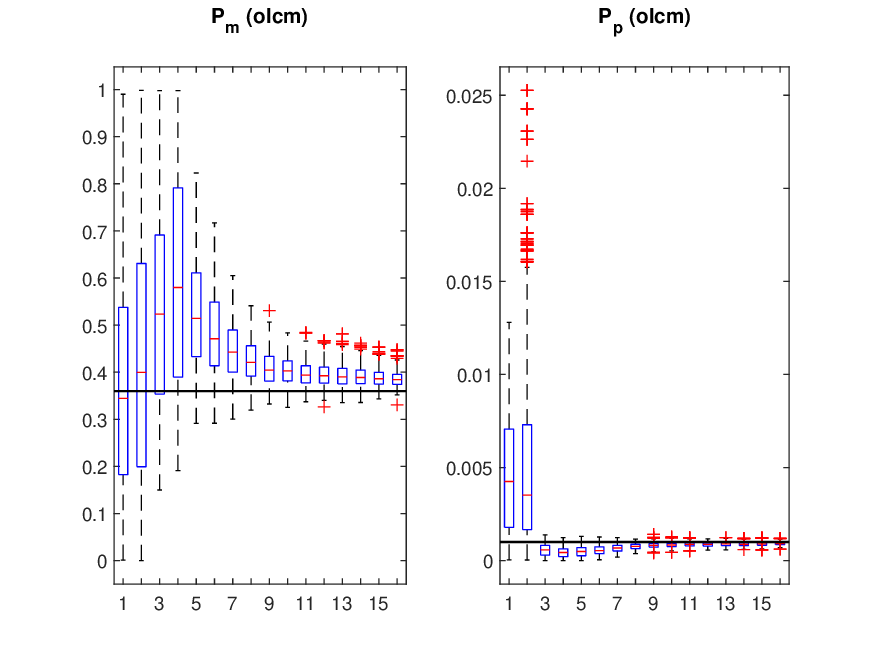}
\includegraphics[width=.5\textwidth]{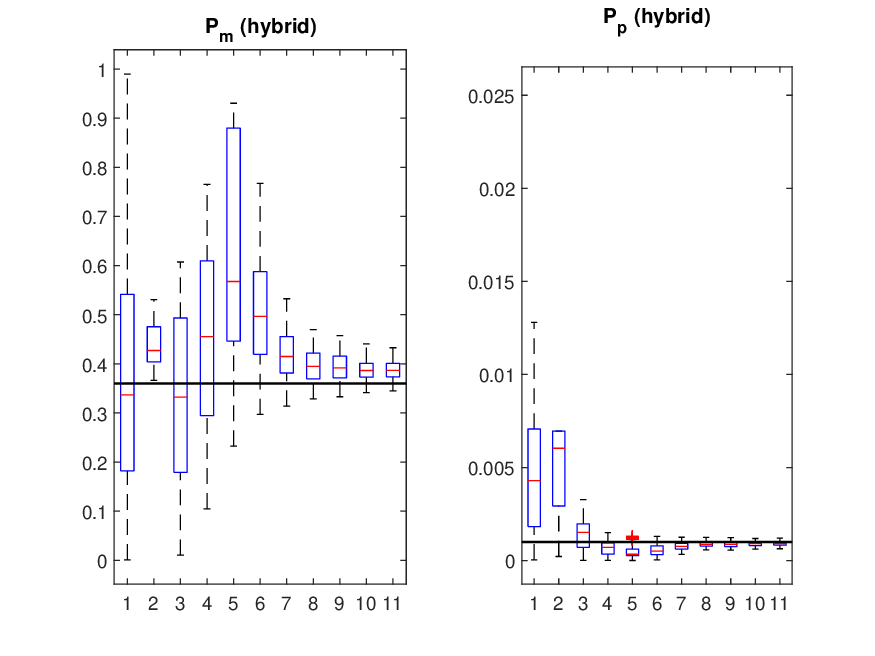}    
\caption{Cell model using 433 summaries:  evolution of the marginal posterior for $P_m$ and $P_p$  when using  \texttt{olcm} (the two leftmost panels) and the guided \texttt{hybrid} (the two rightmost panels). The horizontal lines are the ground-truth parameters. The number of iterations correspond to reaching an acceptance rate of around 2\%.}
    \label{fig:cell_boxplots_433summaries}
\end{figure}

\begin{figure}[h!]
    \centering
    \includegraphics[scale=.5]{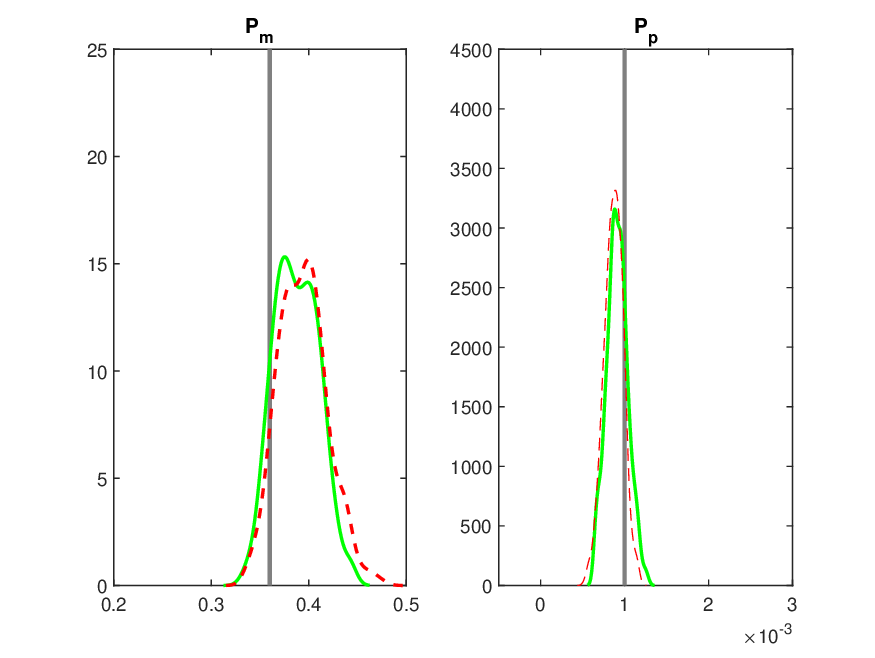}
    \caption{Cell model using 433 summaries:  marginal posteriors at the last iteration for olcm (red lines, $\delta=621.6$) and guided hybrid (blue lines, $\delta=619.4$). Vertical lines mark ground truth parameters.}
    \label{fig:cell_marginals_433summaries}
\end{figure}

Here, we simulate  recording cell positions for a longer time frame than before, considering images recorded every 5 minutes for a total of 24 hours, implying that $s_y$ has dimension 289. The vanilla BSL sampler  of \citealp{price2018bayesian} was again initialised at the same starting parameters as in the previous section, but it was unable to mix even with 10,000 model simulations per MCMC iteration, showing a very high rejection rate with the chains occasionally moving but often stuck for hundreds of iterations in the same position. Hence, no inference was produced, despite the initialization at ground-truth parameters. On the opposite side, sequential ABC strategies were able to handle this and even larger dimensions, namely summary $s_y$ with dimension 433, obtained by tweaking the simulation setup, while using the same positions for the 110 initial cells as before. We still use $N=1,000$ but, since the dimension of the summaries is larger,  here we use $\psi=25$ (the 25th percentile) to employ a more cautious decrease of $\delta$. 
\textcolor{black}{We start with $\delta_1=8,085$, run SMC-ABC with \texttt{olcm} and the guided Gaussian \texttt{hybrid}, reporting the results corresponding to an acceptance rate of around 2\%. This happens at iteration 16 ($\delta_{16}=621.6$) and 11 ($\delta_{11}=619.4$) for \texttt{olcm} and  \texttt{hybrid}, respectively. Also in this case, despite reaching a smaller threshold, \texttt{hybrid} required overall fewer model simulations than \texttt{olcm}, 213,984 vs 241,573 with a runtime of 25.2 minutes (\texttt{hybrid}) vs 32.4 minutes (\texttt{olcm}) for a single run. Inference results are in Figure \ref{fig:cell_boxplots_433summaries}. Especially for $P_p$, we notice that as soon as the guided procedure kicks-in at the second iteration, the bulk of the posterior shrinks much more than for the non-guided \texttt{olcm}. The marginals posteriors obtained at the last iteration are quite similar (Figure \ref{fig:cell_marginals_433summaries}), especially considering that the priors are very vague,  but again, those returned by the guided method have been obtained more efficiently.}

\section{Samplers based on a generic distance kernel}
In the main paper, we focused on considering a uniform kernel to compare simulated and observed summaries, i.e. the indicator function $\mathbb{I}_{||s^*-s_y||<\delta}$ for summaries $s^*$ of simulated data and summaries $s_y$  of observed data. While this is arguably the most popular choice to compare summaries, it is only one of the possibilities. For the reader's benefit, we add here a formulation of the basic ABC-rejection  when a generic kernel is used. For the SIS-ABC and SMC-ABC samplers, the reader is referred to \cite{sissonfan2018}.

A version of ABC-rejection, as given in \cite{wilkinson2013approximate},  proposes parameter $\theta^*$ yielding a distance $||s^*-s_y||$. Then, $\theta^*$  is accepted with probability
$K_\delta(s^*,s_y)/c$,  where $K_\delta$ is an unnormalized kernel and $c>0$ is a normalising constant such that $c\geq \sup K_\delta(||\cdot||)$. 
As discussed in \cite{wilkinson2013approximate}, this acceptance criterion is equivalent to accepting when $||s^*-s_y||<\delta$.
Typically, the kernel is a function of some distance $\rho=||\cdot||$ (e.g. the Euclidean or the Mahalanobis distance), and we expect $K_\delta(\rho)$ to have a modal value at $\rho=0$,  so taking $c= K_\delta(0)$ will make the algorithm valid and also ensure efficiency by maximizing the acceptance rate. 
For full generality, in Algorithm \ref{abc-gen} we report the version given in \cite{bornn2017use}, where at each proposed parameter, $M\geq 1$ datasets are simulated.
\begin{algorithm}
\small
\begin{algorithmic}[1]
  \For{$t=1$ to $N$}
    \Repeat 
   \State Sample $\theta^{*} \sim \pi(\theta)$,
   \State 
     Simulate $z^{i} \sim p(z| \theta^{*})$ iid and compute $s^{i}=S(z^{i})$, $i=1,\dots, M$;
   \State simulate $u \sim \mbox{Uniform}(0,1)$;
    \Until{$u <  \frac{1}{cM} \sum_{i=1}^M K_\delta(s^{i},s_y)$}
   \State Set $\theta^{(t)} = \theta^*$\;
  \EndFor
  \State
\noindent {\it Output:}\\
A set of parameters $(\theta^{(1)},\ldots,\theta^{(N)})\sim \pi_{\delta}(\theta|s_y)$.
  \end{algorithmic}
\caption{Generalized ABC-rejection (\citealp{bornn2017use})\label{abc-gen}}
\end{algorithm}

\section{Review of some proposal samplers for SMC-ABC}\label{sec:filippi-samplers}

Here, we review some of the proposal samplers for SMC-ABC that provide a useful comparison with our novel methods introduced in Section 3 of the main paper. Some of the most notable work for the construction of such samplers is in \cite{filippi2013optimality}. There, for the case where the interest is in constructing samplers based on perturbing a particle randomly sampled from the previous set/iteration, ways to tune the covariance matrix of Gaussian proposals that improve on  \cite{beaumont2009adaptive} are proposed. The goal in both \cite{beaumont2002approximate} and \cite{filippi2013optimality}  is to obtain a proposal sampler that is maximally similar to the targeted ABC posterior, in a Kullback-Leibler (KL) sense, and simultaneously maximizes the acceptance probability. 
See instead \cite{alsing2018optimal} for a different approach based on obtaining an optimal trade-off between a high acceptance rate and the reduction of the variance of the importance-weights of the accepted samples (hence increasing  the  effective  sample  size  for  the  accepted  sample). 

In particular,  \cite{beaumont2009adaptive} construct a ``component-wise'' perturbation kernel that at iteration $t$ proposes $\theta^{**}$ by separately perturbing each $\theta^{*}_k$ (the $k$th components of $\theta^{*}$) by sampling from the univariate Gaussian proposal
\[
q_k(\theta_k|\theta^*_k)\equiv \mathcal{N}(\theta^*_{k},\tau^2_k),
\]
where $\tau^2_k=2\sigma^2_k$, with $\sigma^2_k$ being the weighted empirical variance of the $k$th component of all particles accepted at iteration $t-1$. That this variance should be used in a component-wise approach comes as the optimal solution to the minimization of the KL divergence between the target $\pi_\delta(\theta|s_y)$ and the proposal kernel. Of course, at time $t$, all $d_\theta=\dim(\theta)$ components are perturbed to form $\theta^{**}=(\theta^{**}_1,...,\theta^{**}_{d_\theta})$. 
As expected, the component-wise approach is not particularly appealing when the components of $\theta$ are correlated in the posterior. A first intuitive amelioration considered in \cite{filippi2013optimality} is to propose the full $\theta^{**}$ by generating from the $d_\theta$-dimensional Gaussian $q(\theta|\theta^*)\equiv \mathcal{N}(\theta^*,2\Sigma)$ at iteration $t$, with $\Sigma$ being the empirical weighted covariance matrix from the particles accepted at iteration $t-1$. This results in the importance weights in equation (1) of the main text becoming (with the usual notation abuse) $\tilde{w}^{(i)}_t = \pi(\theta^{(i)}_t)/\sum_{j=1}^N {w}_{t-1}^{(j)}\mathcal{N}(\theta^{(i)}_t;\theta^{(j)}_{t-1},2\Sigma_{t-1})$, $i=1,...,N.$
In our experiments, this proposal embedded into SMC-ABC is denoted \texttt{standard}, being in some way a baseline approach. 
However, by appealing again to KL optimization arguments,  \cite{filippi2013optimality} remark that the \texttt{standard} sampler is optimal only in the limiting case $\delta_t=\delta_{t-1}=0$. As this is commonly not the case, they derived further multivariate Gaussian proposal samplers. Among others, one considers the  ``optimal local covariance matrix'' (\texttt{olcm}). The key feature of this sampler is that each proposed particle $\theta^{**}$ arises from perturbing $\theta^*$ using a Gaussian having a covariance matrix that is specific to $\theta^*$ (hence ``local''), rather than being tuned on all particles accepted at $t-1$. To construct  \texttt{olcm},  \cite{filippi2013optimality} define the following weighted set of particles of size $N_0\leq N$ at iteration $t-1$
\begin{equation}
\{\tilde{\theta}_{t-1,l},\gamma_{t-1,l}\}_{1\leq l\leq N_0}=\biggl\{\biggl({\theta}_{t-1}^{(i)}, \frac{w_{t-1}^{(i)}}{\bar{\gamma}_{t-1}} \biggr), \text{ s.t. } ||s^{i}_{t-1}-s_y||<\delta_t, \quad i=1,...,N \biggr\},
\end{equation}
where $\bar{\gamma}_{t-1}$ is a normalisation constant such that $\sum_{l=1}^{N_0}\gamma_{t-1,l}=1$.
That is, the $N_0$ weighted particles are the subset of the $N$ particles accepted at iteration $t-1$ (when using $\delta_{t-1}$) having generated summaries that produce distances that are \textit{also} smaller than $\delta_t$.
To avoid confusion, here we used the letter $\gamma$ to denote normalized weights associated to the subset of $N_0$ particles, rather than $w$. At iteration $t$, \texttt{olcm} is defined as a multivariate Gaussian sampler centred at $\theta^*$ with
$q_t(\theta|\theta^*)=\mathcal{N}(\theta^*,\Sigma^{\mathrm{olcm}}_{\theta^*})$
where
\[
\Sigma^{\mathrm{olcm}}_{\theta^*} = \sum_{l=1}^{N_0}\gamma_{t-1,l}(\tilde{\theta}_{t-1,l}-\theta^*)(\tilde{\theta}_{t-1,l}-\theta^*)'.
\]
It is clear that each $\theta^*$ is then perturbed using a covariance matrix that is specific to $\theta^*$, unlike the \texttt{standard} sampler which uses the same covariance for all perturbed particles. 
Accepted particles are given unnormalized weights
$$\tilde{w}^{(i)}_t = \pi(\theta^{(i)}_t)/\sum_{j=1}^N {w}_{t-1}^{(j)}\mathcal{N}(\theta^{(i)}_t;\theta^{(j)}_{t-1},\Sigma^{\mathrm{olcm}}_{\theta^{(j)}})$$ 
where 
\[
\Sigma^{\mathrm{olcm}}_{\theta^{(j)}} = \sum_{l=1}^{N_0}\gamma_{t-1,l}(\tilde{\theta}_{t-1,l}-\theta^{(j)}_{t-1})(\tilde{\theta}_{t-1,l}-\theta^{(j)}_{t-1})'.
\]
Note that the calculation of $\Sigma^{\mathrm{olcm}}_{\theta^{*}}$ causes some non-negligible overhead in the computations, since it has to be computed for every proposal parameter, and sanity checks on this matrix have to be performed to ensure that this is a proper covariance matrix, otherwise computer implementations will halt with an error. We discuss this in Section 3.2 in the main text.
For the implementation of \texttt{olcm}, we are required to store the $N$ distances $||s^{i}_{t-1}-s_y||$ accepted at iteration $t-1$, so that it is possible to determine which indeces $i$ have $||s^{i}_{t-1}-s_y||<\delta_t$. \texttt{olcm} implicitly embeds a stopping criterion for SMC-ABC, as clearly the sampler should stop at iteration $t$ when the current updated $\delta_t$ is such that $N_0=0$ (or possibly a positive but ``small enough'' $N_0$, though this aspect is not investigated in \citealp{filippi2013optimality}).

Finally, \cite{filippi2013optimality} introduce also further samplers, including one based on nearest neighbors which provides the highest acceptance rate but is computationally more demanding. Ultimately, they recommend \texttt{olcm} as an acceptable trade-off between sampling efficiency and computational demands.

\end{document}